\documentclass[aps,prc,preprint,superscriptaddress,showpacs,amssymb,amsmath,amsfonts,floatfix,nofootinbib]{revtex4-1}
\usepackage{graphicx}
\usepackage{natbib}
\usepackage[usenames]{color}
\usepackage{epsfig}
\usepackage{amssymb,amsmath}
\usepackage{subfig}
\usepackage{lineno,hyperref}
\usepackage{enumitem}

\definecolor{grey}{rgb}{0.9,0.9,0.9}
\definecolor{black}{rgb}{0,0,0}

\hypersetup{pdfpagemode=FullScreene}

\newcommand{\fpf}[2]{{F}_{#1}^{*}{F}_{#2}}

\def\sl#1{\slash{\hspace{-0.2 truecm}#1}}

\newcommand{\Tuzla}[1]
{\affiliation{University of Tuzla, Faculty of Natural Sciences and Mathematics, Univerzitetska 4, 75000 Tuzla, Bosnia and Herzegovina}}
\newcommand{\Mainz}[1]
{\affiliation{Institut f\"ur Kernphysik, Johannes Gutenberg-Universit\"at Mainz, D-55099
Mainz,Germany}}
\newcommand{\Zagreb}[1]
{\affiliation{Rudjer Bo\v{s}kovi\'{c} Institute, Bijeni\v{c}ka cesta 54, P.O. Box 180, 10002 Zagreb, Croatia}}

\begin{document}

\title{Fixed-t analyticity as a constraint in single energy partial wave analyses
of meson photoproduction reactions}

\author{H.~Osmanovi\'{c}}\thanks{hedim.osmanovic@untz.ba}\Tuzla \\
\author{M.~Had\v{z}imehmedovi\'{c}}\Tuzla \\
\author{R.~Omerovi\'{c}}\Tuzla \\
\author {J.~Stahov}\Tuzla\\
\author{V.~Kashevarov}\Mainz\\
\author{K.~Nikonov}\Mainz\\
\author{M.~Ostrick}\Mainz\\
\author{L.~Tiator}\Mainz \\
\author{A.~\v{S}varc}\Zagreb

\vspace{5cm}
\date{\today}

\begin{abstract}
Partial wave amplitudes of meson photoproduction reactions are an
important source of information in baryon spectroscopy. We
investigate a new approach in single-energy partial wave analyses of
these reactions. Instead of using a constraint to theoretical models
in order to achieve solutions which are continuous in energy, we
enforce the analyticity of the amplitudes at fixed values of the
Mandelstam variable $t$. We present an iterative procedure with
successive fixed-$t$ amplitude analyses which constrain the
single-energy partial wave analyses and apply this method to the
$\gamma p \to \eta p$ reaction. We use pseudo data, generated by the
EtaMAID model, to test the method and to analyze ambiguities.
Finally, we present an analytically constrained partial wave
analysis using experimental data for four polarization observables
recently measured at MAMI and GRAAL in the energy range from
threshold to $\sqrt{s}=1.85$~GeV.
\end{abstract}

\maketitle


\section{Introduction}
\label{sec:intro}
Photoproduction of pseudoscalar mesons is a
powerful tool in hadron spectroscopy. Excited baryons with definite
quantum numbers appear as resonances in the energy dependence of
partial-wave scattering or multipole amplitudes. The complex
interference pattern of these amplitudes manifests itself in four
different, energy and angular dependent helicity amplitudes and,
finally, in 16 single and double spin observables. The inverse
problem of uniquely reconstructing partial wave amplitudes up to an
overall phase from an experimentally measured subset of these
observables is an important prerequisite for a reliable isolation of
resonances and their separation from background.

Experimentally, major progress was observed during the last two
decades due to the extensive developments at ELSA, GRAAL, JLab and
MAMI in beam and target polarization techniques as well as in recoil
polarimetry (see e.g. \cite{Crede:2013} for a recent review).
Several spin observables with full angular coverage and fine energy
binning are already available for $\pi N$, $\eta N$, $\eta' N$ and
$KY$ final states.

Starting with the classical paper of Barker, Donnachie and
Storrow~\cite{Barker} the question was intensively investigated
which set of observables is necessary in order to reconstruct the
helicity amplitudes in a model independent way. Such an approach is
called complete experiment analysis (CEA) and it turned out that a
minimum of 8 observables is necessary, including three types of
polarization, beam, target and recoil. However, even with modern
data, such a CEA is at best possible only in a very limited
kinematic range. The general problem with the CEA is, however, that
it does not directly lead to partial waves, since an angle-dependent
overall phase remains unknown~\cite{Svarc:2017yuo}. Therefore, the
concept of a truncated partial wave analysis (TPWA) has been
extensively studied recently~\cite{Wunderlich:2014xya}, where only
partial waves up to an angular momentum of $L_{max}$ are analyzed
from the data, while all other higher partial waves are simply
expected to be zero. Such TPWA analyses have been studied for kaon
and pion photoproduction as well as for
electroproduction~\cite{Workman:2016irf,Tiator:2017cde}. It was
shown that even a minimum set of four polarization observables are
sufficient to extract all multipoles from numerical data, providing
that the data are practically free of errors and do not contain
higher partial wave contributions. In practical analyses, however,
the higher partial waves can not be completely ignored and at least
some next or next-to-next order needs to be included as a
constraint. In charged pion photoproduction, the well-known
pion-pole contribution gives such a constraint fairly model
independently, for other channels such constraints have to be
considered as model dependent. With such a TPWA analysis one also
ends up with continuum phase ambiguity \cite{Bowcock1975}, however, this phase is only
energy dependent.

In the past, for pion photoproduction model independent partial wave
analyses have been performed by using unitarity constraints. In the
Delta region this has been done in an approach using Watson's
theorem, where all multipole phases were fixed by unitarity to the
well known $\pi N$ phases~\cite{Hanstein:1997tp,Beck:1999ge}. Near
threshold, a PWA of $\pi^0$ photoproduction on the proton target
became possible with unitarity constraints from charged pion
photoproduction and well-known Born terms in the low-energy
region~\cite{Hornidge:2012ca}.

At higher energies, starting with the excitation of the Roper
resonance, such powerful constraints are no longer applicable.
However, analytical constraints from fixed-$t$ dispersion relations
can be applied to much higher
energies~\cite{Kamalov:2002wk,Aznauryan:2002gd,Aznauryan:2003zg}, as
long as the Mandelstam variable $t$ remains in the region
$-1.0$~GeV$^2 \lesssim t < 0$.

In general, unitarity in a coupled channel formalism would allow us
to constrain this phase. In the simplest case of pion
photoproduction at low energies, this is provided by Watson's
theorem. In other cases, this phase has to be fixed, e.g. to the
phase of a reaction model. A detailed study of model independent
single-energy TPWAs was performed for the $\gamma p \to K^+ \Lambda$
reaction ~\cite{Sandorfi:2010uv} using experimental data for 8
observables as well as pseudo-data generated from a model. In
general, at each energy many different, experimentally
indistinguishable, multipole solutions were obtained even if a
global, energy dependent phase was fixed. In order to obtain a
solution, which is continuous in energy, the fit can be constrained
to a parametrization of the amplitudes given by a  preferred
reaction model. This approach is commonly used in single-energy
partial wave analyses, however, it introduces a strong model
dependence.

To avoid or at least reduce this model dependence significantly, we
propose another method where continuity in energy is achieved by
enforcing the full analyticity of the amplitudes not in one, but in
two Mandelstam variables. Amplitudes at different energies are not
fully independent from each other but they are related by
analyticity in Mandelstam $s$ at a fixed value 
 of the Mandelstam variable $t$. The main idea of the
approach is to impose this analyticity in $s$ at fixed values of $t$
in addition to the analyticity in $t$ at fixed energy. The method
was fully developed and applied by H\"ohler in KH80 SE $\pi N$
elastic PWA~\cite{Hohler84}.

We apply our approach to the $\gamma p \to \eta p$ reaction in the
energy range from threshold up to center of mass energies of $W =
1.85$~GeV. In this range high precision data of the differential
cross section,  as well as on photon beam ($\Sigma$), target ($T$)
and beam-target ($F$) asymmetries in particular from MAMI
\cite{McNicoll:2010qk, Annand2014prl, Kashevarov2017}  and
GRAAL~\cite{Bartalini:2007fg} are available. As the $\eta$ meson is
isoscalar, no isospin separation of the amplitudes is necessary.

Different reaction models for eta photoproduction have been
developed. First of all they give an energy-dependent
parametrization of the partial waves, and second, they can then be
used for a (model-dependent) single-energy analysis. Isobar models
introduce nucleon resonances with Breit-Wigner
forms~\cite{etaMAID_2002,etaMAID_2003,Aznauryan:2003zg,Tryasuchev:2003st,Tryasuchev:2016xcr},
T-matrix methods parameterize partial wave amplitudes and search for
resonances in the fitted solutions~\cite{Paris:2010tz}.
Coupled-channels approaches with K-matrix
methods~\cite{Shklyar:2006xw,Anisovich:2012ct,Anisovich:2015tla} and
with meson-baryon dynamics~\cite{Ronchen:2015vfa} involve a series
of hadronic channels, where experimental data has been measured. At
higher energies, outside of the resonance region, Regge models have
been successfully
applied~\cite{etaMAID_2003,Sibirtsev:2010yj,Kashevarov:2017vyl}, and
for an analytical connection between resonance and Regge regions,
finite-energy sum rules have been investigated~\cite{Nys:2016vjz}.

In parts of our studies, in particular for pseudo data simulations,
the isobar model EtaMAID for $\eta$ and $\eta'$ photo and
electroproduction on nucleons~\cite{etaMAID_2002, etaMAID_2003} has
been applied. The model includes nucleon resonances in the $s$
channel parameterized with Breit-Wigner shapes and non-resonant
background. Recently, three updated versions were published and will
be addressed in our current work as I, II and III. In Version I,
EtaMAID-2015~\cite{Kashev2015}, the background consists of nucleon
Born terms in the $s$ and $u$ channels and the $\rho$ and $\omega$
meson exchange in the $t$ channel with pole-like Feynman
propagators. In Version II, EtaMAID-2016~\cite{Kashevarov2016}, the
background from Born terms was excluded because of very small
contribution. The background is described by vector and axial-vector
meson exchanges in the $t$ channel using the Regge cut
phenomenology. Version III, EtaMAID-2017~\cite{Kashevarov2017}, is
very close to Version II. In this version a fit procedure was mainly
improved and currently gives the best description of the
experimental data.

The paper is organized as follows. In section II we will first give
the basic formalism for kinematics, amplitudes and observables. In
section III we describe the method of imposing fixed-$t$ analyticity
to the partial wave analysis in an iterative procedure. As a test of
our new procedure, in section IV we perform an analytically
constrained partial wave analysis with input from pseudo data,
generated from a recent version of the EtaMAID model. And as a real
data application, in section V we analyze recent experimental data
for $\eta$ photoproduction. Finally, in section VI we discuss our
results and the remaining ambiguities due to limitations in the data
base and residual model uncertainties and give an outlook for future
developments.

\section{Formalism}

\subsection{\boldmath Kinematics in $\eta$ photoproduction}

For $\eta$ photoproduction on the nucleon, we consider the reaction
\begin{equation}
\gamma(k)+N(p_i)\rightarrow \eta(q)+N'(p_f)\,,
\end{equation}
where the variables in brackets denote the four-momenta of the
participating particles. These are $k^\mu=(k,\bold{k})$,
$q^\mu=(\omega,\bold{q})$ for photon and eta meson, and
$p_i^\mu=(E_i,\bold{p}_i)$, $p_f^\mu=(E_f,\bold{p}_f)$ for incoming
and outgoing nucleon, respectively. The familiar Mandelstam
variables are given as
\begin{equation}
s=W^2=(p_i+k)^2,\qquad t=(q-k)^2,\qquad u=(p_f-q)^2,
\end{equation}
the sum of the Mandelstam variables is given by the sum of the external masses
\begin{equation}
s+t+u=2m_N^2+m_{\eta}^2\,,
\end{equation}
where $m_N$ and $m_{\eta}$ are masses of proton and $\eta$ meson,
respectively. In the eta-nucleon center-of-mass (c.m.) system, we
have $\bold{p}_i=-\bold{k}$, $\bold{p}_f=-\bold{q}$, and the
energies and momenta can be related to the Mandelstam variable $s$
by
\begin{equation}
k=|\bold{k}|=\frac{s-m_N^2}{2\sqrt{s}},\quad
\omega=\frac{s+m_{\eta}^2-m_N^2}{2\sqrt{s}}\,,
\end{equation}
\begin{equation}
q=|\bold{q}|=\left[\left(\frac{s-m_{\eta}^2+m_N^2}{2\sqrt{s}}\right)-m_N^2\right]^{\frac{1}{2}}\,,
\end{equation}
\begin{equation}
E_i=\frac{s-m_N^2}{2\sqrt{s}},\quad
E_f=\frac{s+m_N^2+m_{\eta}^2}{2\sqrt{s}}\,,
\end{equation}
$W=\sqrt{s}$ is the c.m. energy. Furthermore, we will also refer to
the lab energy of the photon, $E=(s-m_N^2)/(2m_N)$.

Starting from the $s$-channel reaction $\gamma+N\Rightarrow \eta+N$,
using crossing relation, one obtains two other channels:
\begin{eqnarray}
\gamma +\eta\; &\Rightarrow&\; N+\bar{N}\qquad t-\text{channel}\,,\\
\eta + \bar{N}\; &\Rightarrow&\; \gamma + \bar{N}\qquad\;
u-\text{channel}\,.
\end{eqnarray}

All three channels defined above are described by a set of four
invariant amplitudes. The singularities of the invariant amplitudes
are defined by unitarity in $s$, $u$ and $t$ channels:
\begin{eqnarray}
 s-\text{channel cut:}\; &&  (m_N+m_{\eta})^2\le s <  \infty\,, \\
  \text{with unphysical cut:}\; && (m_{\pi}+m_N)^2\le s  \le (m_{\eta}+m_N)^2\,, \\
 u-\text{channel cut:}\; && (m_N+m_{\eta})^2\le u <  \infty\,, \\
  \text{with unphysical cut:}\; && (m_{\pi}+m_N)^2 \le u \le (m_{\eta}+m_N)^2\,,
\end{eqnarray}
and nucleon poles at $s=m_N^2$, $u=m_N^2\,$. The crossing
symmetrical variable is
\begin{equation}
\nu=\frac{s-u}{4m_N}\,.
\end{equation}

\begin{figure}[htb]
\begin{center}
\includegraphics[width=8.0cm]{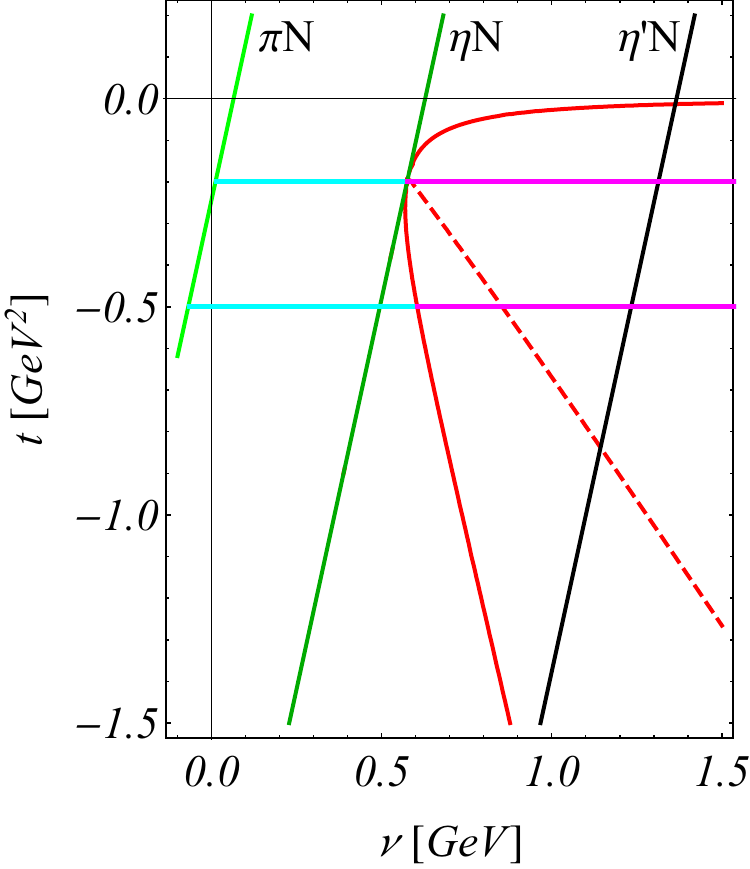}
\end{center}
\vspace{3mm} \caption{\label{KinematicalBuund}  The Mandelstam plane
for eta photoproduction on the nucleon. The red solid curves are the
boundaries of the physical region from $\theta=0$ to $\theta=180^0$
and the red dashed line shows $\theta=90^0$. The tilted vertical
lines from left to right are the thresholds for $\pi N, \eta N,
\eta\prime N$ production, respectively. The (blue) dotted line shows
the $u$-channel threshold $u=(m_N+m_\pi)^2$. The horizontal lines
denote the $t$-values $-0.2,-0.5$~GeV$^2$, where we will show our
fixed-$t$ analyses. The magenta parts give the part inside the
physical region, where the cyan parts have to be evaluated in the
unphysical region. The threshold values for $\gamma,\eta$ in $W$ are
$W_{thr}=1.486$~GeV ($t=-0.2$~GeV$^2$) and $W_{thr}=1.554$~GeV
($t=-0.5$~GeV$^2$). }
\end{figure}

The $s$-channel region is shown in Fig.~\ref{KinematicalBuund}. The
upper and lower boundaries of the physical region are given by the
scattering angles $\theta=0$ and $\theta=180^{\circ}$, respectively.
The horizontal lines at $t=-0.2$ and $-0.5$~GeV$^2$, show the
kinematical regions, where our fixed-$t$ analyses will be discussed
in details. The c.m. energy $W$ and the c.m. scattering angle
$\theta$ can be obtained from the variables $\nu$ and $t$ by
\begin{equation}
W^2=m_N(m_N+2\nu)-\frac{1}{2}(t-m_\eta^2)\,
\end{equation}
and
\begin{equation}
\mbox{cos}\,\theta\,= \frac{t-m_\eta^2 + 2\,k\,\omega}{2\,k\,q}\,.
\end{equation}

\subsection{Cross section and polarization observables}

\begin{figure}[h]
\begin{center}
\includegraphics[width=8.0cm]{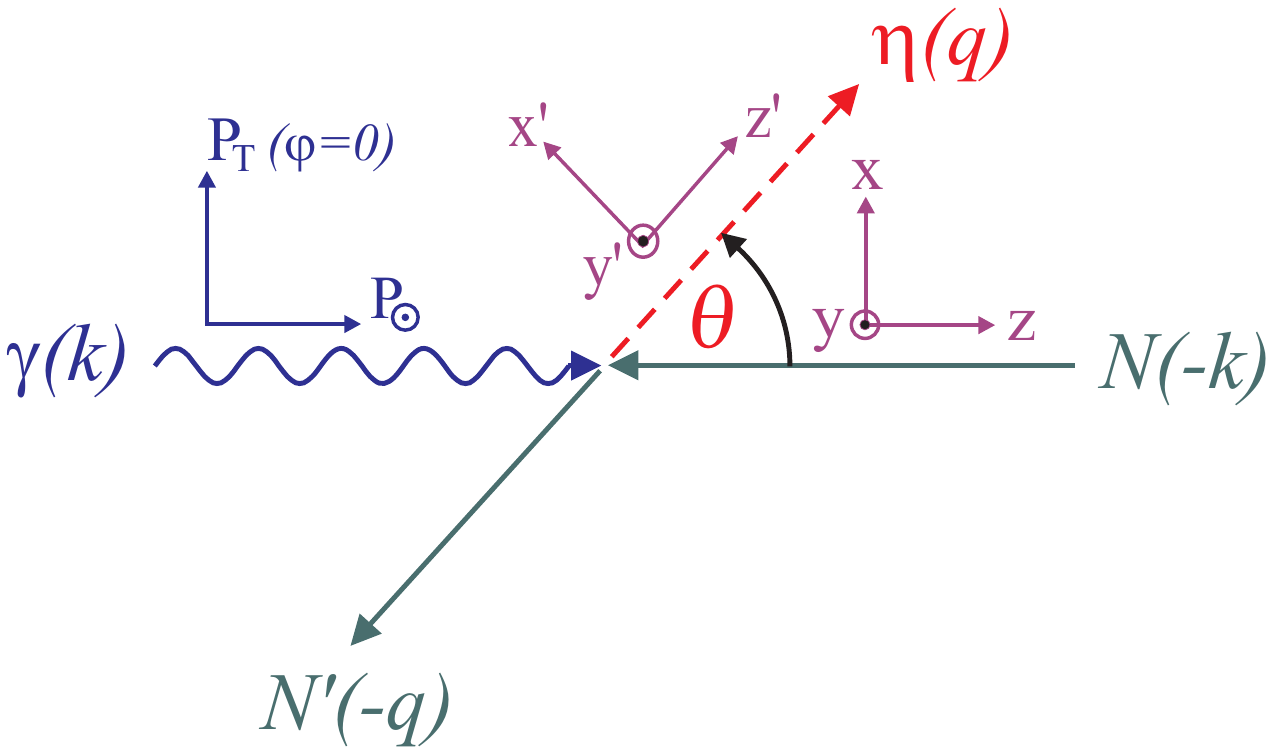} \vspace{3mm}
\caption{Kinematics of photoproduction and frames for polarization
vectors.}\label{fig:kin}
\end{center}
\end{figure}

Experiments with three types of polarization can be performed in
meson photoproduction: photon beam polarization, polarization of the
target nucleon and polarization of the recoil nucleon. Target
polarization will be described in the frame $\{ x, y, z \}$ in
Fig.~\ref{fig:kin}, with the $z$-axis pointing into the direction of
the photon momentum $\hat{ \bold{k}}$, the $y$-axis perpendicular to
the reaction plane, ${\hat{\bold{y}}} = {\hat{\bold{k}}} \times
{\hat{\bold{q}}} / \sin \theta$, and the $x$-axis given by
${\hat{\bold{x}}} = {\hat{\bold{y}}} \times {\hat{\bold{z}}}$. For
recoil polarization we will use the frame $\{ x', y', z' \}$, with
the $z'$-axis defined by the momentum vector of the outgoing meson
${\hat{\bold{q}}}$, the $y'$-axis as for target polarization and the
$x'$-axis given by ${\hat{\bold{x'}}} = {\hat{\bold{y'}}} \times
{\hat{\bold{z'}}}$.

The photon polarization can be linear or circular. For a linear
photon polarization $(P_T=1)$ in the reaction plane
$~\hat{\bold{x}}$ we get $\varphi=0$ and perpendicular, in direction
${\hat{\bold{y}}}$, the polarization angle is $\varphi=\pi/2$. For
right-handed circular polarization $P_{\odot}=+1$.

We may classify the differential cross sections by the three
classes of double polarization experiments and one class of triple
polarization experiments, which, however, do not give additional
information:
\begin{itemize}
\item polarized photons and polarized target
\end{itemize}
\begin{eqnarray}
\frac{d \sigma}{d \Omega} & = & \sigma_0
\left\{ 1 - P_T \Sigma \cos 2 \varphi \right. \nonumber \\
& & + P_x \left( - P_T H \sin 2 \varphi + P_{\odot} F \right)
\nonumber \\
& & + P_y \left( T - P_T P \cos 2 \varphi \right) \nonumber \\
& & \left. + P_z \left( P_T G \sin 2 \varphi - P_{\odot} E \right)
\right\} \, ,
\end{eqnarray}
\begin{itemize}
\item polarized photons and recoil polarization
\end{itemize}
\begin{eqnarray}
\frac{d \sigma}{d \Omega} & = & \sigma_0
\left\{ 1 - P_T \Sigma \cos 2 \varphi \right. \nonumber \\
& & + P_{x'} \left( -P_T O_{x'} \sin 2 \varphi - P_{\odot} C_{x'} \right)
\nonumber \\
& & + P_{y'} \left( P - P_T T \cos 2 \varphi \right) \nonumber \\
& & \left. + P_{z'} \left( -P_T O_{z'} \sin 2 \varphi  - P_{\odot}
C_{z'} \right) \right\} \, ,
\end{eqnarray}
\begin{itemize}
\item polarized target and recoil polarization
\end{itemize}
\begin{eqnarray}
\frac{d \sigma}{d \Omega} & = & \sigma_0 \left\{ 1 + P_{y} T + P_{y'} P
+ P_{x'} \left( P_x T_{x'} - P_{z} L_{x'} \right) \right. \nonumber \\
& & \left. + P_{y'} P_y \Sigma  + P_{z'}\left( P_x T_{z'} + P_{z}
L_{z'}\right) \right\}\,.
\end{eqnarray}

In these equations $\sigma_0$ denotes the unpolarized differential cross
section, the transverse degree of photon polarization is denoted by $P_T$,
$P_{\odot}$ is the right-handed circular photon polarization and $\varphi$ the
azimuthal angle of the photon polarization vector in respect to the reaction plane.
Instead of asymmetries, in the following we will often discuss the
product of the unpolarized cross section with the asymmetries and
will use the notation $\check{\Sigma}=\sigma_0\Sigma\,,
\check{T}=\sigma_0T\,,\cdots\,$. In the appendix we give expressions
of the observables in terms of CGLN and helicity amplitudes.

\subsection{\boldmath Invariant amplitudes and fixed-$t$ dispersion
relations}

The nucleon electromagnetic current for pseudoscalar meson
photoproduction can be expressed in terms of four invariant
amplitudes $A_i$~\cite{Chew:1957zz},
\begin{eqnarray}
J^\mu = \sum_{i=1}^4 A_i(\nu,t)\, M^\mu_i,
\end{eqnarray}
with the gauge-invariant four-vectors $M^\mu_i$ given by
\begin{eqnarray}
M^\mu_1&=&
-\frac{1}{2}i\gamma_5\left(\gamma^\mu\sl{k}-\sl{k}\gamma^\mu\right)\,
,
\nonumber\\
M^\mu_2&=&2i\gamma_5\left(P^\mu\, k\cdot(q-\frac{1}{2}k)-
(q-\frac{1}{2}k)^\mu\,k\cdot P\right)\, ,\nonumber\\
M^\mu_3&=&-i\gamma_5\left(\gamma^\mu\, k\cdot q
-\sl{k}q^\mu\right)\, ,\nonumber\\\
M^\mu_4&=&-2i\gamma_5\left(\gamma^\mu\, k\cdot P
-\sl{k}P^\mu\right)-2m_N \, M^\mu_1\, ,
 \label{eq:tensor}
\end{eqnarray}
where $P^\mu=(p_i^\mu+p_f^\mu)/2$ and the gamma matrices are defined
as in Ref.~\cite{Bjo65}.

Invariant amplitudes have definite crossing symmetry. For $\pi^0$
and $\eta$ photoproduction the amplitudes $A_{1,2,4}$ are crossing
even and $A_3$ is crossing odd.

In the work of Aznauryan~\cite{Aznauryan:2002gd,Aznauryan:2003zg} on
pion and eta photoproduction, a complete set of 8 invariant
amplitudes $B_i(\nu,t)$ and Dirac operators $N^\mu_i$ are
introduced, allowing any arbitrary current to be expanded in this
set of amplitudes:
\begin{eqnarray}
J^\mu = \sum_{i=1}^8 B_i(\nu,t)\, N^\mu_i\,.
\end{eqnarray}

If current conservation is implied, the 8 amplitudes are reduced to
6 amplitudes in electroproduction. In photoproduction, the set of
amplitudes is further reduced to four amplitudes, e.g. $B_1, B_2,
B_6, B_8$, which are then simple linear combinations of the four
$A_i(\nu,t)$ amplitudes $A_1, A_2, A_3, A_4$:
\begin{eqnarray}
B_1(\nu,t) &=& A_1(\nu,t)-2m_N A_4(\nu,t)\, ,\nonumber\\
B_2(\nu,t) &=& \frac{1}{2}(t-\mu^2) A_2(\nu,t)\, ,\nonumber\\
B_6(\nu,t) &=& -2 A_4(\nu,t)\, ,\nonumber\\
B_8(\nu,t) &=& -A_3(\nu,t)\, .
\end{eqnarray}

For the set of crossing symmetric amplitudes, $B=\{B_1, B_2, B_6,
B_8/\nu\}$, fixed-$t$ dispersion relations can be written in the
following form
\begin{equation}
{\rm Re} B(\nu,t) = B^{N}(\nu,t) +\frac{2}{\pi}\;{\cal
P}\hspace{-6pt}\int_{\nu_{thr}}^{\infty}{\rm d}\nu'\;
\frac{\nu'\,{\rm Im}B(\nu',t)}{\nu'^2-\nu^2}\,, \label{eq:dr1}
\end{equation}
where $B^{N}$ is the nucleon pole contribution that can be
calculated from the Born terms in pseudoscalar coupling, and
$\nu_{thr}$ corresponds to the $\pi N$ photoproduction threshold.

\subsection{CGLN and helicity amplitudes}

In partial wave analysis of pseudoscalar meson photoproduction it is
convenient to work with CGLN amplitudes  giving simple
representations in terms of electric and magnetic multipoles and
derivatives of Legendre polynomials
\begin{equation}
\begin{aligned}
F_{1} & =  \sum_{l=0}^{\infty}[(lM_{l+}+E_{l+})P'_{l+1}(x)+((l+1)M_{l+}+E_{l-})P'_{l-1}(x)]\,,\\
F_{2} & =  \sum_{l=1}^{\infty}[(l+1)M_{l+}+lM_{l-}]P_{l}'(x)\,,\\
F_{3} & =  \sum_{l=1}^{\infty}[(E_{l+}-M_{l+})P''_{l+1}+(E_{l-}+M_{l-})P''_{l-1}(x)]\,,\\
F_{4} & =
\sum_{l=2}^{\infty}[M_{l+}-E_{l+}-M_{l-}-E_{l-}]P_{l}''(x)\,,
\end{aligned}
\end{equation}
where $x=\cos\theta$ is the cosine of the scattering angle. Another
common set of amplitudes, which we will use in our current work, are
helicity amplitudes, linearly related to the CGLN amplitudes by
\begin{equation}\label{Ahel}
\begin{aligned}
H_{1} & =  -\frac{1}{\sqrt{2}}\sin\theta\cos\frac{\theta}{2}(F_{3}+F_{4})\,,\\
H_{2} & =  \sqrt{2}\cos\frac{\theta}{2}[(F_{2}-F_{1})+\frac{1-\cos\theta}{2}(F_{3}-F_{4})]\,,\\
H_{3} & =  \frac{1}{\sqrt{2}}\sin\theta\sin\frac{\theta}{2}(F_{3}-F_{4})\,,\\
H_{4} & =
\sqrt{2}\sin\frac{\theta}{2}[(F_{1}+F_{2})+\frac{1+\cos\theta}{2}(F_{3}+F_{4})]\,.
\end{aligned}
\end{equation}


\section{\boldmath Imposing fixed-$t$ analyticity}\label{Fixed-t}
This is the central part of our paper. Partly, it contains lost
knowledge in partial wave analysis of scattering data.

\subsection{Pietarinen's expansion method}
To introduce our method we follow Pietarinen's
papers~\cite{Pietarinen,Pietarinen1, Pietarinen2} and the review by Hamilton and
Petersen~\cite{HamiltonPetersen}. Consider a scattering amplitude
having the following analytic structure at a fixed-$t$ value in the
complex $\nu$-plane.
\begin{itemize}
 \item [i)] $F(\nu)$ is a real analytic function having a cut from $\nu_{th}$ to $\infty$ (physical
 cut).
 \item [ii)] $F(\nu)$ is bounded on the $\nu$ plane.
\end{itemize}
In most problems only information on $|F(\nu)|^2$ or some other
bilinear form are available, but, for simplicity of the description
of the method, let us suppose that we have the following information
on the amplitude $F(\nu)$:
\begin{itemize}
 \item [a)] Real parts $Re F(\nu_1)$, $\ldots$, $Re F(\nu_{M})$ at $M$ points $\nu_1$,$\ldots$, $\nu_M$ with errors $\varepsilon_1,\ldots,\varepsilon_M$.
 \item [b)] Imaginary parts $Im F(\nu_{M+1})$, $\ldots$, $Im F(\nu_{M+N})$  at $N$ points $\nu_{M+1},\ldots,\nu_{M+N}$ with errors $\varepsilon_{M+1},\ldots,\varepsilon_{M+N}$.\
\end{itemize}
The task is to find an approximant $\varphi(\nu)$ of the function
$F(\nu)$ having the same analytic structure. The standard procedure
is to find a minimum of the quadratic form
\begin{equation}\label{eqChi1}
\chi^2(\varphi)=\sum_{i=1}^M\left(\frac{Re \varphi(\nu_i)-Re
F(\nu_i)}{\varepsilon_i}\right)^2+\sum_{i=M+1}^{N+M}\left(\frac{Im
\varphi(\nu_i)-Im F(\nu_i)}{\varepsilon_i}\right)^2.
\end{equation}
There are  many approximants giving  very small $\chi^2$. Most of
them are non-smooth inside or outside of the region where the data
on $F(\nu)$ are available. The problem to find an optimal one
consists of finding a compromise between a good fit to the data and
the smoothness of the approximant. The standard approach is to
introduce a penalty function $\Phi(\varphi)$ (also known as a
convergence test function) which makes a choice of the smoothest
approximant that has an acceptable $\chi^2$. Such an approximant is
obtained by finding a minimum of the quadratic form
\begin{equation}\label{CHIC}
\mathalpha{X}^2=\chi^2(\varphi) +\Phi(\varphi),
\end{equation}
where $\chi^2(\varphi)$ is defined by Eq.~(\ref{eqChi1}). The form
of the penalty function $\Phi(\varphi)$ is not unique. To find
$\Phi(\varphi)$, Pietarinen \cite{Pietarinen} proceeded as follows
\begin{enumerate}
 \item Conformal mapping \begin{equation}\label{confmapp}
                          z=\frac{\alpha -\sqrt{\nu_{th}-\nu}}{\alpha +\sqrt {\nu_{th}-\nu}},\quad \alpha  \in \mathbb{R},\alpha > 0
                         \end{equation}
transforms a complex $\nu$ plane with physical cut along
$\nu_{th}\le\nu <\infty$ into unit circle $|z|=1$. The physical cut
is mapped on $|z|=1$.
\item Any function having properties i) and ii) may be represented in the complex  $z$-plane by a convergent series
\begin{equation}\label{zed}
\varphi (z)=\sum_{n=0}^{\infty}c_n z^n,
\end{equation}
which preserves the analytic properties of the approximant $\varphi$
in the $\nu$ plane. As a consequence of real analyticity of the
amplitude $F(\nu)$, the coefficients $c_n$ are real.
\end{enumerate}
Using arguments from complex analysis (theory of functions) and
probability theory, Pietarinen found a penalty function for the
approximant (\ref{zed}) in the form
\begin{equation}\label{penfun}
\Phi(\varphi)=\lambda \sum_{n=0}^{\infty}(n+1)^3c_n^2\,,
\end{equation}
where $\lambda$ is a real scaling parameter (weight factor). It can
be shown that higher coefficients $c_n$ in expansion (\ref{penfun})
are suppressed and behave as
\begin{equation}
|c_i|\le \frac{1}{n^{\frac{3}{2}}\lambda}\,.
\end{equation}
Therefore, expansion~(\ref{zed}) may be truncated at some finite
order $N_{max}$. With the approximant in the form~(\ref{zed}) and
penalty function $\Phi$ in the form~(\ref{penfun}), the minimization
of $\mathalpha{X}^2$ is a compromise between fitting the data and
keeping higher coefficients in (\ref{zed}) small. With finite
numbers of coefficients in expansion (\ref{zed}) and data on
$F(\nu)$ as described in a) and b), the minimization of
$\mathalpha{X}^2$ consists of solving a system of $N_{max}$ linear
equations, which may be performed fast and reliably. Following
H\"ohler \cite{Hohler84} (Appendix A6.3.4, page 536), the representation (\ref{zed}) with
penalty (\ref{penfun}) is called Pietarinen's expansion and the
method Pietarinen's expansion method. As concluding remarks
concerning Pietarinen's expansion we want to point out:
\begin{enumerate}
 \item Due to conformal mapping in the form (\ref{confmapp}),
 a polynomial expansion of any order $n\ge 1$ reconstructs the analytic
 structure of the amplitude at a fixed value of $t$, i.e. fixed-$t$ analyticity.
 \item The conformal variable $z$ can be defined in such a way to assure the correct
  crossing property of the scattering amplitudes.
 \item The asymptotical behavior of the amplitudes can be imposed
 explicitly.
 \item Technically, the expansion (\ref{zed}) can be evaluated fast and reliably using nested multiplication.
\end{enumerate}
Due to the bilinear structure of the relations between observables
and invariant amplitudes, the minimization of $\mathalpha{X}^2$
becomes nonlinear, therefore much more demanding. Observables are
expressed in terms of several amplitudes (depending on spin and
isospin structure of the particular process). Each of the amplitudes
is represented by its own representation (\ref{zed}) and all
coefficients are to be determined simultaneously. According to
Pietarinen, one has to minimize a quadratic form
\begin{equation}
 \mathalpha{X}^2 = \chi^2_{data}+\Phi,
\end{equation}
where the penalty function $\Phi$ consists of a sum of forms
(\ref{penfun}), one for each amplitude.

\subsection{\boldmath Representation of invariant amplitudes in $\eta$ photoproduction fixed-$t$ amplitude analysis}

\begin{figure}[!h]
\includegraphics[width=8.0cm]{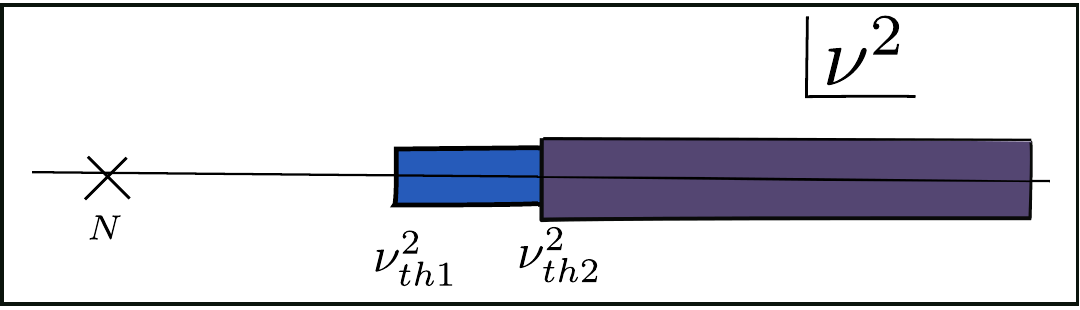}
\caption{\label{Fig:nu2} Analytic structure of invariant amplitudes
for a fixed-$t$ value in the complex $\nu^2$-plane.}
\end{figure}

As stated above, $\eta$ photoproduction can be described by four
independent, crossing symmetric amplitudes
 $B_{1}$, $B_{2}$, $B_{6}$, and $B_{8}/\nu$.
Their analytic structure for fixed-$t$ values (fixed-$t$
analyticity) in the complex $\nu^2$ plane is shown in
Fig.~\ref{Fig:nu2}.
The values $\nu_{th1}$ and $\nu_{th2}$ correspond to the thresholds
of $\pi p$ ($W_{\pi,th}=1.077$~GeV) and $\eta p$
($W_{\eta,th}=1.486$~GeV), respectively. $N$ is the nucleon pole. To
represent an analytic function having two branch points and two
corresponding cuts we use two conformal mappings
\begin{equation}
z_{1}=\frac{\alpha_{1}-\sqrt{\nu_{th1}^{2}-\nu^{2}}}{\alpha_{1}+\sqrt{\nu_{th1}^{2}
-\nu^{2}}}\,,\qquad
z_{2}=\frac{\alpha_{2}-\sqrt{\nu_{th2}^{2}-\nu^{2}}}{\alpha_{2}+\sqrt{\nu_{th2}^{2}-\nu^{2}}}\,,\label{2}
\end{equation}
where $\alpha_{1},\alpha_{2}$ are real positive parameters. First,
$z_1$ maps the $\nu^2$ plane into the unit circle $|z_{1}|\le1$.
Values on the physical cut $(\nu^2_{th1},\infty)$ are mapped onto
the unit circle $|z_{1}|=1$. Second, $z_2$ maps the values on the
physical cut $(\nu^2_{th2},\infty)$ onto the unit circle $|z_2|=1$.
The invariant amplitudes are represented by two Pietarinen
expansions
\begin{eqnarray}\label{InvAmp}
\overline{B}_{1}=B_{1}-B_1^N&=&(1+z_{1})\cdot\sum_{i}b_{1i}^{(1)}z_{1}^{i}+(1+z_{2})\cdot\sum_{i}b_{1i}^{(2)}z_{2}^{i}\,,\\
\overline{B}_{2}=B_{2}-B_2^N&=&(1+z_{1})\cdot\sum_{i}b_{2i}^{(1)}z_{1}^{i}+(1+z_{2})\cdot\sum_{i}b_{2i}^{(2)}z_{2}^{i}\,,\\
\overline{B}_{6}=B_{6}-B_6^N&=&(1+z_{1})\cdot\sum_{i}b_{6i}^{(1)}z_{1}^{i}+(1+z_{2})\cdot\sum_{i}b_{6i}^{(2)}z_{2}^{i}\,,\\
\frac{\overline{B}{}_{8}}{\nu}=\frac{B_{8}}{\nu}-\frac{B_8^N}{\nu}&=&(1+z_{1})\cdot\sum_{i}b_{8i}^{(1)}z_{1}^{i}+(1+z_{2})\cdot\sum_{i}b_{8i}^{(2)}z_{2}^{i}\,,\label{eq:bb8}
\end{eqnarray}
where $B_i^N$ are the nucleon pole contributions. The factors
$(1+z_1)$ and $(1+z_2)$ in front of the corresponding power series
assure that the amplitudes $B_i$ go to zero at infinity. By
construction, the expansions (\ref{InvAmp}-\ref{eq:bb8}) represent
crossing even amplitudes, having analytic structure required by the
fixed-$t$ analyticity from Mandelstam hypothesis applied to $\eta$
photoproduction. The coefficients in (\ref{InvAmp}-\ref{eq:bb8}) are
obtained by fitting the quadratic form
\begin{equation}
 \mathalpha{X}^2=\chi^2_{FTdata}+\Phi_{conv}
\end{equation}
to the fixed-$t$ data with $\chi^2_{FTdata}$ in the form
\begin{eqnarray}
\chi_{FTdata}^{2} & = & \sum_{a=1}^{N^O}\;
\sum_{n=1}^{N^{E}}\left(\frac{O_a(s_{n},t)^{exp}-O_a(s_{n},t)^{fit}}{\Delta
O_a(s_{n},t)^{exp}}\right)^{2}\,,
\end{eqnarray}
where $N^O$ is number of observables, {$O_a = \{\sigma_0,\check{\Sigma},\check{T},\check{F},...\} $}  with corresponding errors $\Delta
O_a(\theta_{i})$.

$\Phi_{conv}$ is a Pietarinen's convergence test
function~\cite{Pietarinen1}, \cite{Hohler84} (Chapter 2.1.7, page 405 in \cite{Hohler84}),
\begin{equation}
\Phi_{conv}=\Phi_{1}+\Phi_{2}+\Phi_{3}+\Phi_{4}
\end{equation}
with
\begin{equation}
\Phi_{k}=\lambda_{1k}\sum_{i=0}^{N_{1}}(b_{1i}^{(k)})^{2}(i+1)^{3}
+\lambda_{2k}\sum_{i=0}^{N_{2}}(b_{2i}^{(k)})^{2}(i+1)^{3}\,,
\end{equation}
where $\lambda_{1k}$ and $\lambda_{2k}$ are weight factors. For
large numbers of coefficients ($N>20$) the weight factors $\lambda$
can be calculated using a simplified formula
\begin{equation}
\lambda_{1k}=\frac{N_{1}}{\sum_{i=0}^{N_{1}}(b_{1i}^{(k)})^{2}(i+1)^{3}},
\quad\lambda_{2k}=\frac{N_{2}}{\sum_{i=0}^{N_{2}}(b_{2i}^{(k)})^{2}(i+1)^{3}}
\end{equation}
in an iterative procedure, starting from some initial values of
coefficients in expansions (\ref{InvAmp}-\ref{eq:bb8}). The
procedure described above is known as a fixed-$t$ amplitude analysis
(FT AA). For a given $t$-value the result is a set of coefficients
$\{b_{1i}^{(k)}\}$, $\{b_{2i}^{(k)}\}$, $\{b_{6i}^{(k)}\}$,
$\{b_{8i}^{(k)}\}$. Invariant and helicity amplitudes at
predetermined $t$-values may be calculated at any energy $W$ and any
scattering angle inside the physical region using the formula
\begin{equation}
cos\theta_{i}=\frac{t_{i}-m_{\eta}^{2}+2k\omega}{2kq}\,,\quad
|cos\theta_{i}|\le1,\quad t_{i}\in[t_{min},t_{max}]\,.
\end{equation}

\subsection{\boldmath Single energy partial wave analysis}

In single energy partial wave analysis (SE PWA) we minimize the
quadratic form
\begin{equation}
 X^2=\chi^2_{SEdata}+\Phi_{trunc}\,.
\end{equation}
$\chi^2_{SEdata}$ contains all experimental data at a given energy
$W$
\begin{eqnarray}
\chi_{SEdata}^{2} & = & \sum_{a=1}^{N^{O}}\;
\sum_{i=1}^{N_{1}^{D}}\left(\frac{O_a(\theta_{i})^{exp}-O_a(\theta_{i})^{fit}}{\Delta
O_a(\theta_{i})^{exp}}\right)^{2}\,.
\end{eqnarray}
As before, $N^O$ is the number of observables, and
$O_{a}(\theta_{i})^{exp}$ are experimental values of observable
$O_a$ with corresponding errors $\Delta O_a(\theta_{i})$.
$O_a(\theta_{i})^{fit}$ are values of the observable $O_a$,
evaluated from partial waves obtained in the fit. $\Phi_{trunc}$
makes a soft cut-off of higher partial waves and is effective at low
energies close to threshold~\cite{Hohler84}. It is given by the
formula
\begin{equation}\label{Phitrunc}
\Phi_{trunc}=\lambda_{trunc}\sum_{\ell=0}^{\ell_{max}}\left(\,(ReT_{\ell\pm})^{2}R_{1}^{2\ell}+
(ImT_{\ell\pm})^{2}R_{2}^{2\ell}\,\right)\,.
\end{equation}

The form of $\Phi_{trunc}$ arises form the general behavior of
partial waves. An expansion of invariant amplitudes in terms of
Legendre polynomials (PW expansion) converges in an ellipse in the
$\cos\theta$ plane having foci at $-1$ and $+1$ and semi-axes
$y_{0}(s)$ and $(y_{0}^{2}(s)-1)^\frac{1}{2}$, where $y_0(s)$ is
determined by the edge of the nearest double spectral region. In
$\eta$ photoproduction, due to the unphysical cut in the $t$ channel
starting at $t=4m_{\pi}^2$, the edge of the double spectral region
approaches $t=4m_{\pi}^2$ as $s\rightarrow \infty$. We make a
simplest choice, $y_0(s)=\left|cos(\theta(t=4m_{\pi}^2))\right|$
($|y_0|>1$ outside of the physical region). In the simplest
(spinless) case, the PWA expansion converges if $Im T_{\ell}$
behaves as
\begin{equation}
(Im T_e)^{2} \leqslant \left[y_0+\sqrt{y_{0}^2-1}\right]^{-2\ell}\,.
\end{equation}

We assume that  electric and magnetic multipoles $E_{\ell\pm}$,
$M_{\ell\pm}$ behave roughly in the same way. As additional
simplification, we take $R_1=R_2=R=y_0+\sqrt{y_0^2-1}>1\,$.

In order to get small values of $\chi^2$, the minimizer has to keep
$\Phi_{trunc}$ small imposing small values of higher multipoles. As
mentioned above, $\Phi_{trunc}$ is effective at low energies and
makes a soft cut-off of higher multipoles.

\subsection{Iterative minimization scheme}

The method consists of two separate analyses, the fixed-$t$
amplitude analysis (FT AA) and the single energy partial wave
analysis (SE PWA). The two analyses are coupled in such a way that
the results from FT AA are used as a constraint in SE PWA and vice
versa in an iterative procedure. It can not be proven, but it is
extensively tested in  $\pi N$ elastic, fixed-$t$ constrained SE
PWA~\cite{Hohler84}, and since then recommended for other processes.

\begin{itemize}[itemsep=1.ex,leftmargin=2.5cm]
\item[\bf Step 1:] Constrained FT AA is performed by minimizing the form
\begin{equation}\label{chi2_AA}
X^{2}=\chi_{FTdata}^{2}+\chi_{cons}^2+\Phi_{conv}\,,
\end{equation}

where $\chi_{cons}^2$ is a constraining term given by
\begin{eqnarray}
\chi_{cons}^{2} & = &
q_{cons}\sum_{k=1}^{4}\sum_{i=1}^{N^{E}}\left(\frac{Re\:
H_{k}(E_i)^{fit}-Re\:
H_{k}(E_i)^{cons}}{\varepsilon_{k,i}^{Re}}\right)^{2}\\
\nonumber &  &
+q_{cons}\sum_{k=1}^{4}\sum_{i=1}^{N^{E}}\left(\frac{Im\:
H_{k}(E_i)^{fit}-Im\:
H_{k}(E_i)^{cons}}{\varepsilon_{k,i}^{Im}}\right)^{2}. \nonumber
\end{eqnarray}
$H_{k}^{cons}$ are helicity amplitudes from SE PWA in the previous
iteration. In a first iteration, $H_{k}^{cons}$ are calculated from
the initial PWA solution (MAID). $H_{k}^{fit}$  are values of
helicity amplitudes $H_{k}$ calculated from coefficients in
Pietarinen's expansions, which are parameters of the fit. $N_E$ is
the number of energies for a given value of $t$, and $q_{cons}$ is
an adjustable weight factor. $\varepsilon_{k,i}^{Re}$ and
$\varepsilon_{k,i}^{Im}$ are errors of real and imaginary parts of
the corresponding helicity amplitudes. In our analysis we take
$\varepsilon_{k,i}^{Re}=\varepsilon_{k,i}^{Im}=1.$

\item[\bf Step 2:] Constrained  SE PWA
is performed by minimizing the form
\begin{equation}\label{chi2-SEPWA}
X^{2}=\chi_{SEdata}^{2}+\chi_{FT}^2+\Phi_{trunc}\,,
\end{equation}
where the additional term $\chi_{FT}^2$ contains the helicity
amplitudes from the FT AA in step 1:
\begin{eqnarray}
\chi_{FT}^{2} & = &
q_{cons}\sum_{k=1}^{4}\sum_{i=1}^{N^{C}}\left(\frac{Re\:
H_{k}(\theta_{i})^{FT}-Re\:
H_{k}(\theta_{i})^{fit}}{\varepsilon_{k,i}^{Re}}\right)^{2}\\
\nonumber &  &
+q_{cons}\sum_{k=1}^{4}\sum_{i=1}^{N^{C}}\left(\frac{Im\:
H_{k}(\theta_{i})^{FT}-Im\:
H_{k}(\theta_{i})^{fit}}{\varepsilon_{k,i}^{Im}}\right)^{2}.
\nonumber
\end{eqnarray}

\item[\bf Step 3:] Use resulting multipoles obtained in step 2, and  calculate helicity amplitudes which serve
as a constraint in step 1.
\end{itemize}

\begin{figure}[htb]
\begin{center}
\includegraphics[width=10cm]{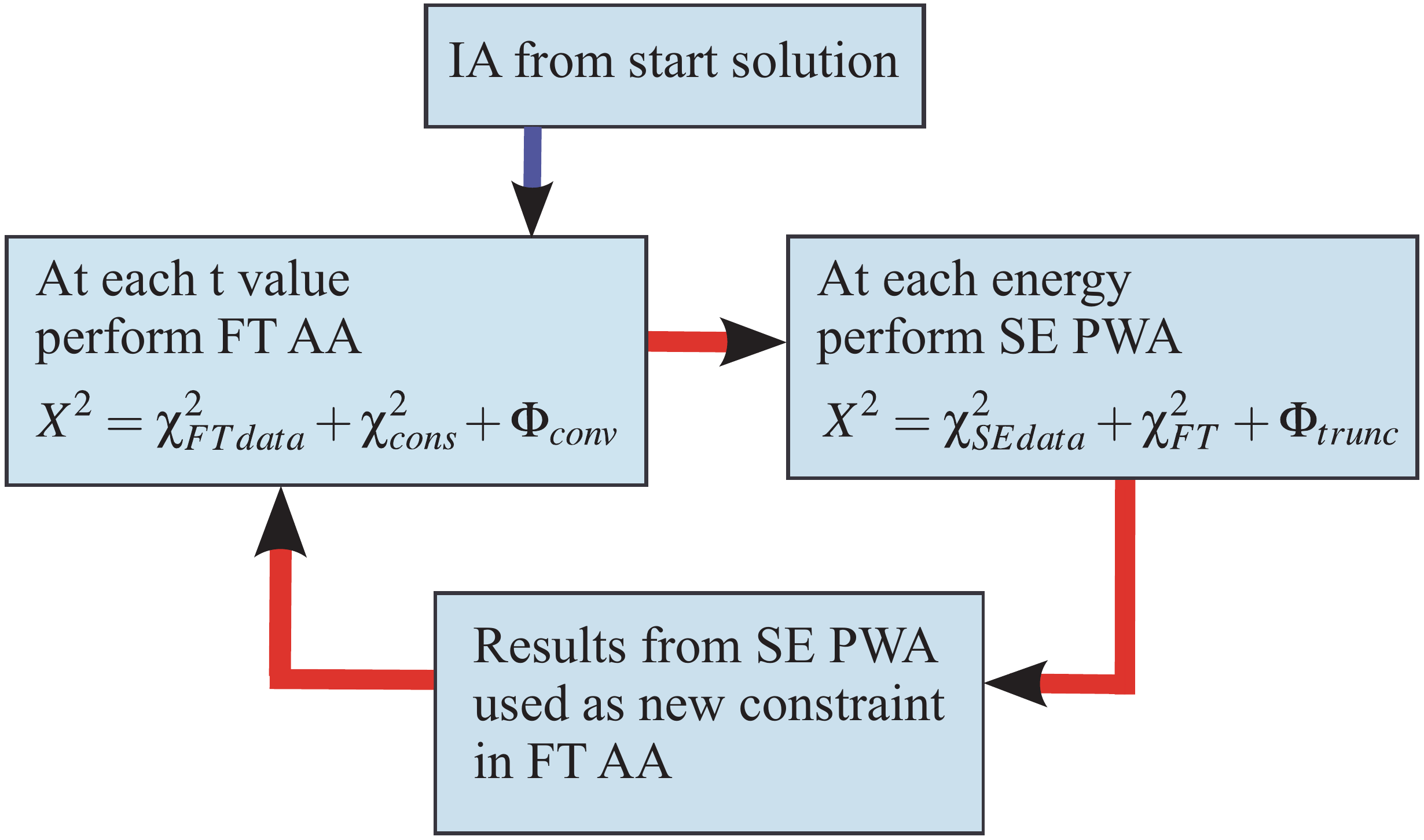}
\vspace{3mm} \caption{\label{Fig:Scheme} Iterative minimization
scheme which achieves point-to-point continuity in energy using
fixed-$t$ analyticity as a constraint. (IA: invariant amplitudes,
FT~AA: fixed-$t$ amplitude analysis, SE~PWA: single-energy partial
wave analysis) }
\end{center}
\end{figure}

An iterative minimization scheme which accomplishes point-to-point
continuity in energy is given in Fig.~\ref{Fig:Scheme}.

\pagebreak

\section{Studies with pseudo data}

In this section we study the potential impact of our method by
analyzing pseudo-data, which were generated numerically from
multipoles of a recent EtaMAID model. Here we know exactly the input
amplitudes and can compare them directly to the results obtained
from fits of different sets of observables. We follow the following
strategy:
\begin{itemize}
\item As a first step, we demonstrate that a fully unconstrained SE fit of even
a complete set of observables does not give a unique solution. At
each energy, we obtain a band of equivalent solutions, depending on
the choice of the initial parameter values.
\item We then perform a fit with fixed-$t$ constraints according to the procedure described
   in Section \ref{Fixed-t} using the same complete set of observables from pseudo-data
   as in the first step.
\item Finally, we reduce the number of observables to four and use the same set which is
available from real measurements at MAMI and GRAAL.
\end{itemize}

\subsection{Input from pseudo data} \label{Pseudo-data}

We have generated all 16 observables from multipoles predicted by
the EtaMAID-2015 model \cite{Kashev2015} (solution I). We randomize
the unpolarized cross section by a normal distribution with a
standard deviation of $0.1\%$. For the polarization observables we
firstly randomize the polarized cross sections again by a normal
distribution with a standard deviation of 0.1\% and calculate the
observables as the difference between two polarization directions.
The asymmetries are then obtained by division and error propagation.
By this procedure, the unpolarized cross section obtains the highest
precision, whereas polarization asymmetries obtain larger errors,
especially when the asymmetries are small. This reflects better the
situation of real experiments. However, here we still investigate
the more or less ideal case with a precision which will not be
reached in real measurements of spin-observables. We start our
analysis using a complete set of observables which has to include
double polarization observables with beam, target  and recoil
polarization ~\cite{Chiang:1996em,Workman:2016irf}. We have chosen
the following set:
$\{\sigma_0,\check{\Sigma},\check{T},\check{P},\check{F},\check{G},\check{C}_{x'},\check{O}_{x'}\}$.

For our procedure we need the data at two different
kinematic grids: energy and $t$ $\{W_i,t_j\}$ for the fixed-t
amplitude analysis, and energy and polar angle $\{W_i,\theta_k\}$ for the
single energy multipole fit. Our pseudo data sets were
generated at 140 energies, each at 50 $t$ values and 18 angles.
Examples of these observables at 2 different energies can be found in
Fig.~\ref{FigPseudoFitSE1}.

\subsection{Unconstrained fit with pseudo data}
\begin{figure}[htbp]
\begin{center}
\includegraphics[width=8.0cm]{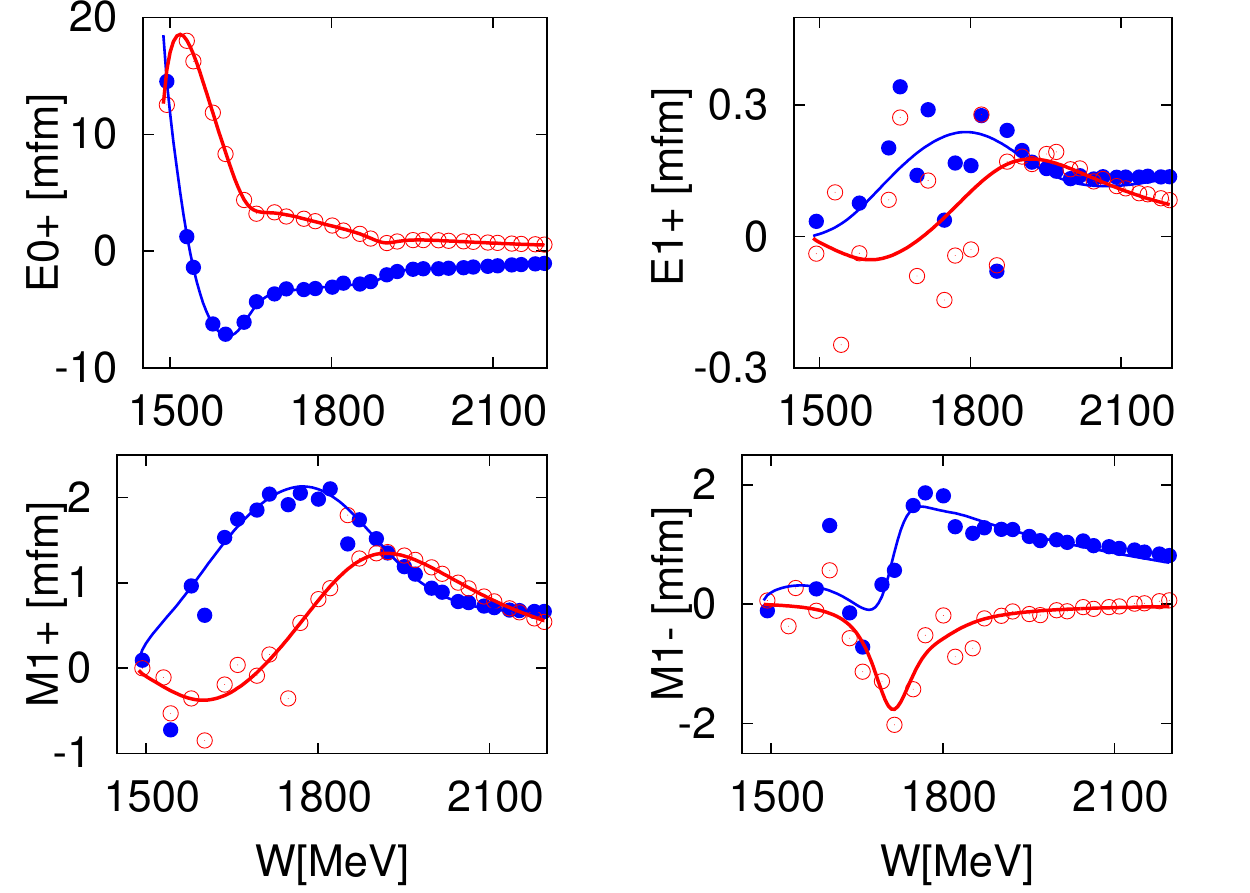}
\includegraphics[width=8.0cm]{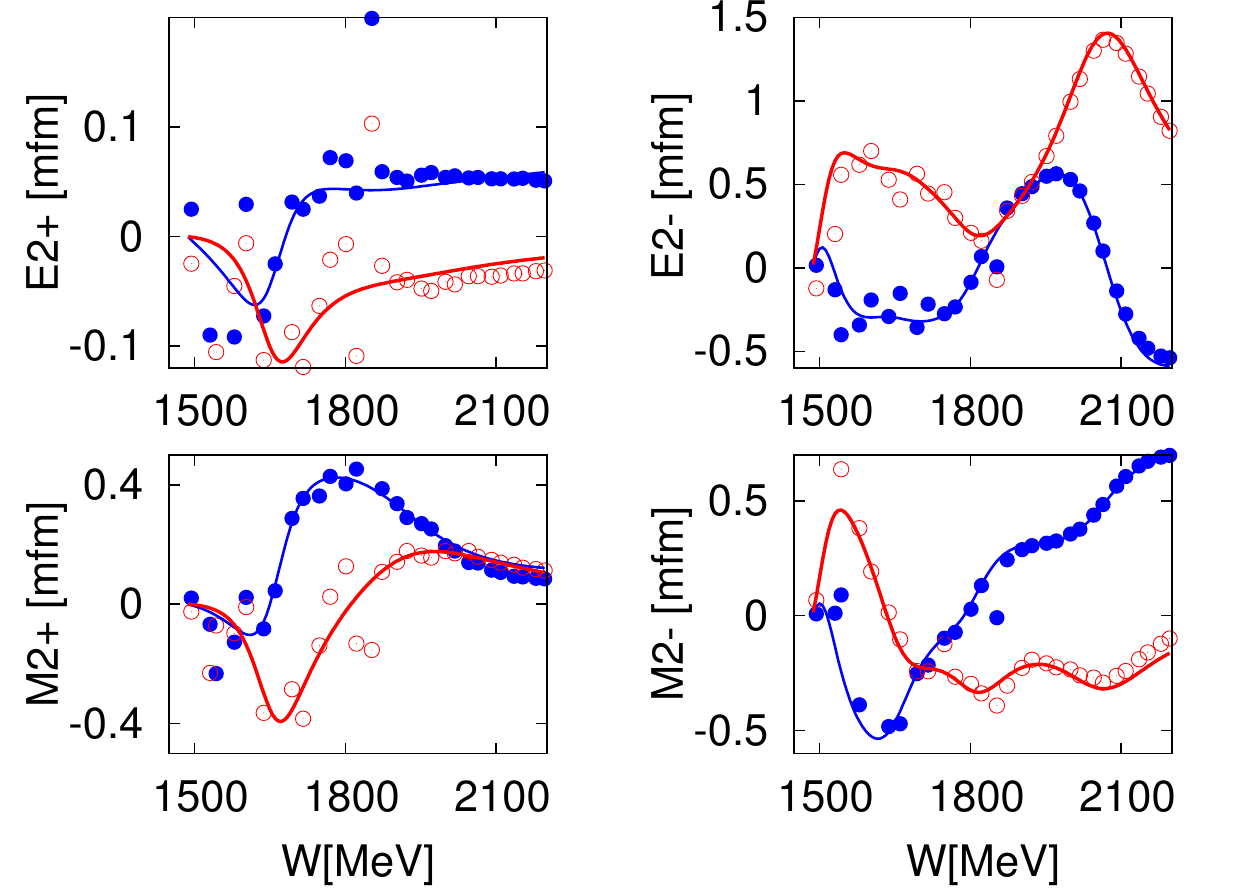}
\includegraphics[width=8.0cm]{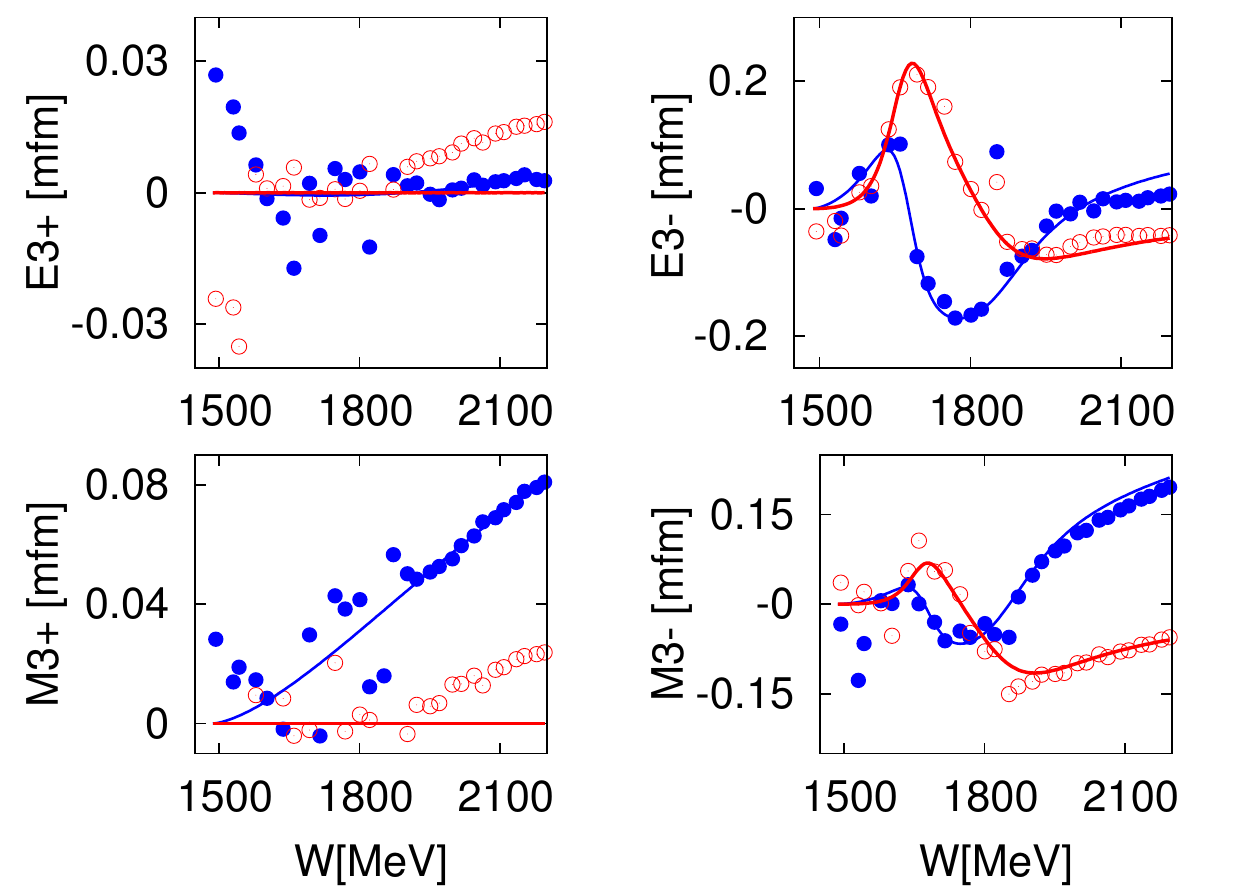}
\includegraphics[width=8.0cm]{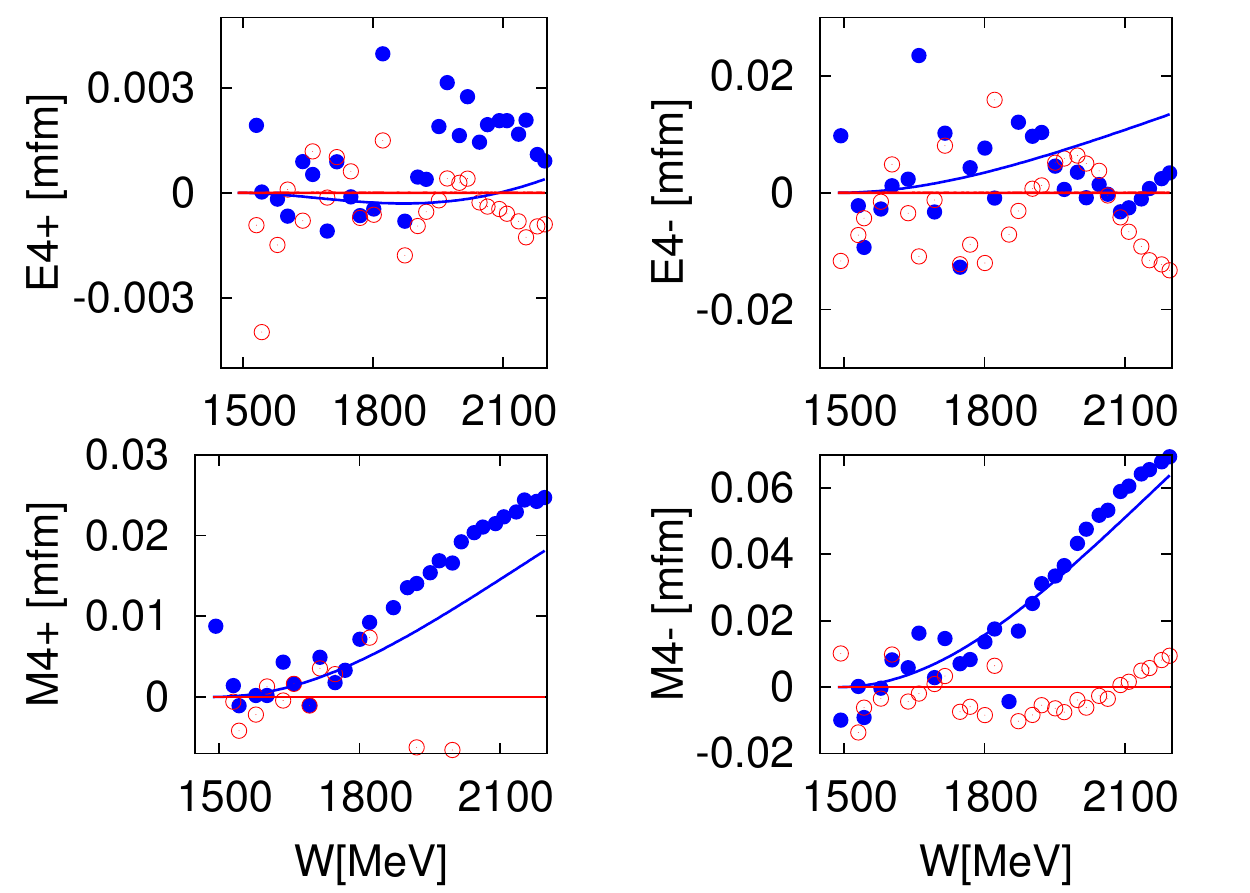}
\vspace{3mm} \caption{\label{FigMultFixedPhase} Result of on
unconstrained single-energy fit described in the text. The blue and
red points show the real and imaginary parts of the multipoles
obtained in the fit compared to the "true" multipoles from the
underlying EtaMAID-2015 model (blue and red solid lines). }
\end{center}
\end{figure}

As first step we have performed a fully unconstrained single energy
multipole fit, truncated to $L_{max}=5$, using the complete set of
observables from pseudo-data defined in Section \ref{Pseudo-data}.
We start the minimization procedure at initial values with were
randomly distributed by 50\% around the true solution. In order to
fix the overall energy-dependent phase, we fixed the phase of the
$E_{0+}$ multipole to the phase of the EtaMAID model, leaving 39
real parameters from 20 complex multipoles. The results of one of
these fits are shown in Fig.~\ref{FigMultFixedPhase}. Even in this
ideal case, a fully unconstrained fit does not result uniquely in
the correct multipoles. At each energy, we find a band of equivalent
solutions (similar $\chi^2$) depending on the particular values
chosen for the starting parameters of the fit. As the fits at each
energy are independent from each other, the energy dependence of
these solutions is discontinuous.
For any other choice of initial parameter values, we obtain another,
completely different, but still discrete set of multipoles. This
observation can be explained in the following way. It can be shown,
that in case of truncated data with highest angular momentum being
$L_{max}$ and very high precision, a SE PWA (TPWA) truncated to the
same $L_{max}$  gives a unique and continuous solution. If the data
are not truncated and higher partial waves do contribute, a unique
solution is not obtained, even using a complete set of infinitely
precise data. A unique solution can, however, be restored if the
higher partial waves, $L > L_{max}$, are known, e.g. from background
terms. In practice, this ideal situation is approximately fulfilled
only for charged pion photoproduction, where a sizeable contribution
of the background arises from the well known pion pole contribution
which can be calculated in a model independent way. For other
reactions like kaon or eta photoproduction, the couplings of the
Born terms are fairly unknown and also the $t$-channel contributions
from vector mesons are important and model dependent.

Therefore, in eta photoproduction higher partial waves from background
cannot be used as model independent input in SE fits.
However, a model independent relation between SE fits at different energies
can be provided by fixed-$t$ analyticity.

\subsection{\boldmath Pseudo data analysis with fixed-$t$ analyticity constraints}
 \label{sec:results}
In the following we describe the iterative procedure of successive
amplitude analyses at fixed-$t$ and multipole fits at fixed energy
introduced in Section~\ref{Fixed-t}.
\\
\underline{\emph{{\bf Step 1}}: Fixed-$t$ amplitude analysis }

\begin{figure*}[htb]
\begin{center}
\includegraphics[width=8.0cm]{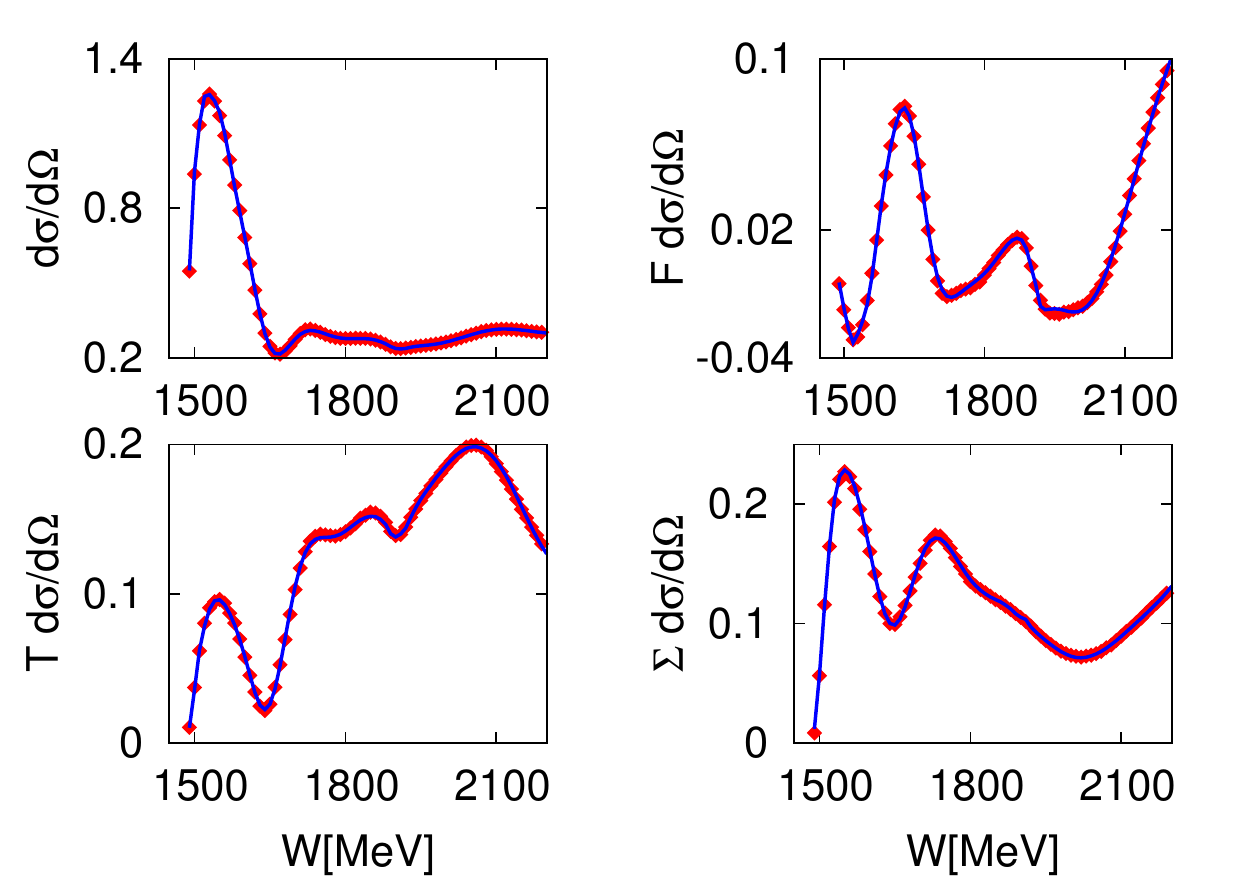}
\includegraphics[width=8.0cm]{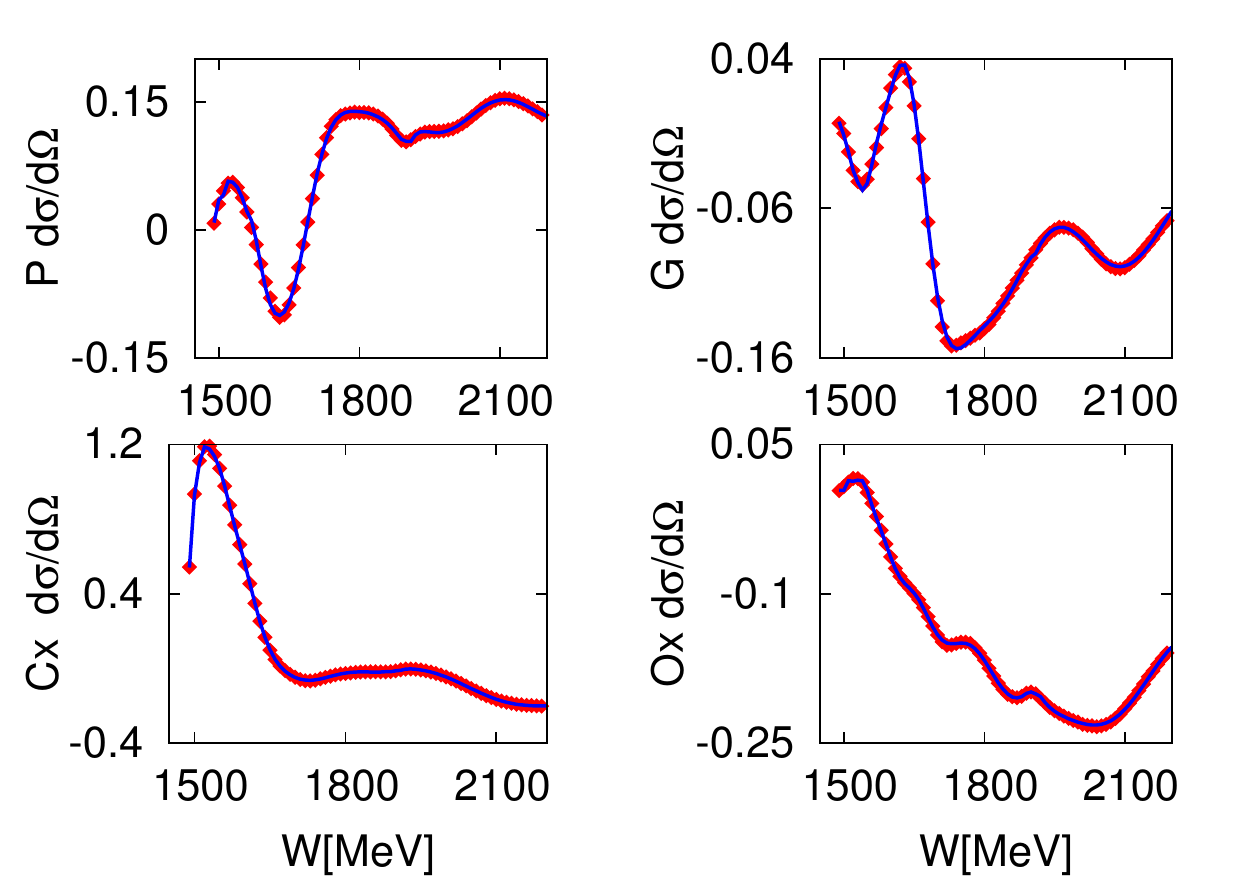}
\includegraphics[width=8.0cm]{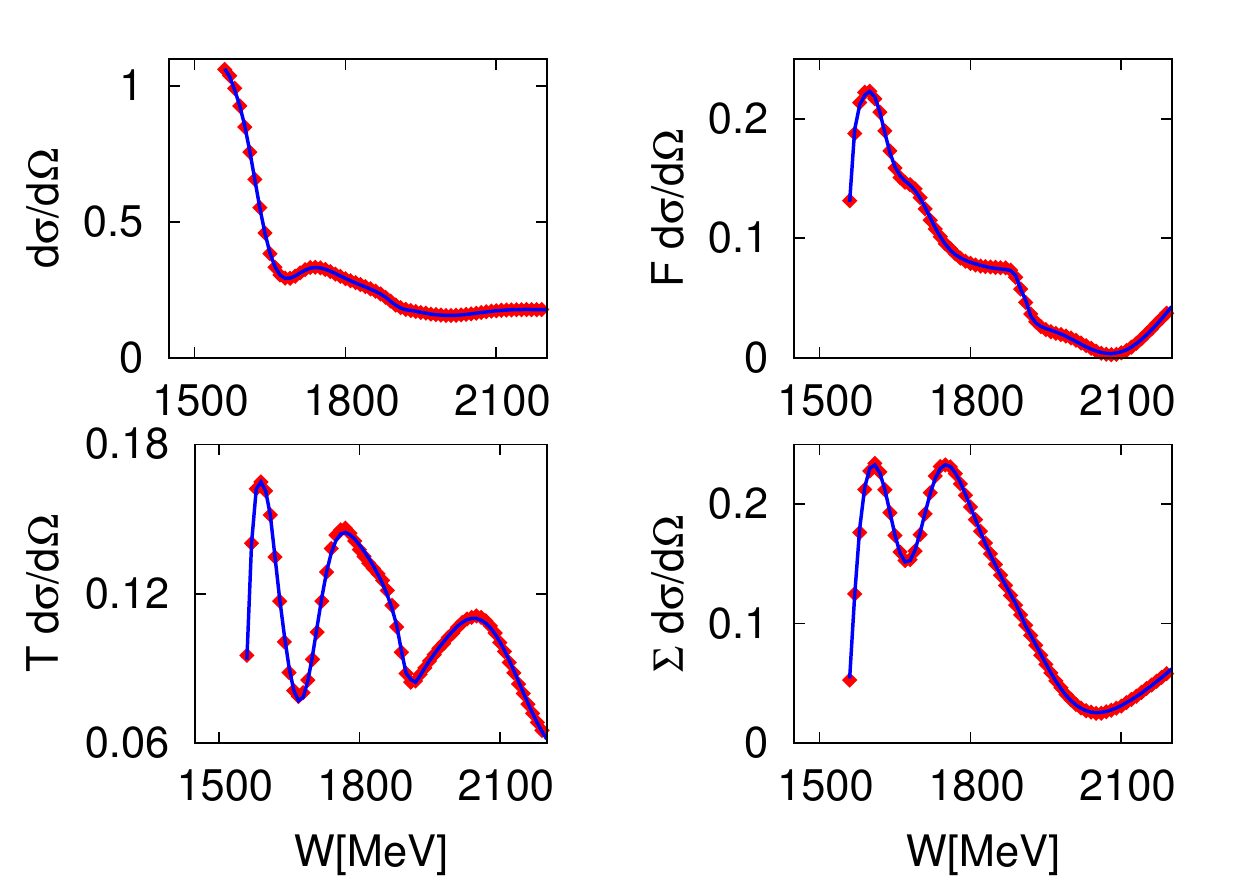}
\includegraphics[width=8.0cm]{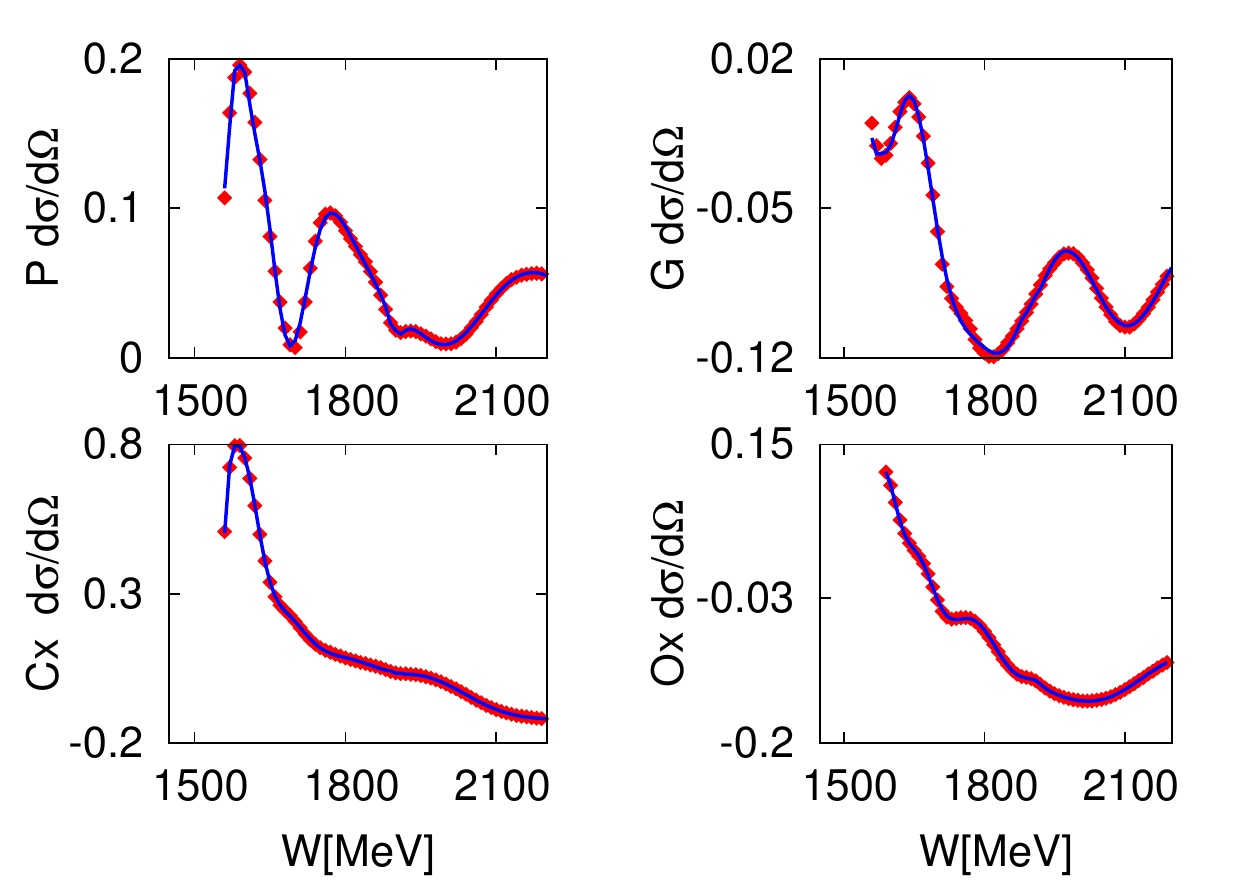}
\vspace{3mm} \caption{\label{FigPseudoFitFT02} Fixed-$t$ amplitude
analysis of a complete set of pseudo data at $t=-0.2\,$GeV$^2$ (top)
and $t=-0.5\,$GeV$^2$ (bottom). The minimization is performed
according to Eq.~(\ref{chi2_AA}). The data points are pseudo data
with a precision of $0.1\%$. The lines are obtained from the
Pietarinen expansion with fitted coefficients.}
\end{center}
\end{figure*}

The complete set of 8 observables $\sigma_0,
\check{\Sigma},\check{T},\check{P},\check{F},\check{G},\check{C}_{x'},\check{O}_{x'}$
at 50 fixed-$t$ values in the range of $-2.0$~GeV$^2\le t \le
-0.1$~GeV$^2$ were fitted. The minimization is performed in terms of
Pietarinen expansions according to Eq.~(\ref{chi2_AA}). We start in
minimization procedure at initial values with were randomly
distributed by 50\% around the true solution. Examples of these fits
at $t=-0.2$~GeV$^2$ and $t=-0.5$~GeV$^2$ are shown in
Fig.~\ref{FigPseudoFitFT02}.
With a typical number of $N\approx 20$ coefficients for each
Pietarinen, a very precise description of all observables can be
achieved. As a consequence, also all four underlying complex
amplitudes, CGLN and helicity amplitudes are perfectly reproduced up
to an overall, energy- and angle-dependent, phase. Furthermore, it
is important to note, that in this case all possible 16 observables
are described, including those, which were not fitted. The helicity
amplitudes obtained in this fit are now used as constraint in a SE
PWA.
\\
\underline{\emph{{\bf Step 2}}: A SE PWA fit is performed using
constraints from Step 1.}

\begin{figure}[htb]
\begin{center}
\includegraphics[width=8.0cm]{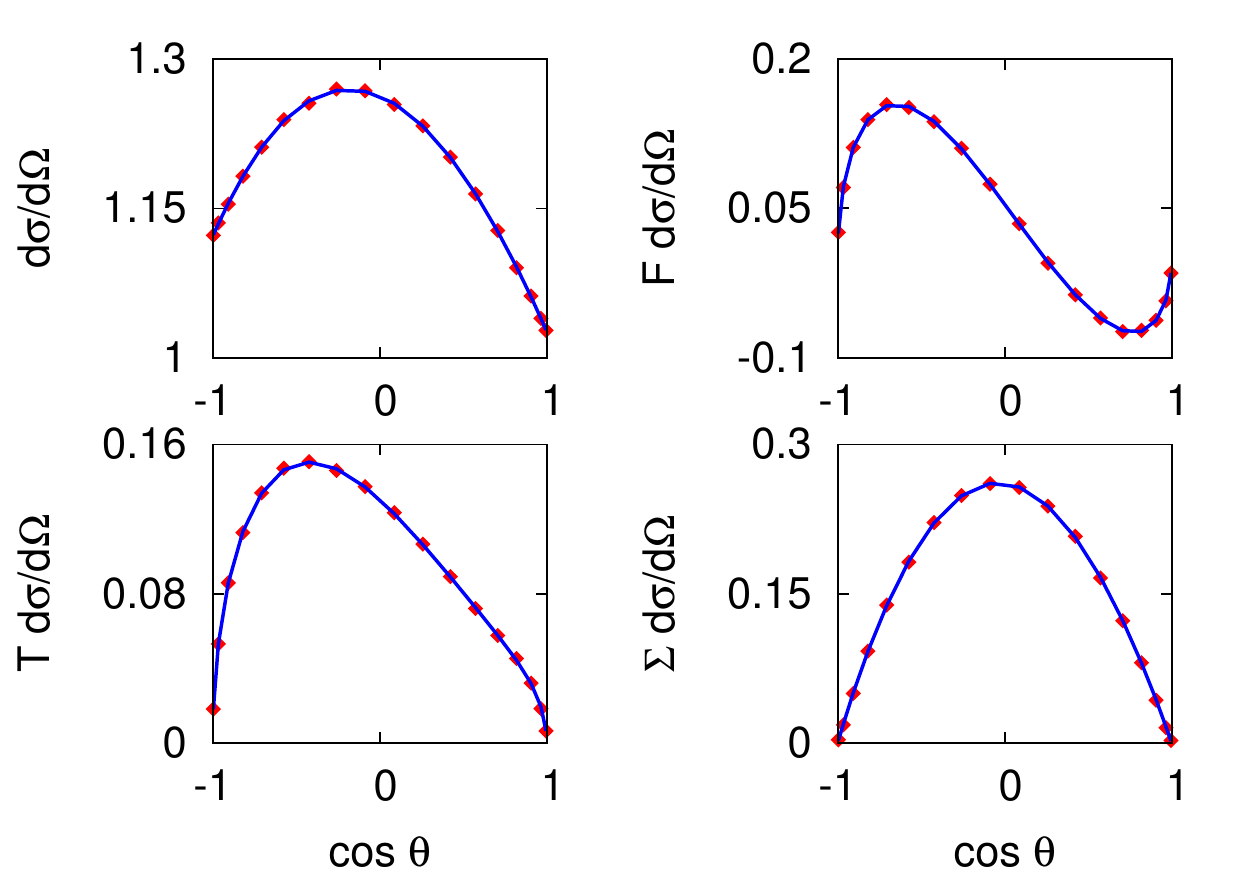}
\includegraphics[width=8.0cm]{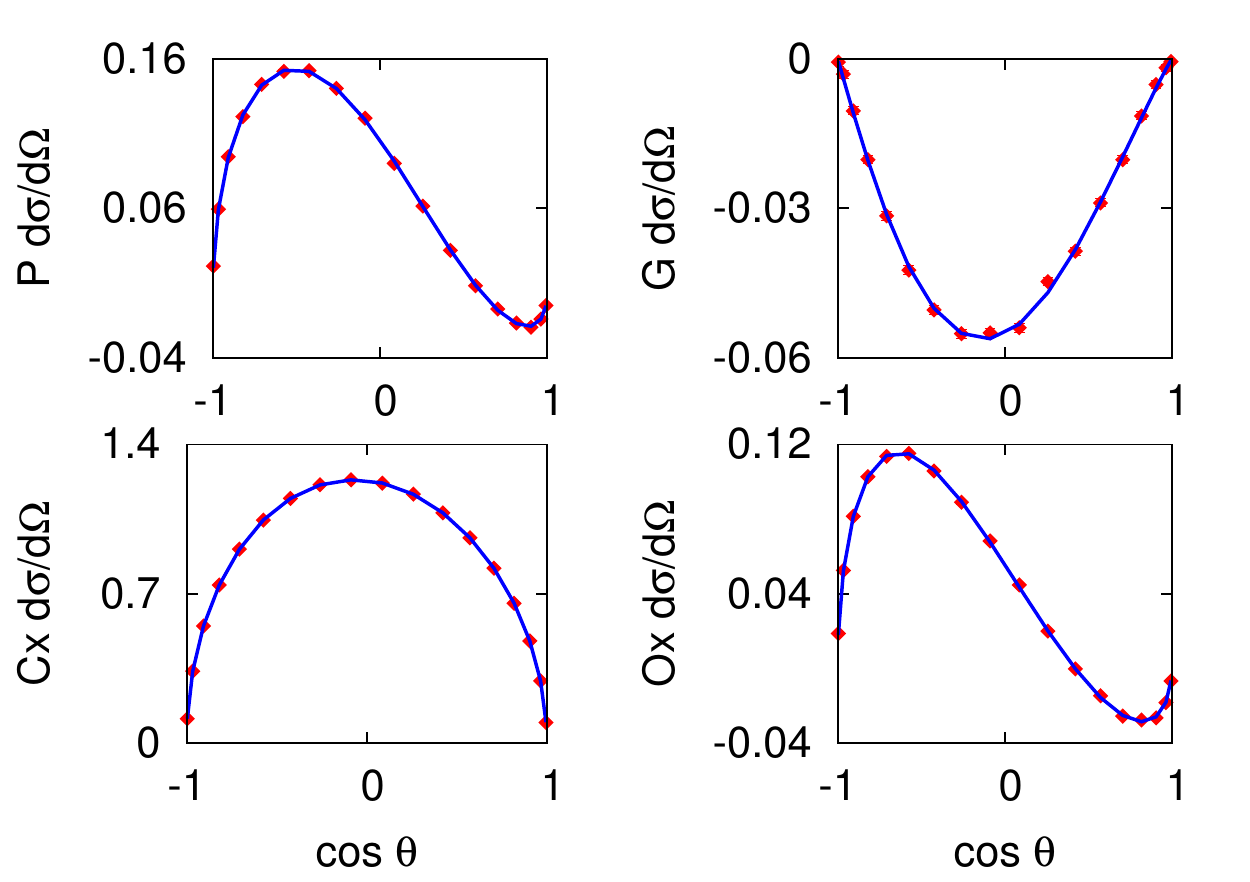}
\includegraphics[width=8.0cm]{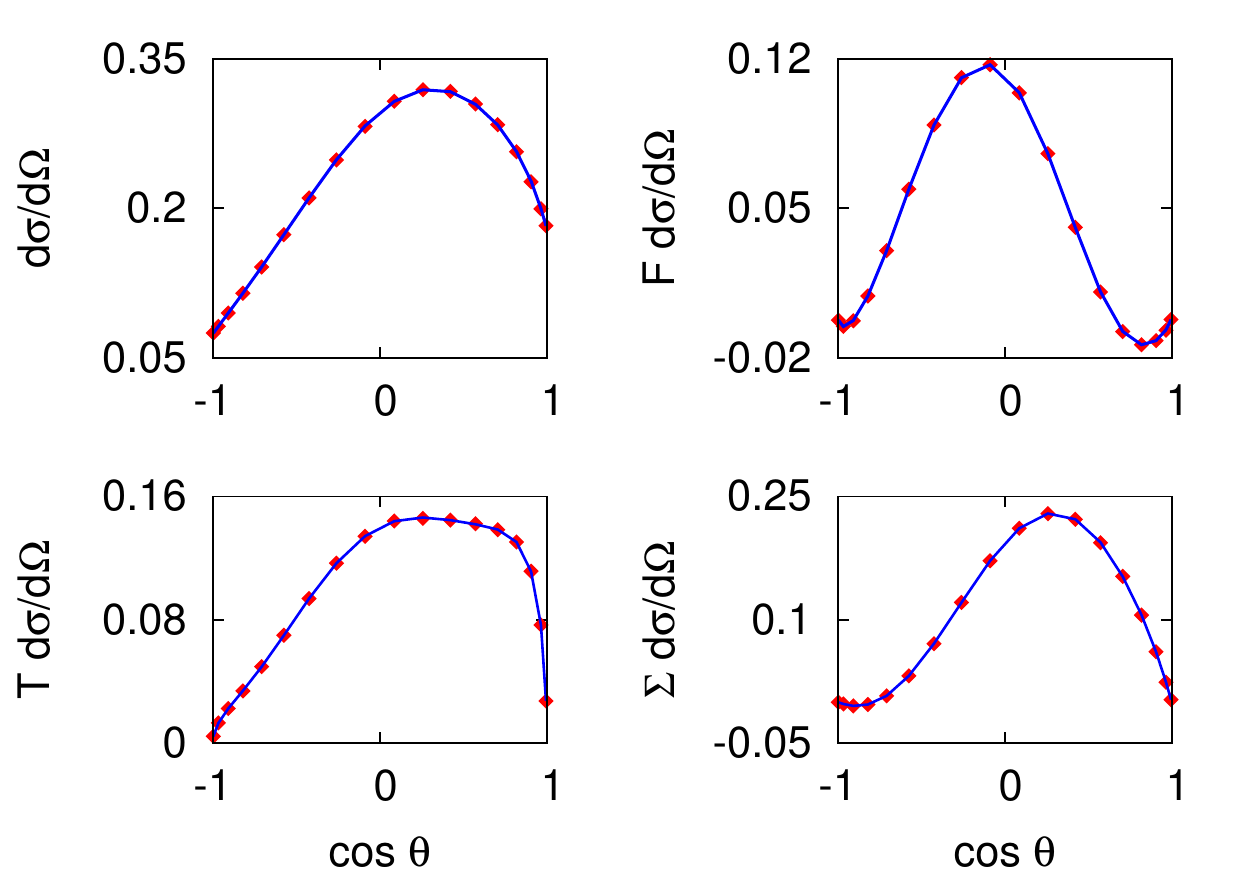}
\includegraphics[width=8.0cm]{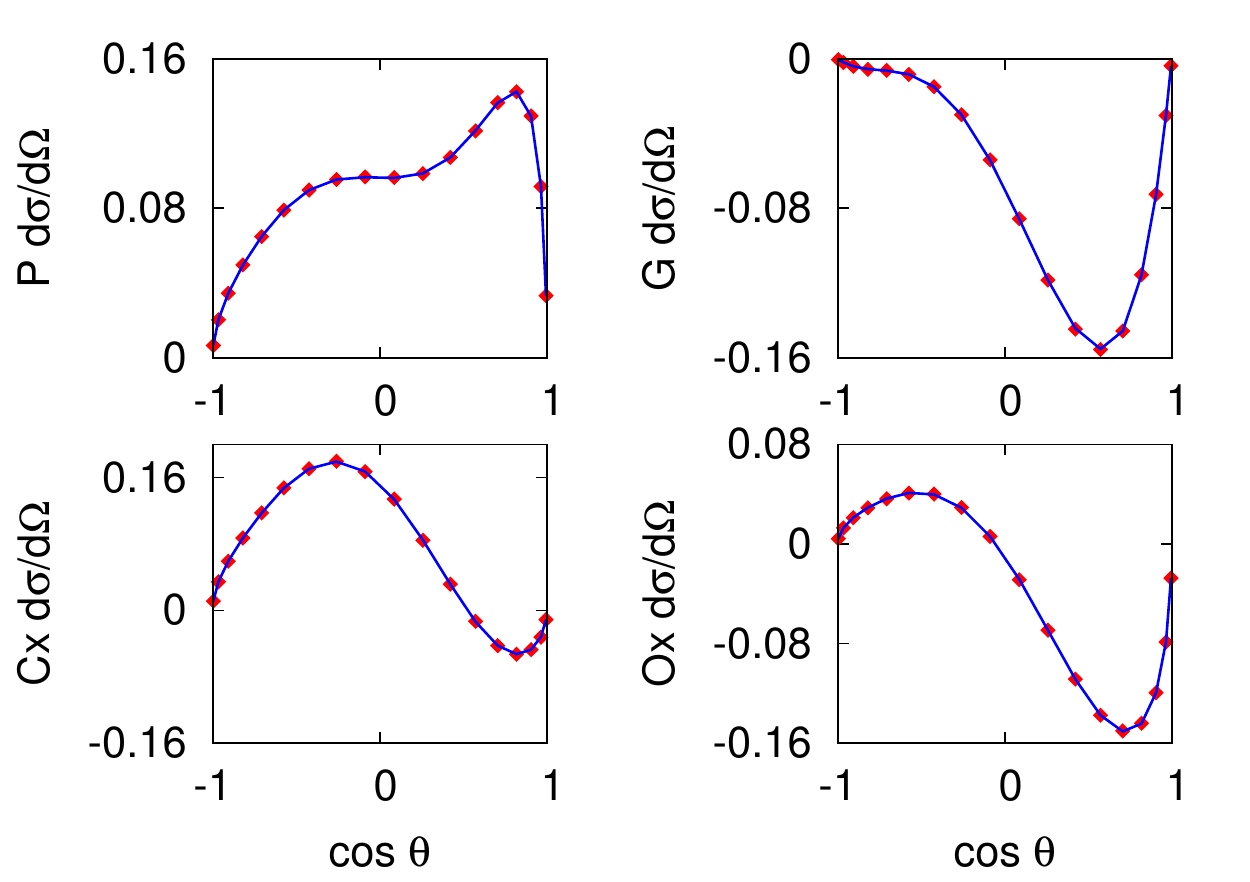}
\vspace{3mm} \caption{\label{FigPseudoFitSE1} Single energy partial
wave analysis of a complete set of pseudo data at $W=1543$~MeV (top)
and $W=1770$~MeV (bottom). The minimization is performed according
to Eq.~(\ref{chi2-SEPWA}). The data points are pseudo data with a
precision of $0.1\%$. The lines are obtained from the CGLN
amplitudes with fitted multipoles up to $L_{max}=5$.}
\end{center}
\end{figure}

The conditions for the single energy partial wave analysis are
identical to those used in the unconstrained analysis, however, we
did not fix the $E_{0+}$ phase. We fit up to $L_{max}=5$ which
corresponds to 20 complex multipoles or 40 real parameters. All
multipoles for $L>5$ are set to zero. The starting values of the fit
parameters are again randomly chosen in a 50\% range around the true
solution. The fits of the complete set of observables from pseudo
data at $E=800$~MeV ($W=1543$~MeV) and $E=1200$~MeV ($W=1770$~MeV)
are shown in Fig.~\ref{FigPseudoFitSE1}. Again, as in step 1, all
observables are perfectly described. As a consequence, again all
four underlying complex amplitudes, CGLN and helicity amplitudes are
perfectly reproduced up to an overall, energy- and angle-dependent,
phase. However, this phase has now changed compared to the solution
from Step 1.
\\
\underline{\emph{{\bf Step 3}}: Further iterations and final
solution.}

\begin{figure}[hp]
\begin{center}
\includegraphics[width=8.0cm]{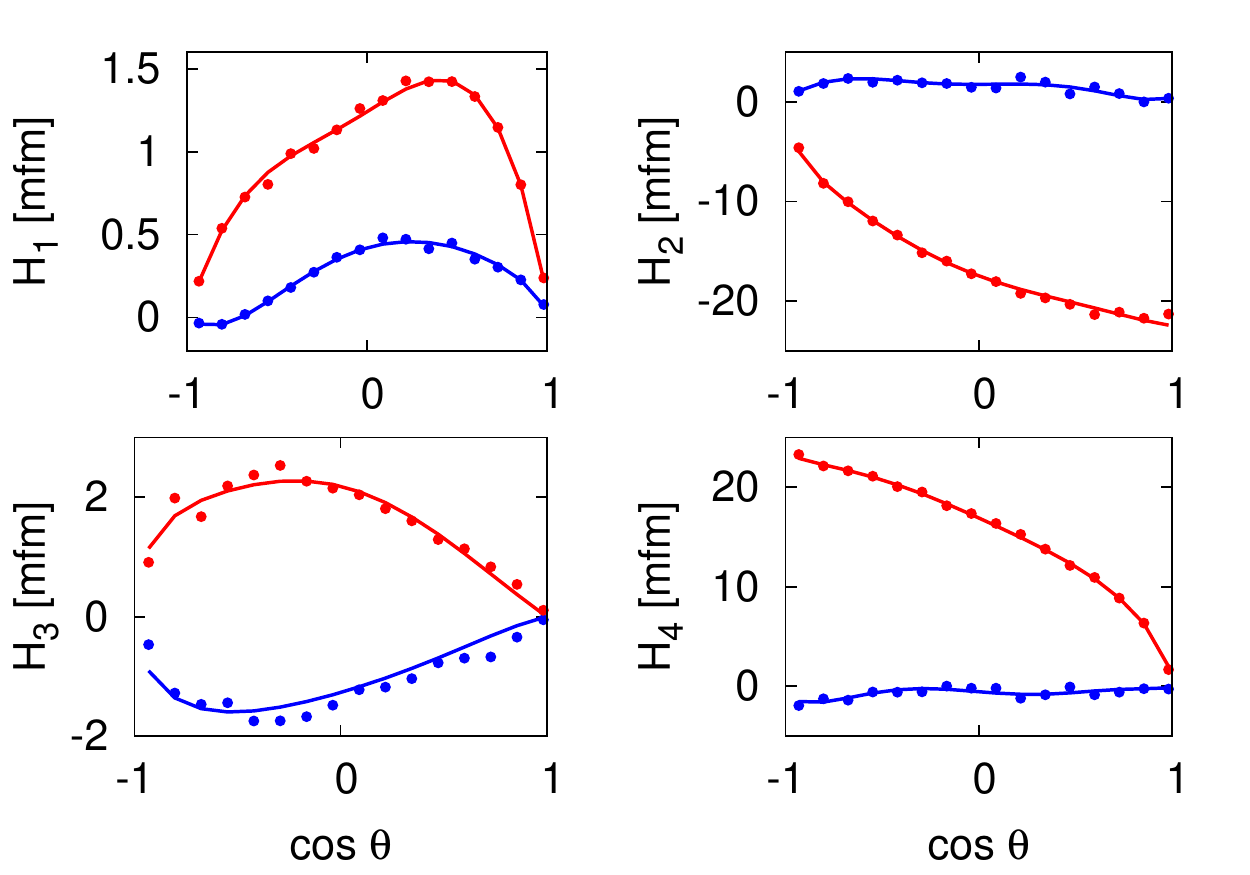}
\includegraphics[width=8.0cm]{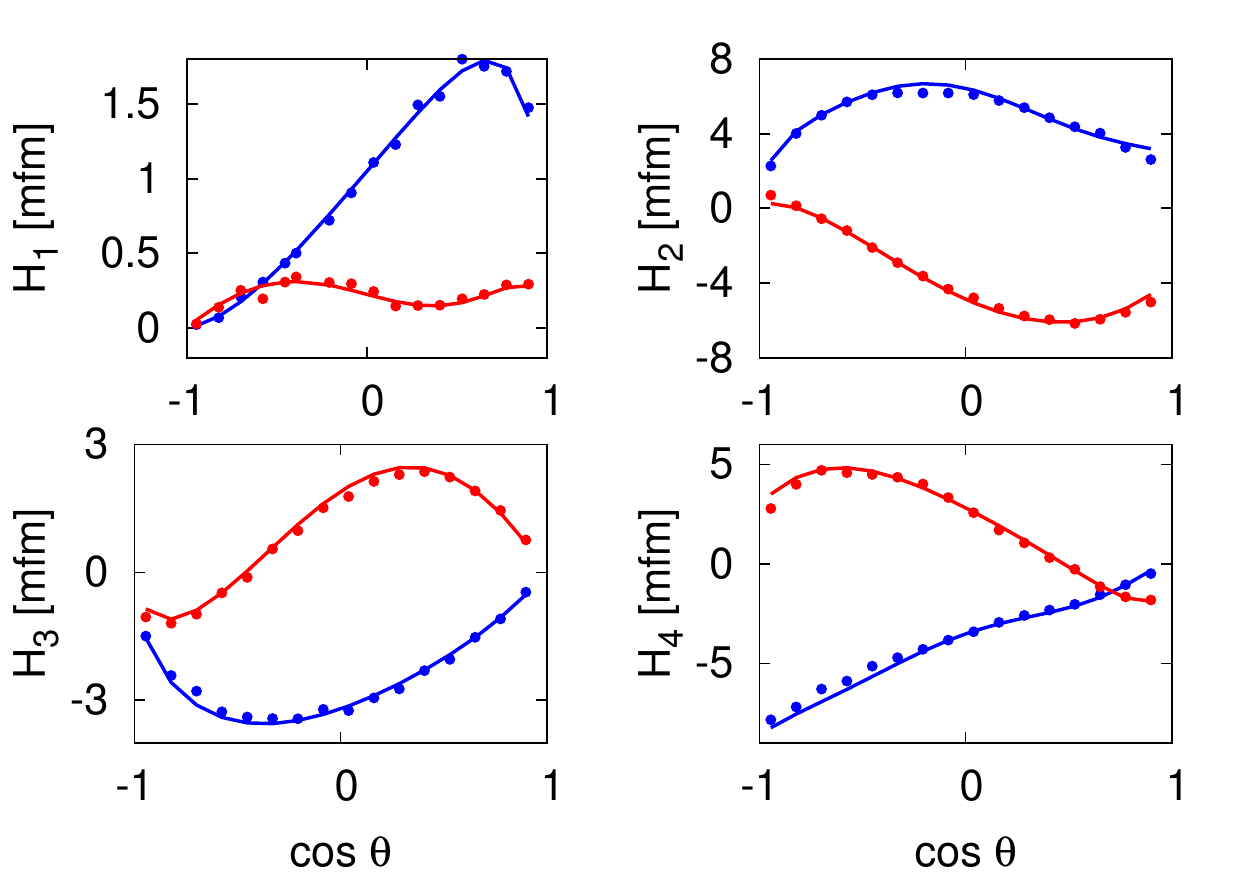}
\vspace{3mm} \caption{\label{FigHelFitSEpseudo} Helicity amplitudes
$H_{k}(W,\cos\theta)$ after final (2nd) iteration at $W=1543\,$MeV
(left) and $W=1660\,$MeV (right). Real and imaginary parts of the
helicity amplitudes (blue and red dots) are obtained from
independent fixed-$t$ AA at different $t$ values in previous (1st)
iteration.  The full lines are the helicity amplitudes from final
iteration in SE PWA.}
\end{center}
\end{figure}

\begin{figure}[htbp]
\begin{center}
\includegraphics[width=8.0cm]{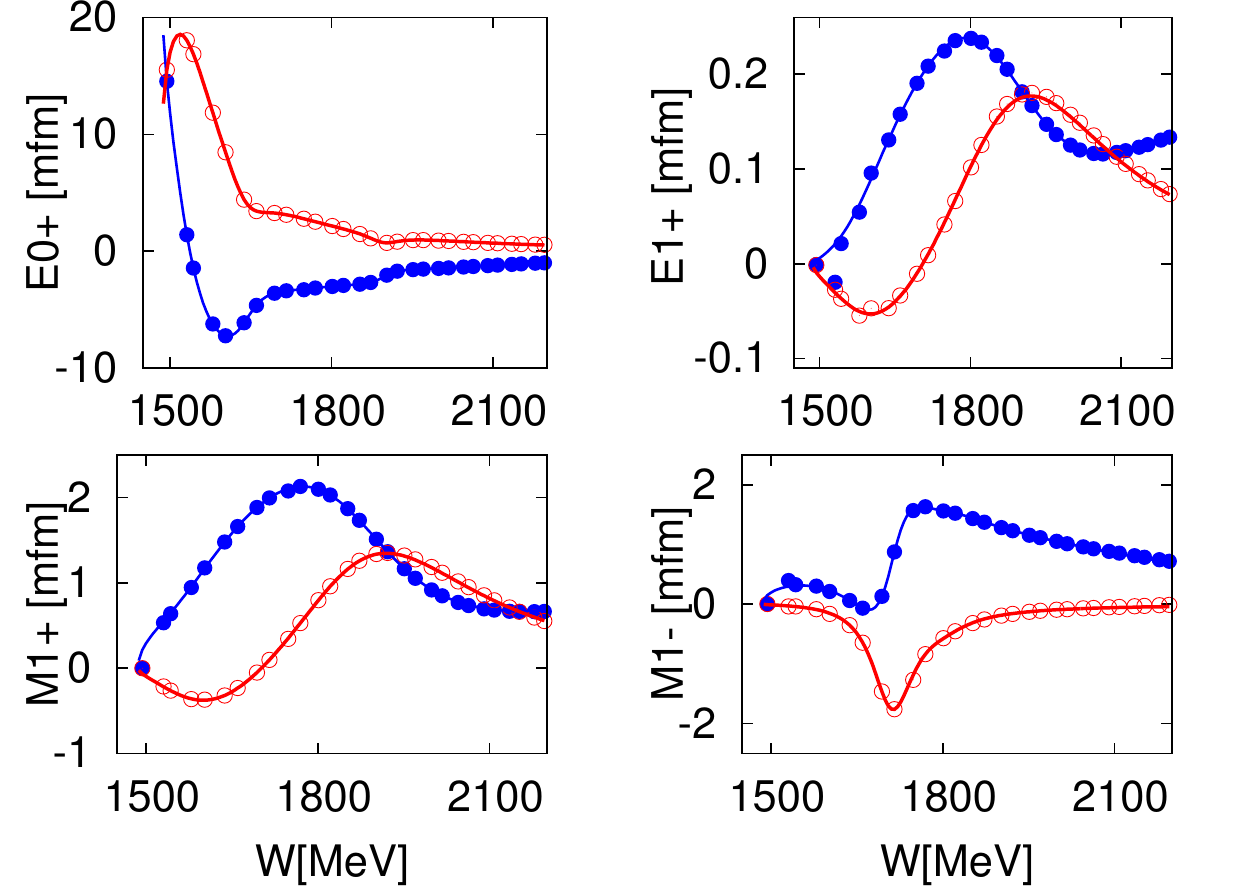}
\includegraphics[width=8.0cm]{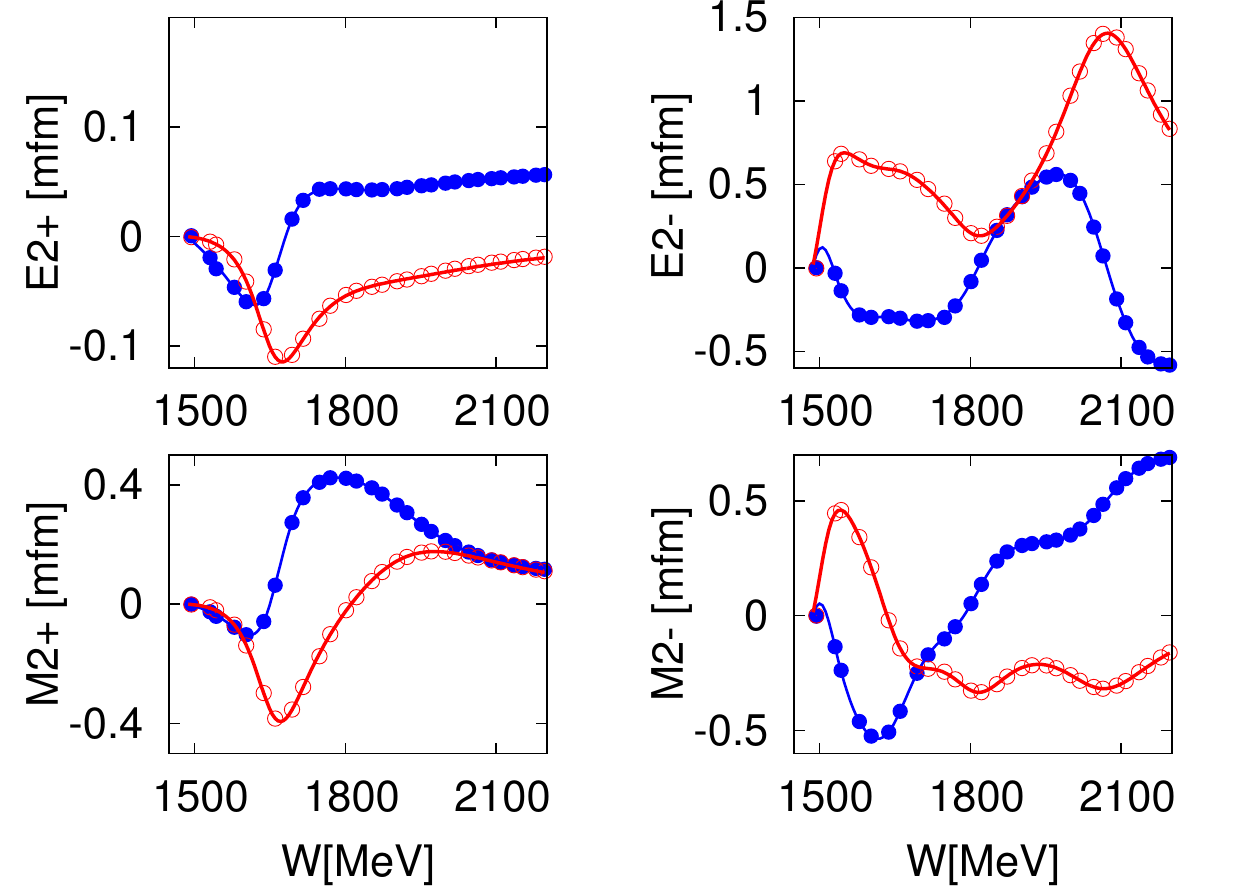}
\includegraphics[width=8.0cm]{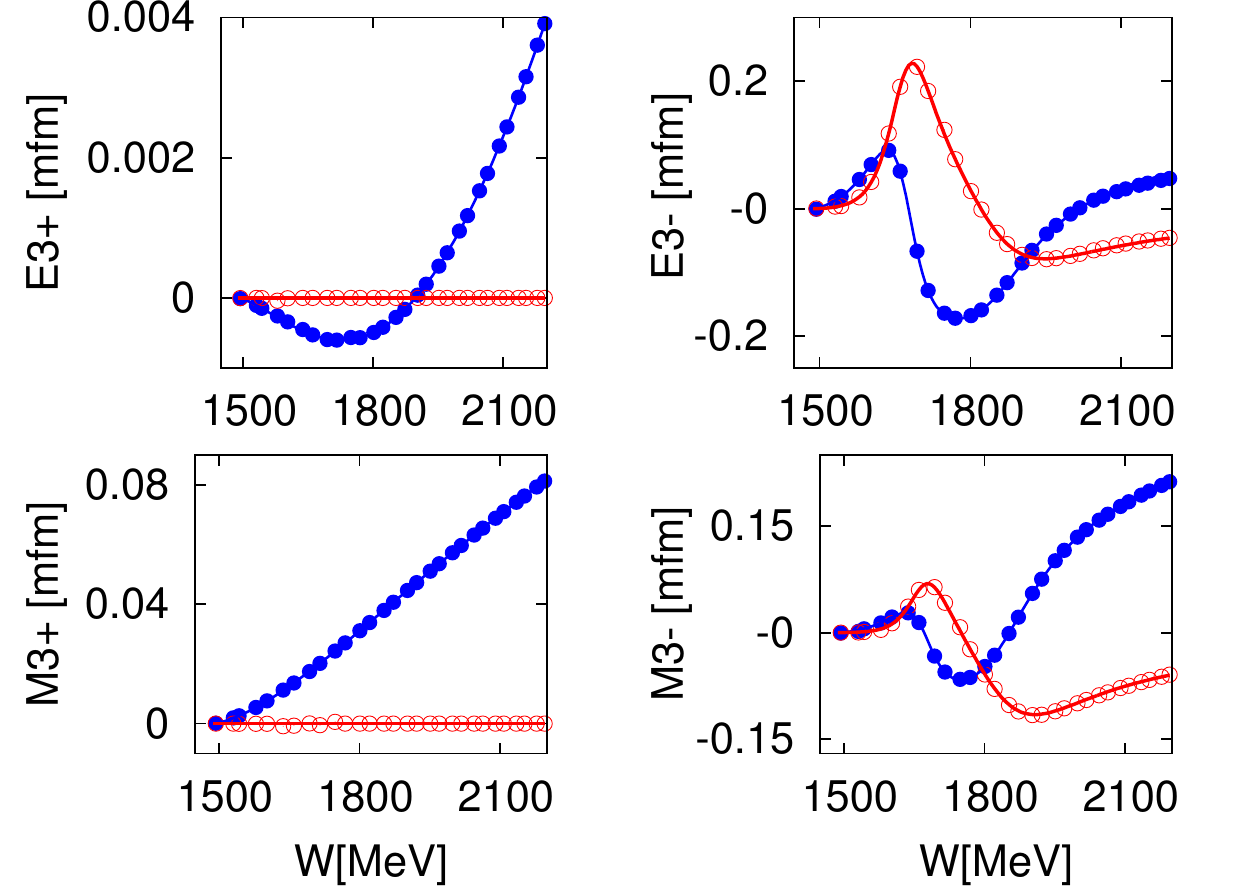}
\includegraphics[width=8.0cm]{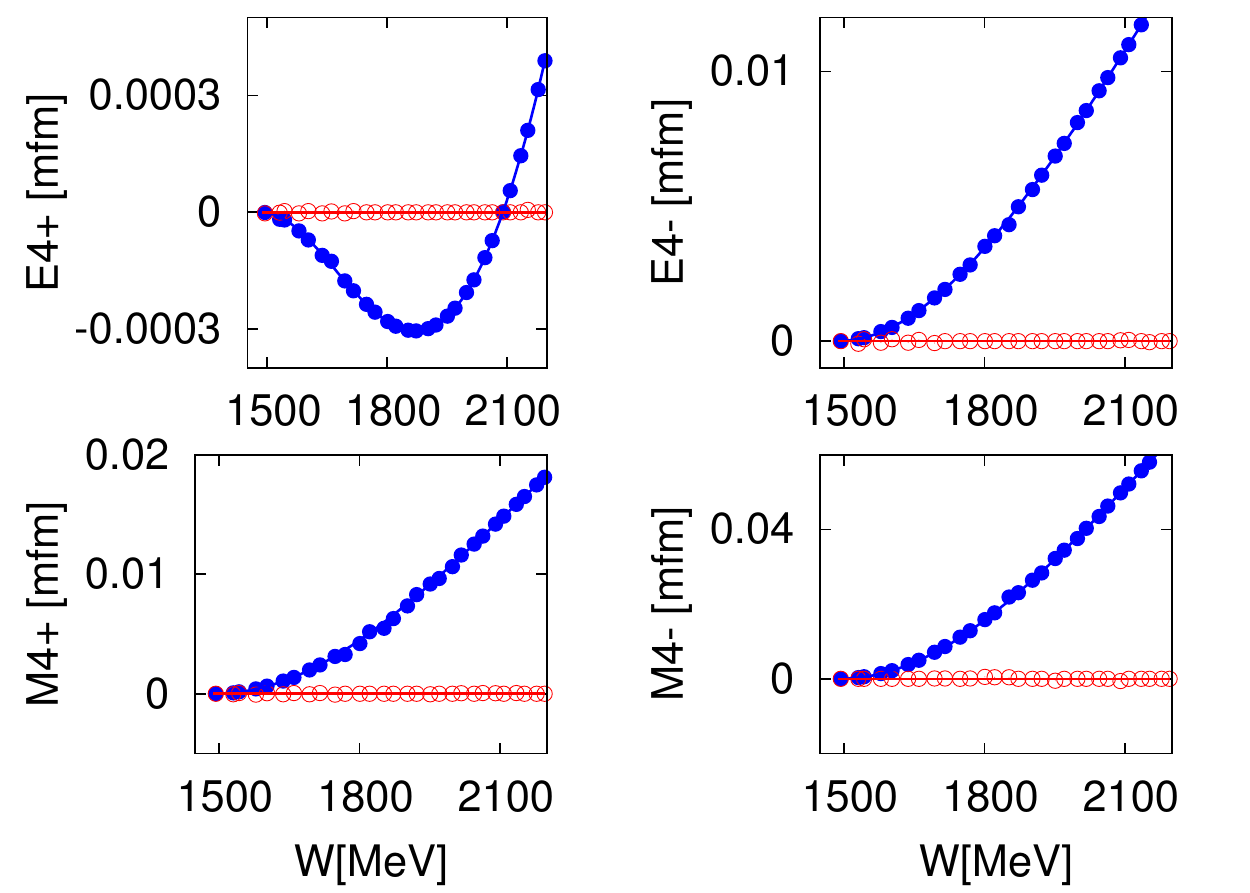}
\vspace{3mm} \caption{\label{FigMultSEpseudo} Real (blue) and
imaginary (red) parts of electric and magnetic multipoles up to
$L=4$. The points are the result of the analytically constrained
single-energy fit to the pseudo data and are compared to the
multipoles of the underlying EtaMAID-2015 model, shown as solid
lines.}
\end{center}
\end{figure}

After only two iterative steps no further change in the helicity
amplitudes is observed. We conclude that the final solution is
obtained. In Fig.~\ref{FigHelFitSEpseudo}, the helicity amplitudes
$H_{k}(W,\cos\theta)$ of the final solution at $W=1543$~MeV and
$W=1660$~MeV are compared to corresponding helicity amplitudes from
previous FT amplitude analysis.
All helicity amplitudes of the generating model are restored. The
multipoles obtained in this fixed-$t$ constrained SE PWA are
compared to the input multipoles of the EtaMAID model in
Fig.~\ref{FigMultSEpseudo}. Indeed, a unique solution was obtained,
which is in perfect agreement with the underlying model.


\subsection{\boldmath Pseudo data analysis with 4 observables and fixed-$t$ analyticity constraints}

As final test with pseudo data, we reduce the number of observables
to 4. We repeat the iterative fitting procedure now using only
pseudo data for $\sigma_0,\check{\Sigma},\check{T},\check{F}$, i.e.
the same observables which have been measured experimentally. This
set does not correspond to a complete experiment and, furthermore,
it does not fulfill the completeness requirements of a TPWA as was
found in Ref.~\cite{Workman:2016irf}. Therefore, we cannot expect a
unique solution.
\\
\underline{\emph{{\bf Step 1}}: Fixed-$t$ amplitude analysis }

\begin{figure*}[htb]
\begin{center}
\includegraphics[width=8.0cm]{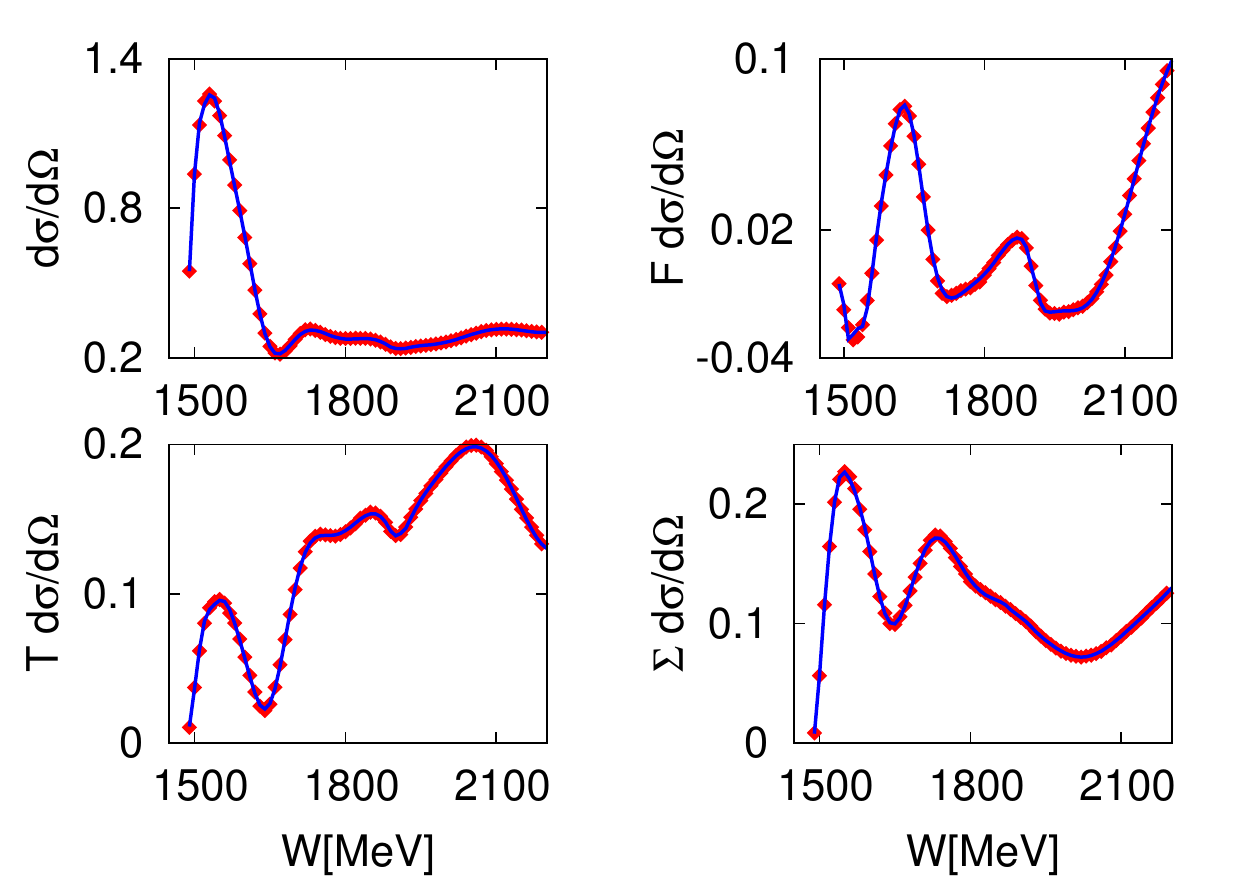}
\includegraphics[width=8.0cm]{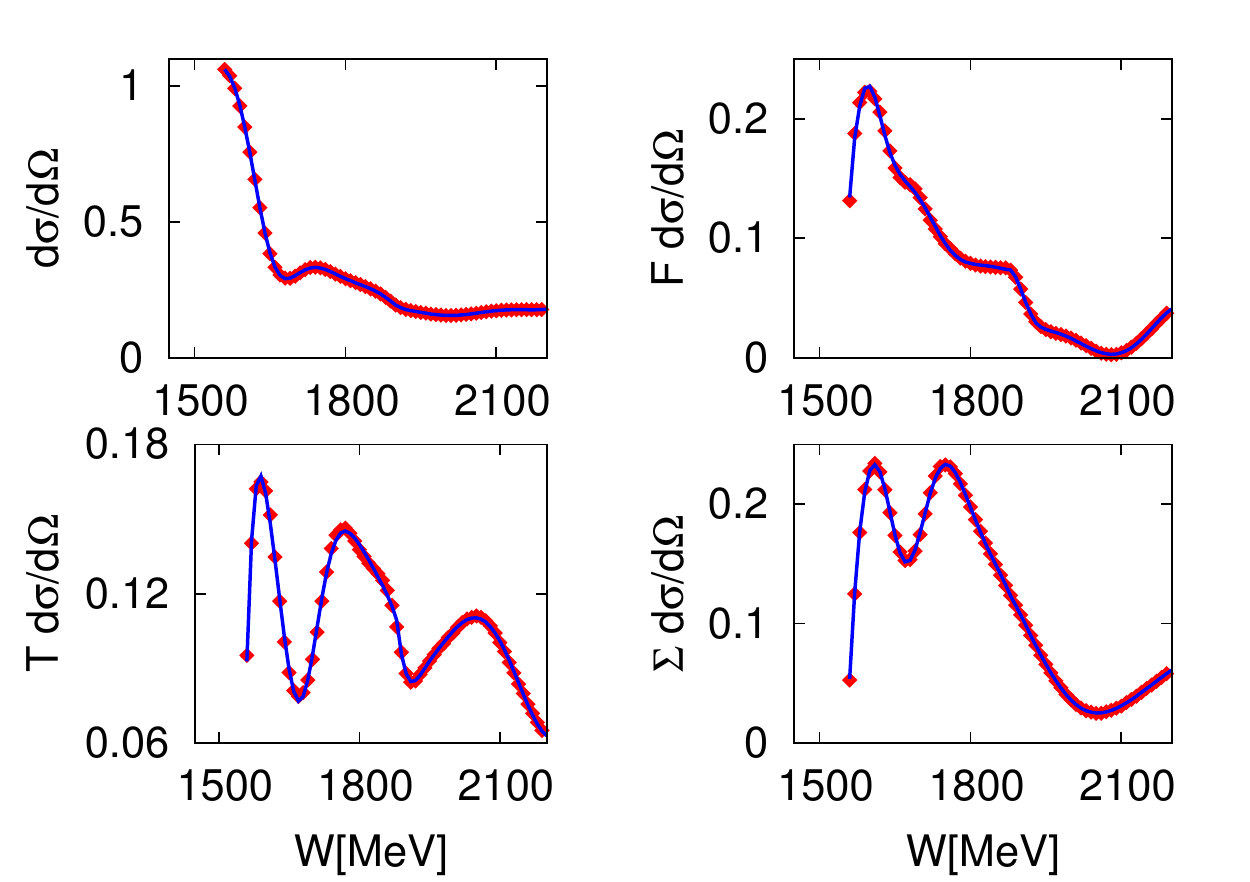}
\vspace{3mm} \caption{\label{FigPseudoFitFT02_4Obs} Fixed-$t$
amplitude analysis of four observables of pseudo data at
$t=-0.2\,$GeV$^2$ (left) and $t=-0.5\,$GeV$^2$ (right). The data
points are pseudo data with a precision of $0.1\%$. The lines are
obtained from the Pietarinen expansion with fitted coefficients.}
\end{center}
\end{figure*}

In the first step again all four observables can be well described
in the fixed-t amplitude analysis. Examples of fits at
$t=-0.2$~GeV$^2$ and $t=-0.5$~GeV$^2$ are shown in
Fig.~\ref{FigPseudoFitFT02_4Obs}.

However, as the set of four observables does not form a complete
experiment, we can now not expect that all other observables are
described as well. Therefore, the helicity amplitudes,  which are
used as  constraint in the following step, are not unique up the an
overall phase as it was the case in the previous study with 8
observables.
\\
\underline{\emph{{\bf Step 2}}: A constrained SE fit
using constraints of Step 1.}

\begin{figure}[htb]
\begin{center}
\includegraphics[width=8.0cm]{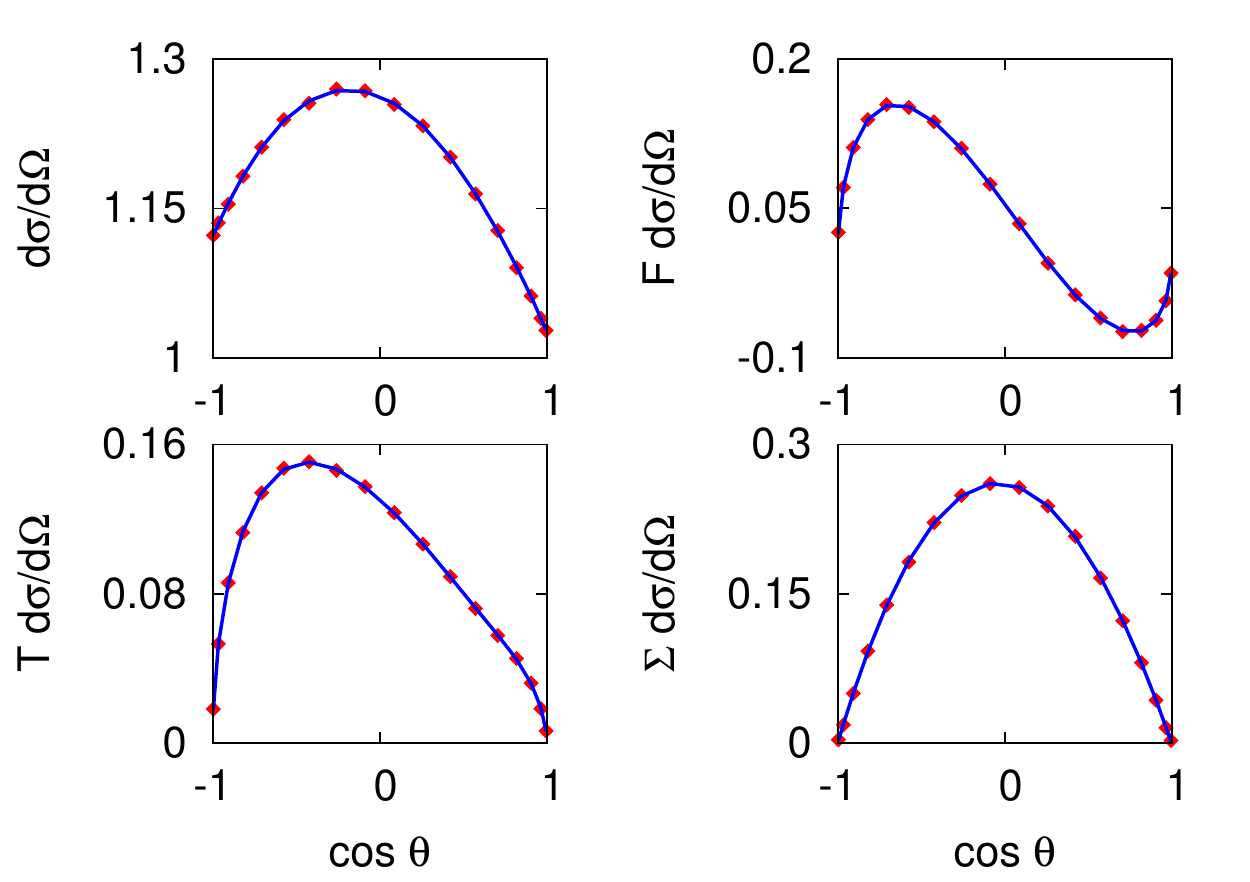}
\includegraphics[width=8.0cm]{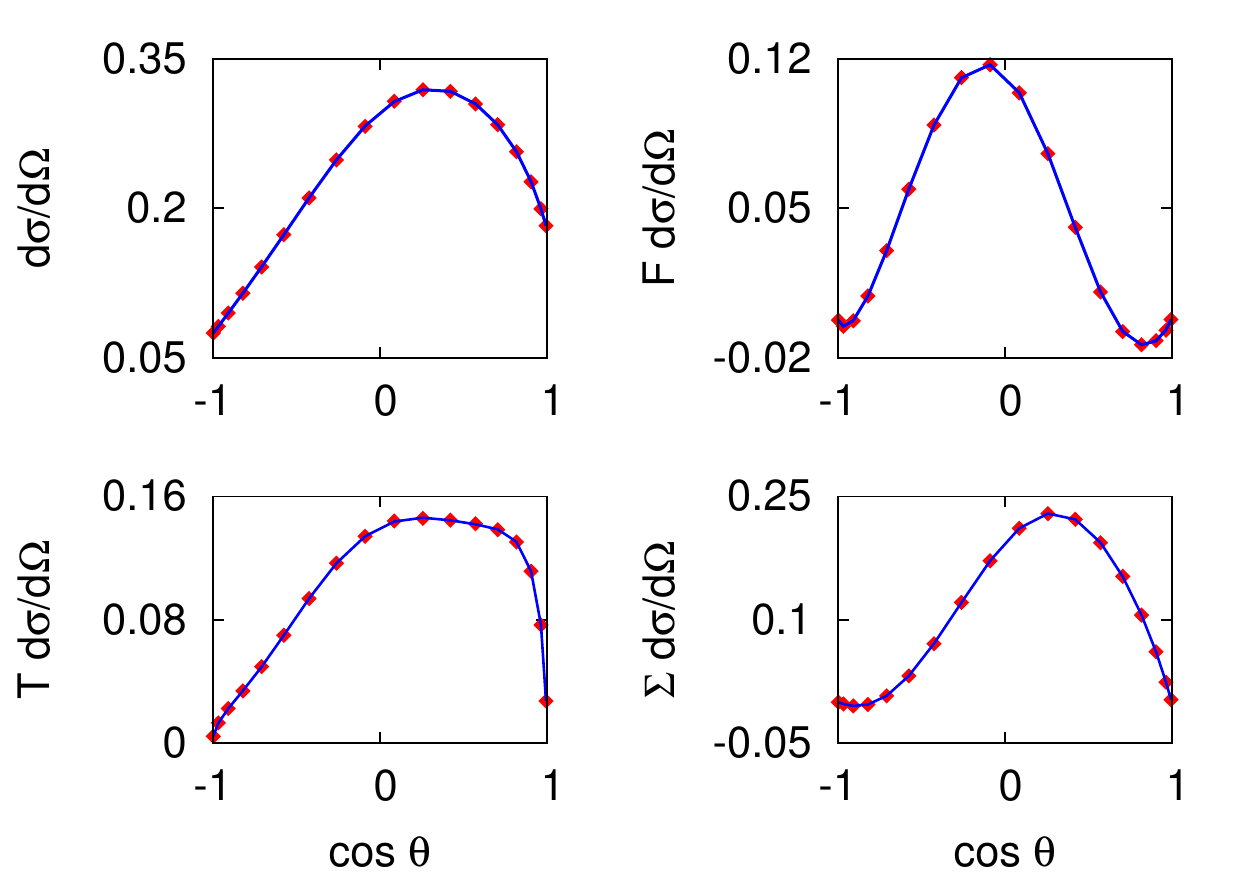}
\vspace{3mm} \caption{\label{FigPseudoFitSE1_4Obs} Single energy
partial wave analysis of a set with four observables of pseudo data
at $W=1543$~MeV (left) and $W=1770$~MeV (right). The data points are
pseudo data with a precision of $0.1\%$. The lines are the results
of the fixed-$t$ constrained single-energy fit describes in the text.}
\end{center}
\end{figure}

In single energy partial wave analysis we again fitted 40 real
parameters up to $L_{max}=5$. The results at  $E=1200$~MeV
($W=1770$~MeV) and $E=1460$~MeV ($W=1900$~MeV) are shown in
Fig.~\ref{FigPseudoFitSE1_4Obs}.

\newpage

\begin{figure}[htb]
\begin{center}
\includegraphics[width=8.0cm]{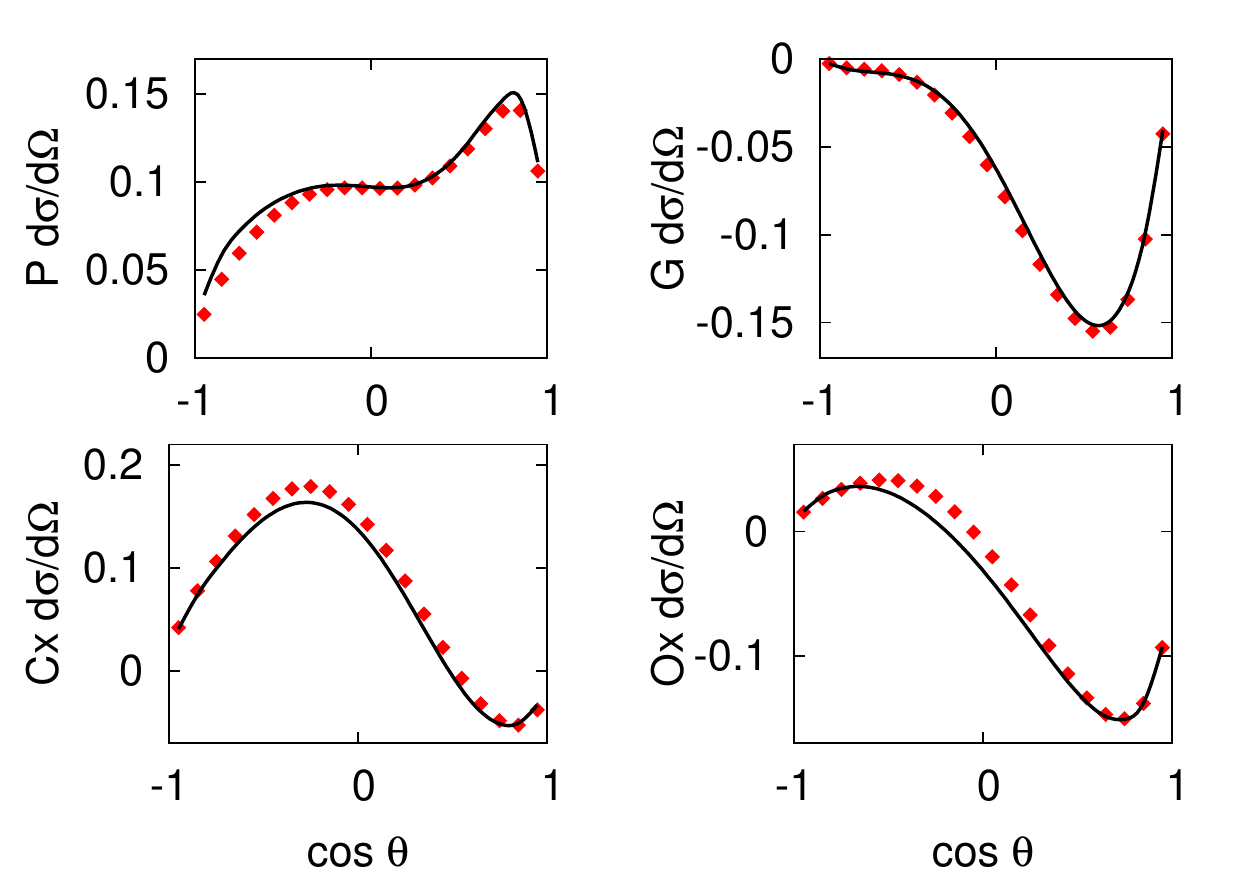}
\includegraphics[width=8.0cm]{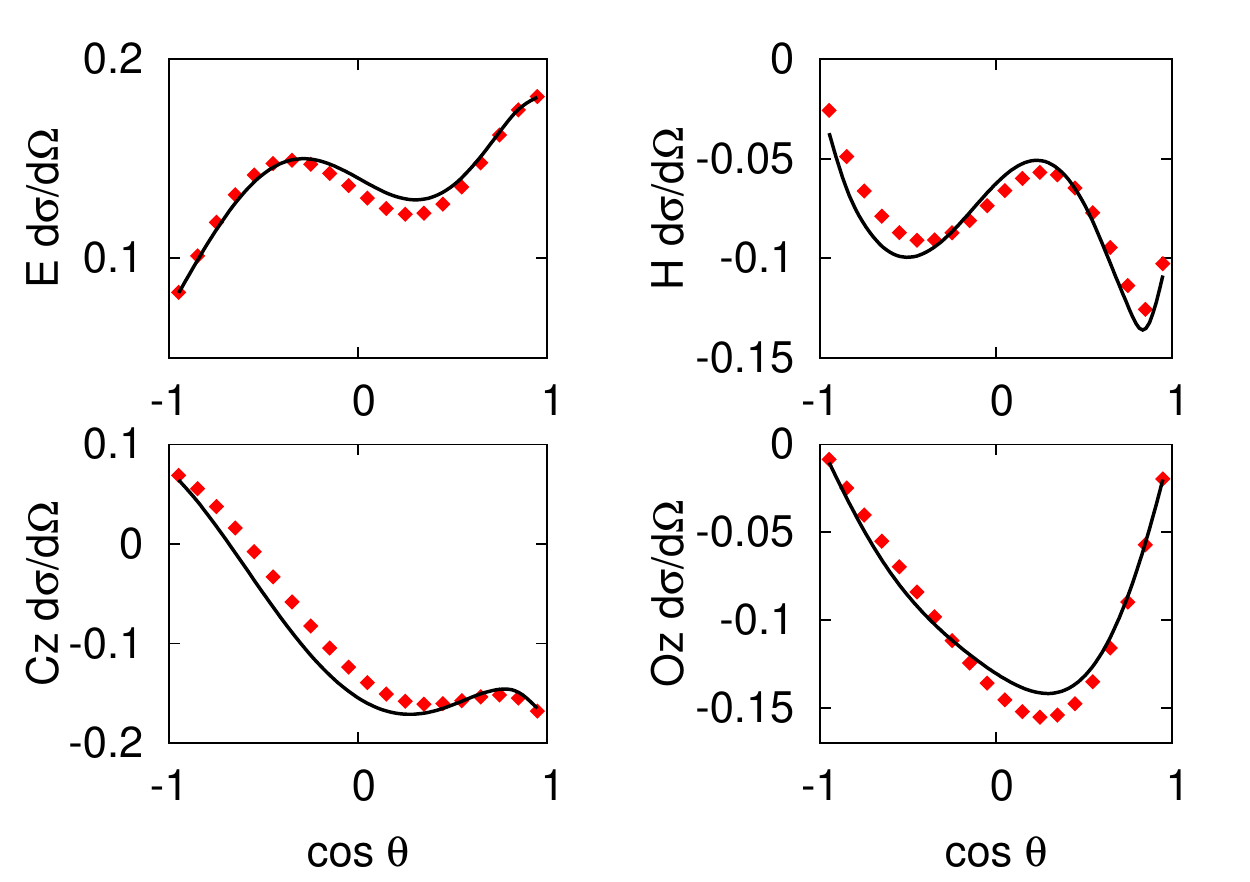}
\includegraphics[width=8.0cm]{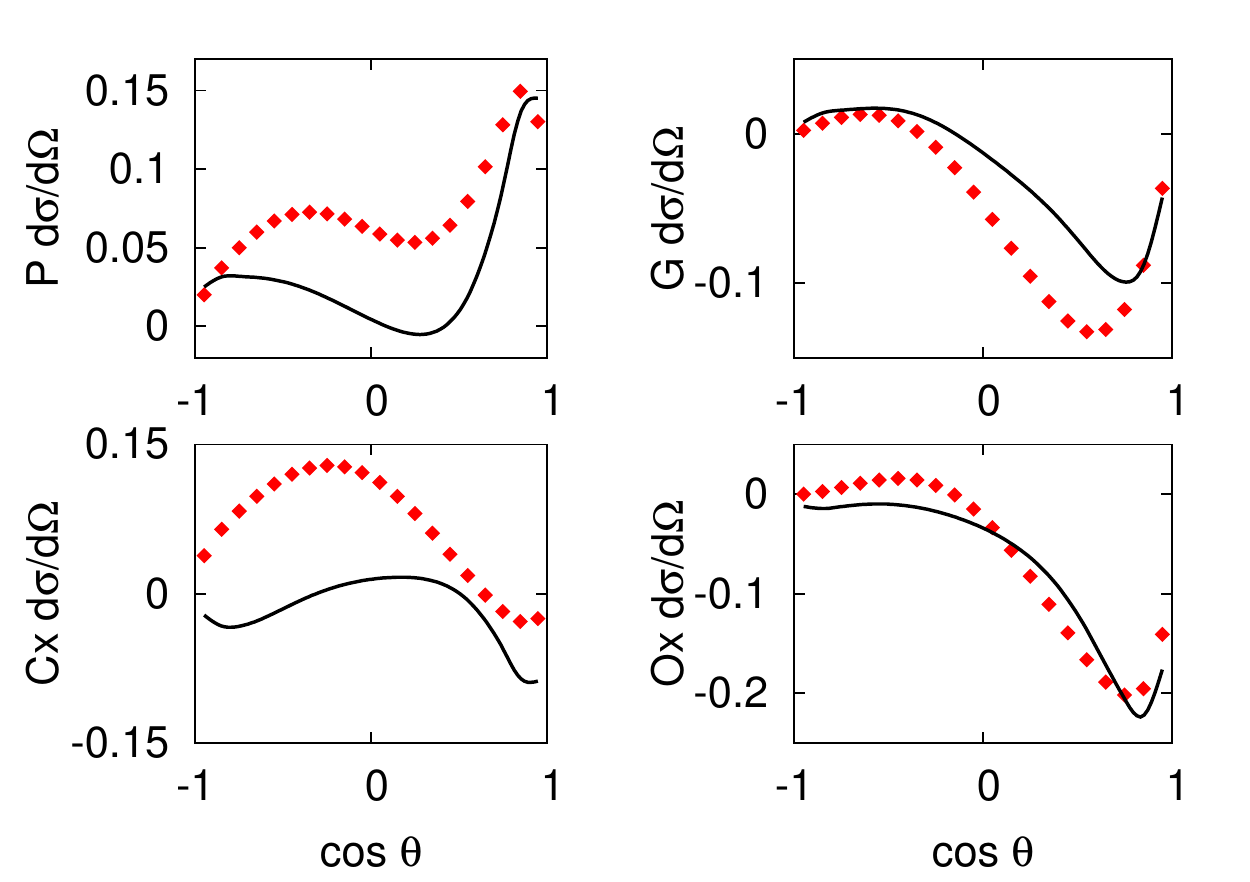}
\includegraphics[width=8.0cm]{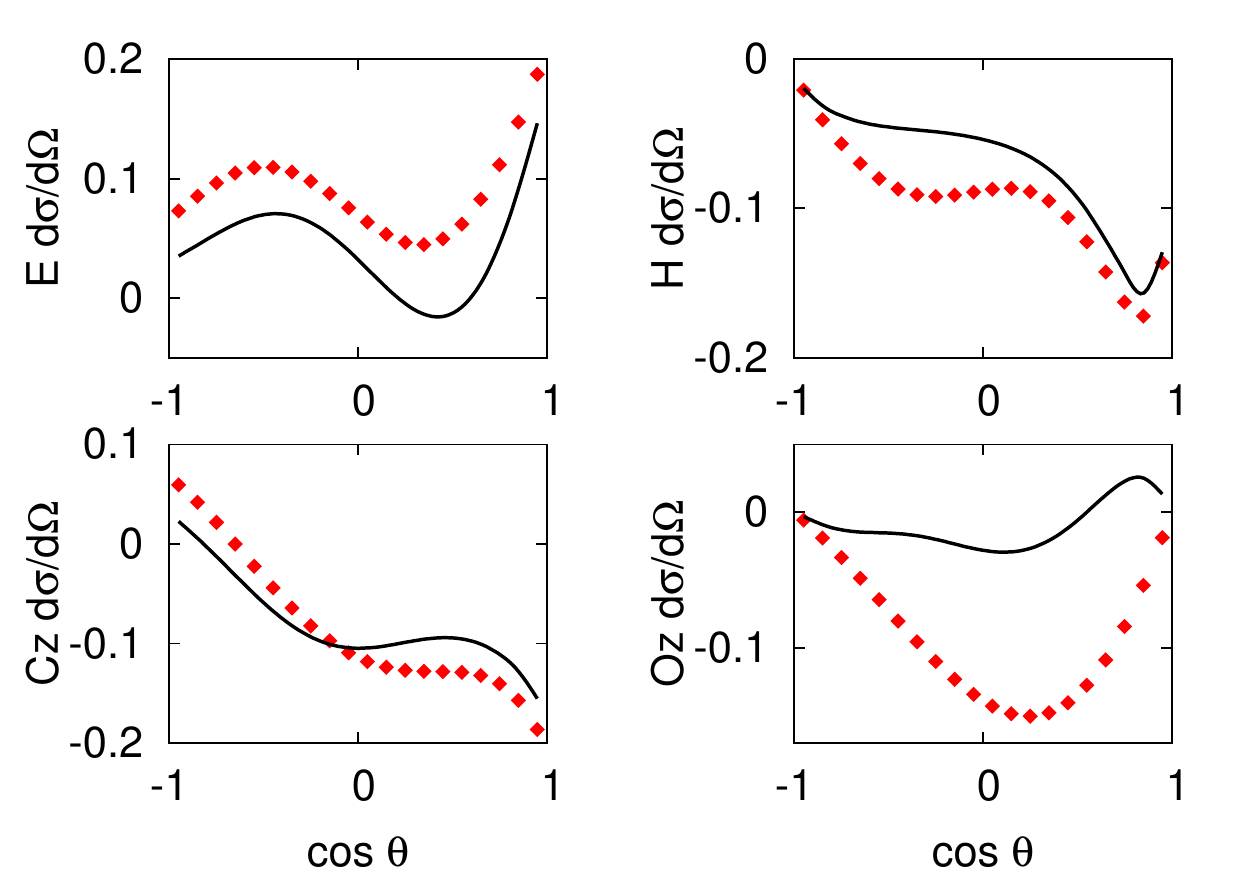}
\vspace{3mm} \caption{\label{Predictions-pseudo1} Predictions for
polarization observables from groups of type ${\cal S, BT}$ and
${\cal BR}$, that are not fitted; shown at $W=1770$~MeV (top) and
$W=1900$~MeV (bottom). The data points are pseudo data with a
precision of $0.1\%$ (red points). The lines are obtained in the fit
with 4 observables $\sigma_0, \check{\Sigma}, \check{T}$ and
$\check{F}$, from the CGLN amplitudes with fitted multipoles up to
$L_{max}=5$.}
\end{center}
\end{figure}

\begin{figure}[hp]
\begin{center}
\includegraphics[width=8.0cm]{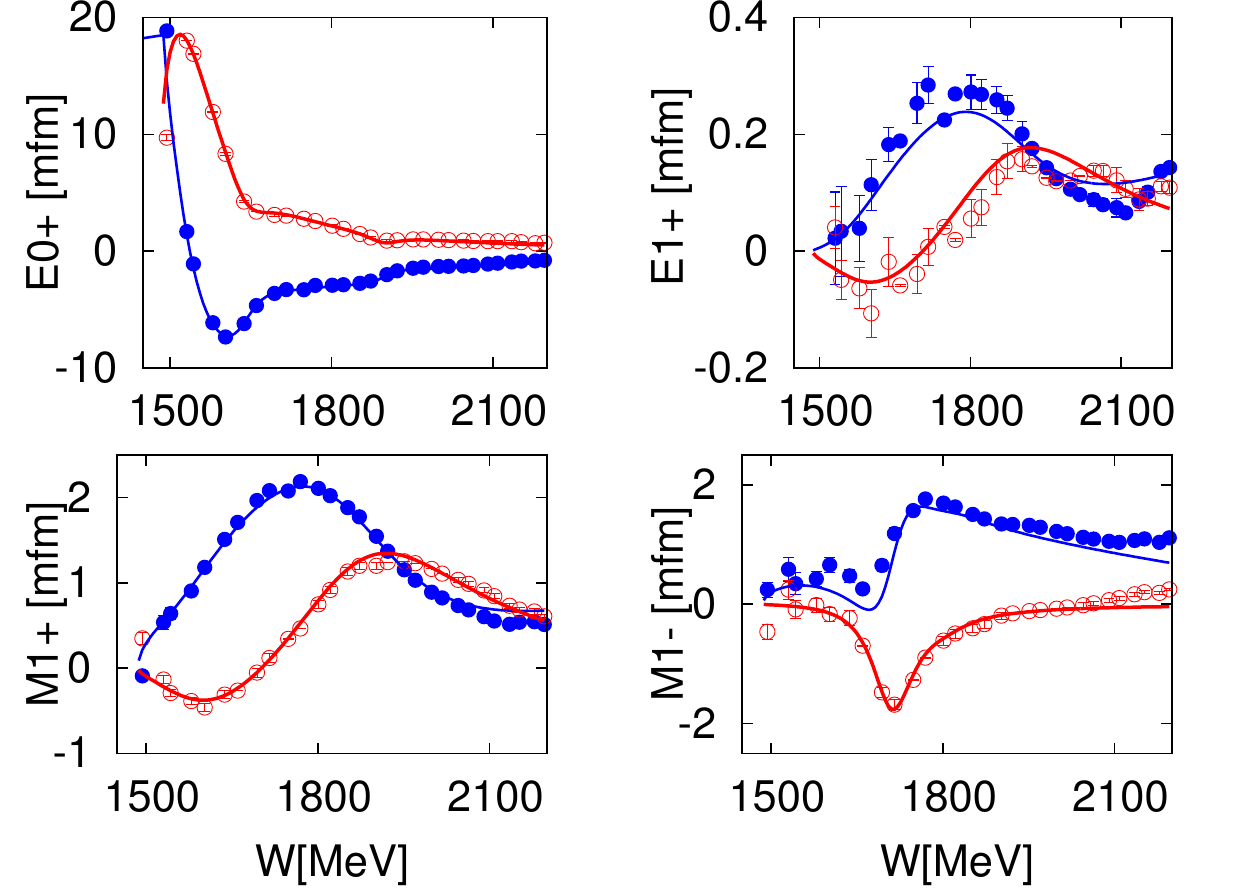}
\includegraphics[width=8.0cm]{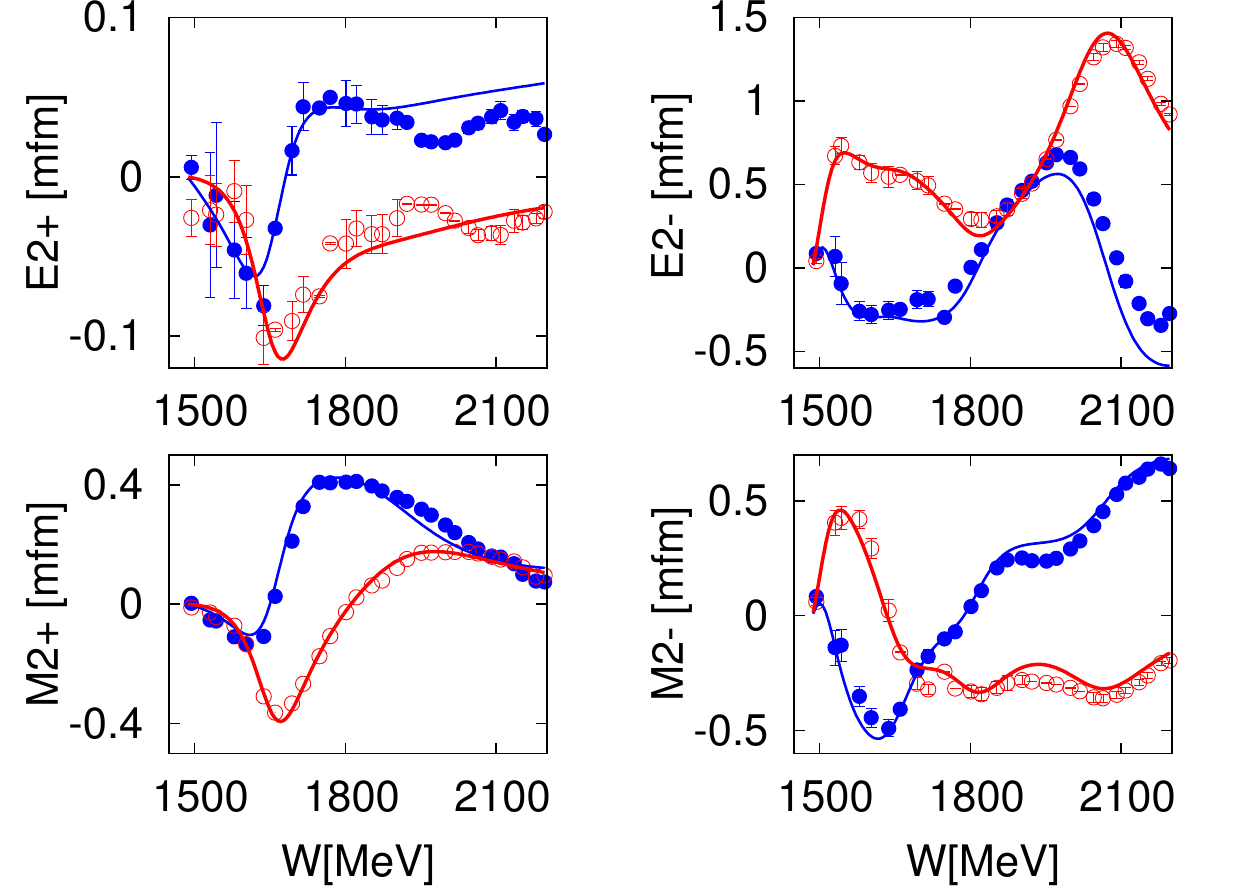}
\includegraphics[width=8.0cm]{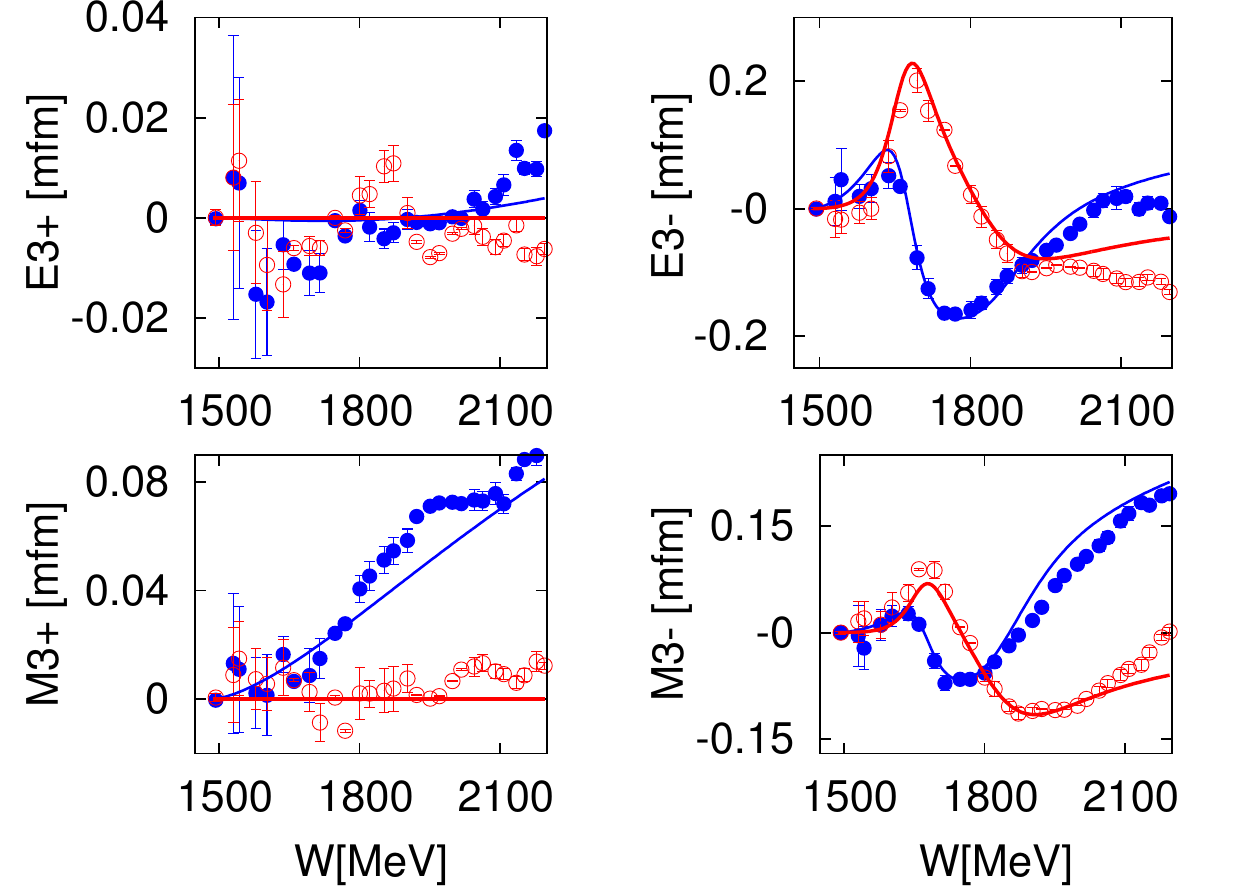}
\includegraphics[width=8.0cm]{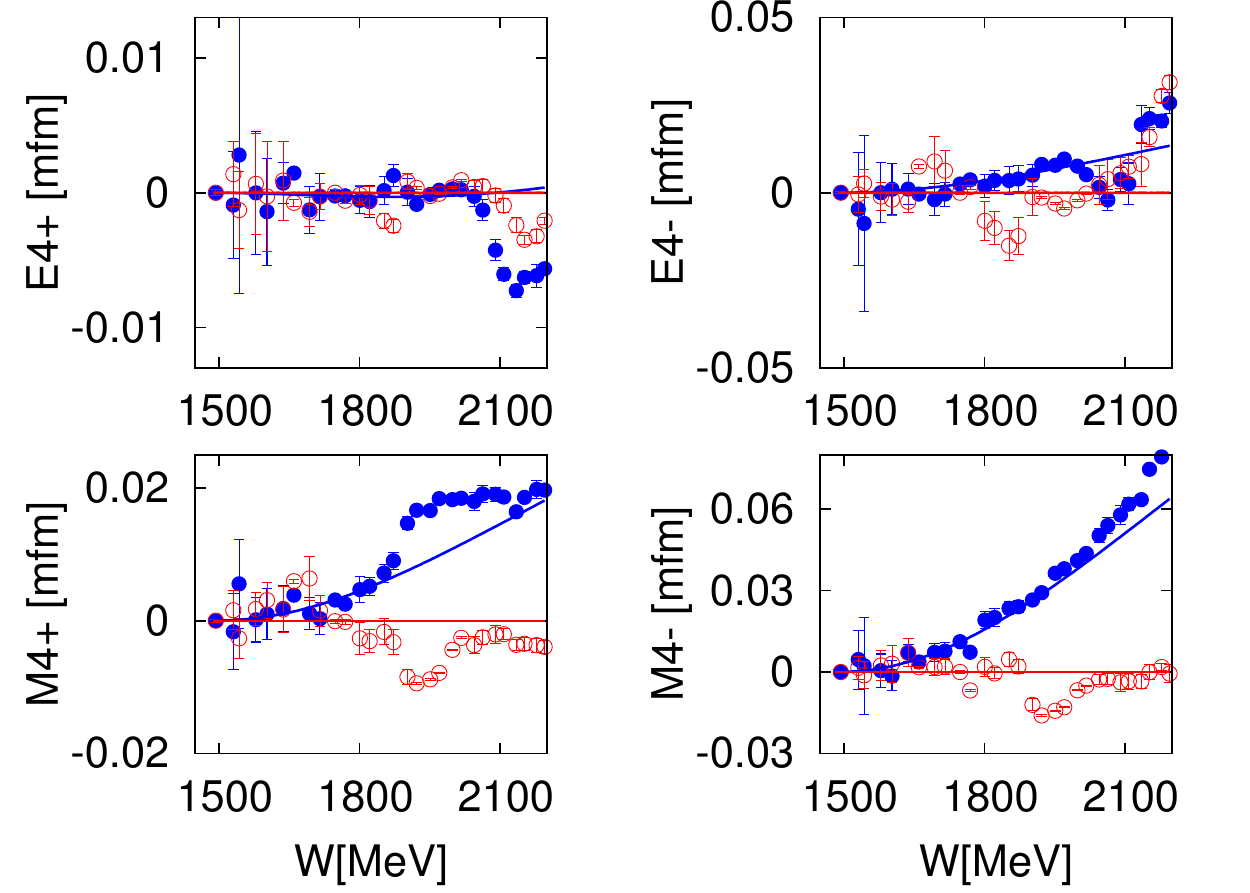}
\vspace{3mm} \caption{\label{FigMultSEpseudo_4Obs} Real (blue) and
imaginary (red) parts of electric and magnetic multipoles up to
$L=4$ after the final iteration. The points are the result of the
analytically constrained single-energy fit to the pseudo data and
are compared to the multipoles of the underlying EtaMAID-2015 model,
solution I, shown as solid \vspace*{-0.5cm} lines.}
\end{center}
\end{figure}

Again, the fitted observables,
$\sigma_0,\check{\Sigma},\check{T},\check{F}$, are perfectly
described. However, we cannot expect that also the remaining 12
observables, which were not fitted, are described. Fig.
~\ref{Predictions-pseudo1} shows this comparison of these remaining
12 observables to the fit result.   While at lower energies, the
prediction calculated from the fitted multipoles are quite good, at
higher energies clear discrepancies are observed. The multipoles,
obtained in the final iteration, are compared to the underlying
model in Fig.~\ref{FigMultSEpseudo_4Obs}. In contrast to fully
unconstrained fits, we do find a solution, which is smooth in
energy. However, in some regions, the multipoles differ from the
"true" values of the input model.

At lower energies, $W\le 1800$~MeV, the SE PWA converges after the
final iteration to the true solution, while at higher
energies some multipoles deviate significantly.
However, in principle these ambiguities can be resolved by adding further
polarization observables as $P$ or $O_z$ (see Fig.~\ref{Predictions-pseudo1}).


\section{Results with experimental data}

In this section we will apply our iterative procedure with fixed-$t$
analyticity constraints to a partial wave analysis of experimental
data on eta photoproduction. Unlike the very successful test with
pseudo data, we have to consider some limitations by using real
data. First of all, we do not have a complete data set with 8
polarization observables available. This is, however, not
stringently required. For a truncated partial wave analysis it was
shown~\cite{Workman:2016irf} that already a minimal set of 4
observables, e.g. $\{\sigma_0, \Sigma, F, H\}$ or $\{\sigma_0, T, P,
F\}$ would be sufficient, providing that the observables are free of
uncertainties. Even very small errors below $0.1\%$ can lead to
multiple solutions. And this becomes more problematic, when the
truncation level increases. In the ambiguity studies of
Omelaenko~\cite{Omelaenko:1981cr,Wunderlich:2014xya} sets of 5
observables, as $\{\sigma_0, \Sigma, T, P, F\}$ were discussed,
which, in numerical analysis, show up more robust than the minimal
sets of four. Such sets will soon become available, when all data
that are presently in progress, are finally analyzed. Currently, we
only have an incomplete data set $\{\sigma_0, \Sigma, T, F\}$
available, for which the limitations have already been discussed in
the previous section by using pseudo data. A further limitation will
always be the statistical and systematical uncertainties of real
data, which will be much larger than in our pseudo data test. But
with the expected data base of all eight observables obtained with
beam and target polarization, part of this deficiency can be
compensated with more than four observables. The same is true for
the limitation in kinematical coverage. The best coverage in energy
and angles is obtained for unpolarized cross sections. But a very
good coverage is also available for the polarization data, that we
will use, at least up to a total c.m. energy of $W=1.85$~GeV.

\subsection{\boldmath $\eta$ photoproduction data base}
\label{sec:data-base}

In our partial wave analysis with experimental data we have used
recent A2@MAMI data for unpolarized differential cross section
$\sigma_0$, single target polarization asymmetry $T$ and double
beam-target polarization with circular polarized photons $F$. In
addition we have used the GRAAL data for single beam polarization
$\Sigma$. For details, see Table~\ref{tab:expdata}.

\begin{table}[htb]
\begin{center}
\caption{\label{tab:expdata} Experimental data from A2@MAMI and
GRAAL used in our PWA.}
\bigskip
\begin{tabular}{|c|c|c|c|c|c|c|}
\hline
 Obs        & $N$ & $E_{lab}$~[MeV] & $N_E$  & $\theta_{cm}$~[$^0$] & $N_\theta$ & Reference    \\
\hline
 $\sigma_0$ & $2400$ & $710 - 1395$ & $120$  & $18 - 162$ & $20$ & A2@MAMI(2010,2017)~\cite{McNicoll:2010qk,Kashevarov2017}  \\
 $\Sigma$   & $ 150$ & $724 - 1472$ & $ 15$  & $40 - 160$ & $10$ & GRAAL(2007)~\cite{Bartalini:2007fg} \\
 $T$        & $ 144$ & $725 - 1350$ & $ 12$  & $24 - 156$ & $12$ & A2@MAMI(2014)~\cite{Annand2014prl} \\
 $F$        & $ 144$ & $725 - 1350$ & $ 12$  & $24 - 156$ & $12$ & A2@MAMI(2014)~\cite{Annand2014prl} \\
\hline
\end{tabular}
\end{center}
\end{table}

In our database we have much more energy points for differential
cross sections than for polarization observables (120 vs 15, 12 and
12). In order to avoid the differential cross section dominance in
our fitting procedure we use the same kinematical points also for
the polarization data. So, experimental values of
double-polarization asymmetry $F$, target asymmetry $T$, and beam
asymmetry $\Sigma$ for given angles have to be interpolated at the
energies where $\sigma_0$ are available. We have used a spline
smoothing method~\cite{deBoor} which was similarly applied in the
Karlsruhe-Helsinki analysis KH80~\cite{Hohler84}. Errors of
interpolated data are taken to be equal to errors of nearest
measured data points.

To impose fixed-$t$ analyticity we also have to create a data base
at fixed $t$ using measured angular distributions at fixed $s$. A
fixed-$t$ data point requires a special combination of energy and
angular points that in general can not be found among directly
measured data points. To obtain a data point in fixed $t$ we have to
interpolate between near-by measured observable values and estimate
the corresponding error. This is again done by the spline smoothing
and interpolation method as described before.

\begin{figure}[htb]
\begin{center}
\includegraphics[width=8.cm]{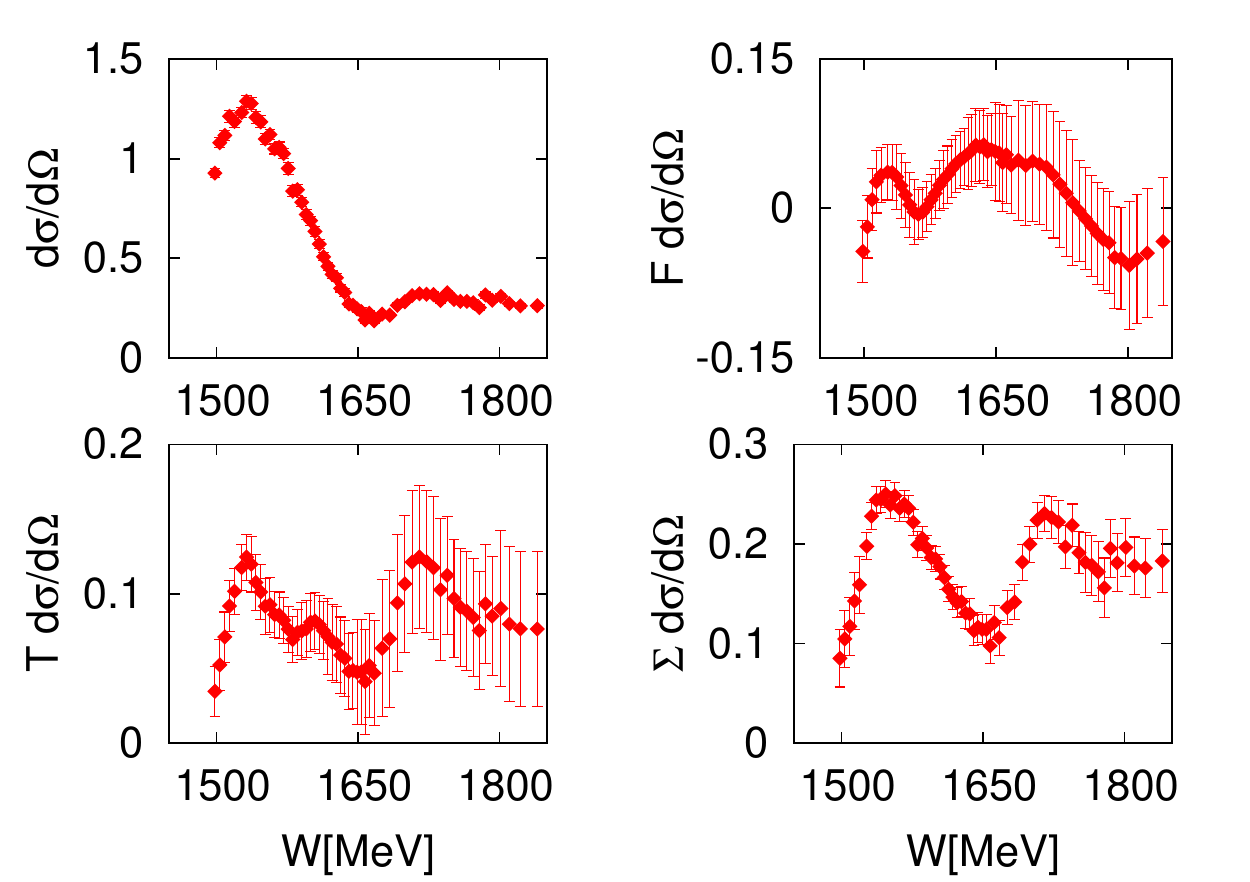}
\includegraphics[width=8.cm]{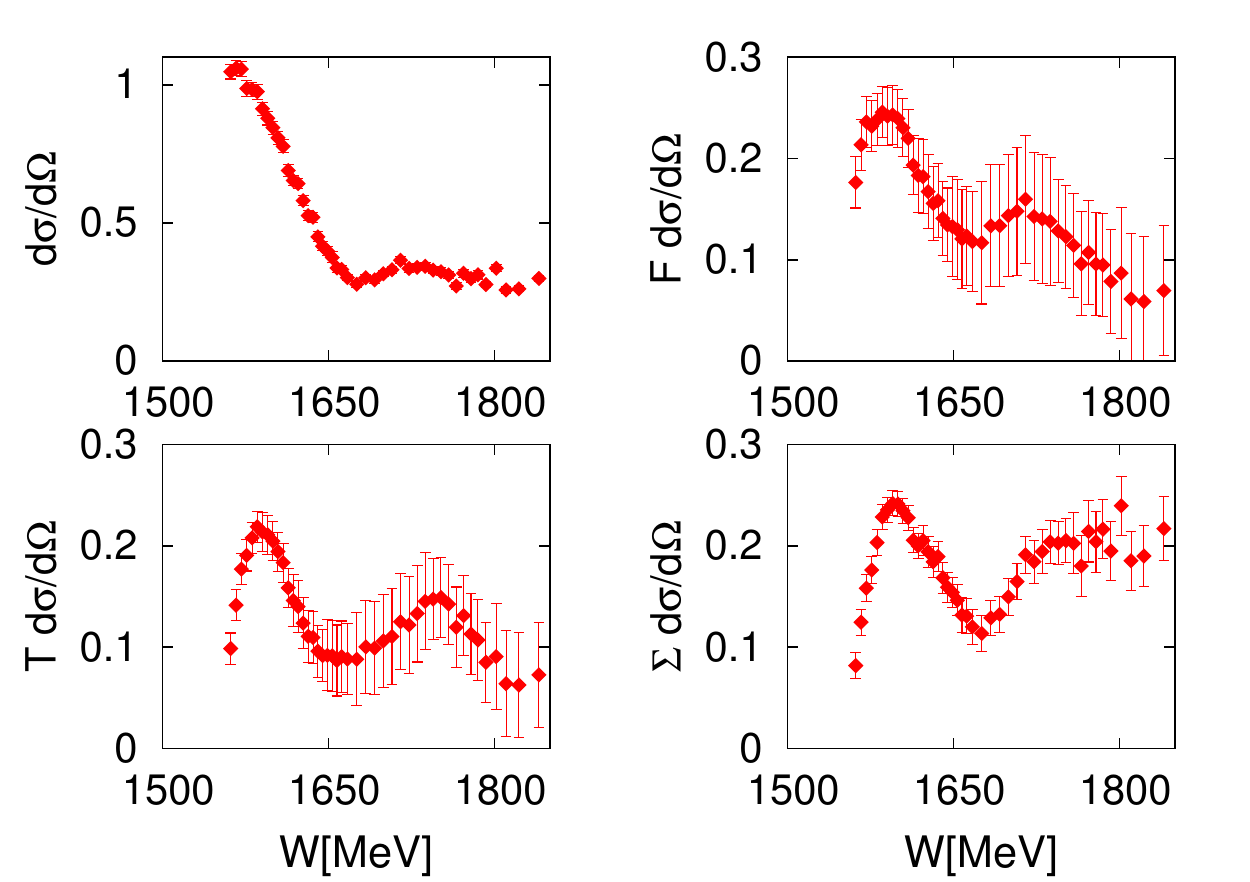}
\vspace{3mm} \caption{\label{FigFTData} Example of our interpolated
fixed-$t$ data base for $t = -0.2\,$GeV$^2$ and  $t =
-0.5\,$GeV$^2$. The threshold values of the physical region are
$W_{thr}=1486$~MeV for $t=-0.2\,$GeV$^2$ and $W_{thr}=1554$~MeV for
$t=-0.5\,$GeV$^2$.}
\end{center}
\end{figure}

Our fixed-$t$ amplitude analysis is performed at $t$ values in the
range $-1.00\,$GeV$^{2}<t<-0.09\,$GeV$^{2}$ with 20 equidistant $t$
values. Examples of interpolated data at $t=-0.2$~GeV$^2$ and
$-0.5$~GeV$^2$ are shown in Fig. \ref{FigFTData}. We note that the
error bars for the polarization observables must be considered as
error bands and the results of our minimization in the fixed-$t$
analysis must not be taken as values of a $\chi^2$ distribution.

\begin{figure}[htb]
\begin{center}
\includegraphics[width=8.cm]{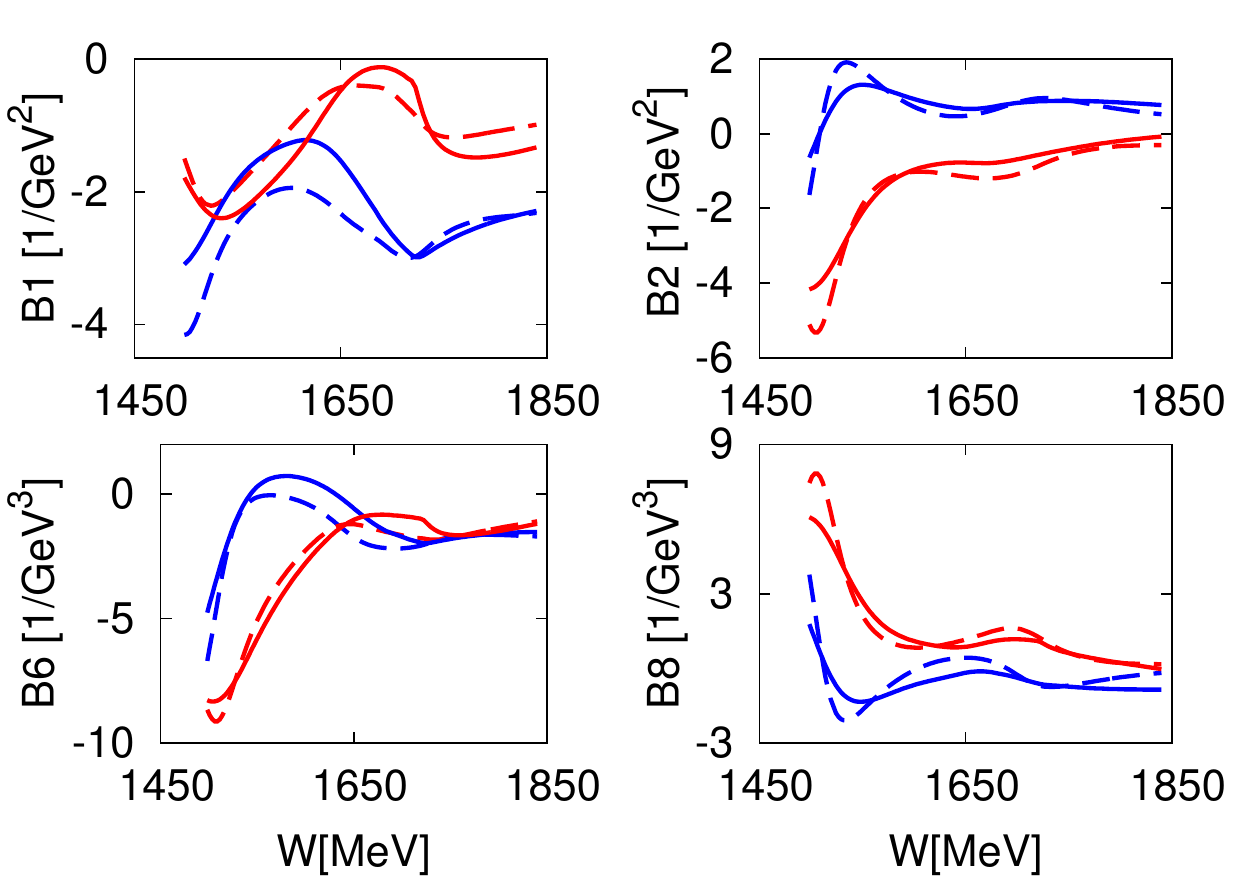}
\includegraphics[width=8.cm]{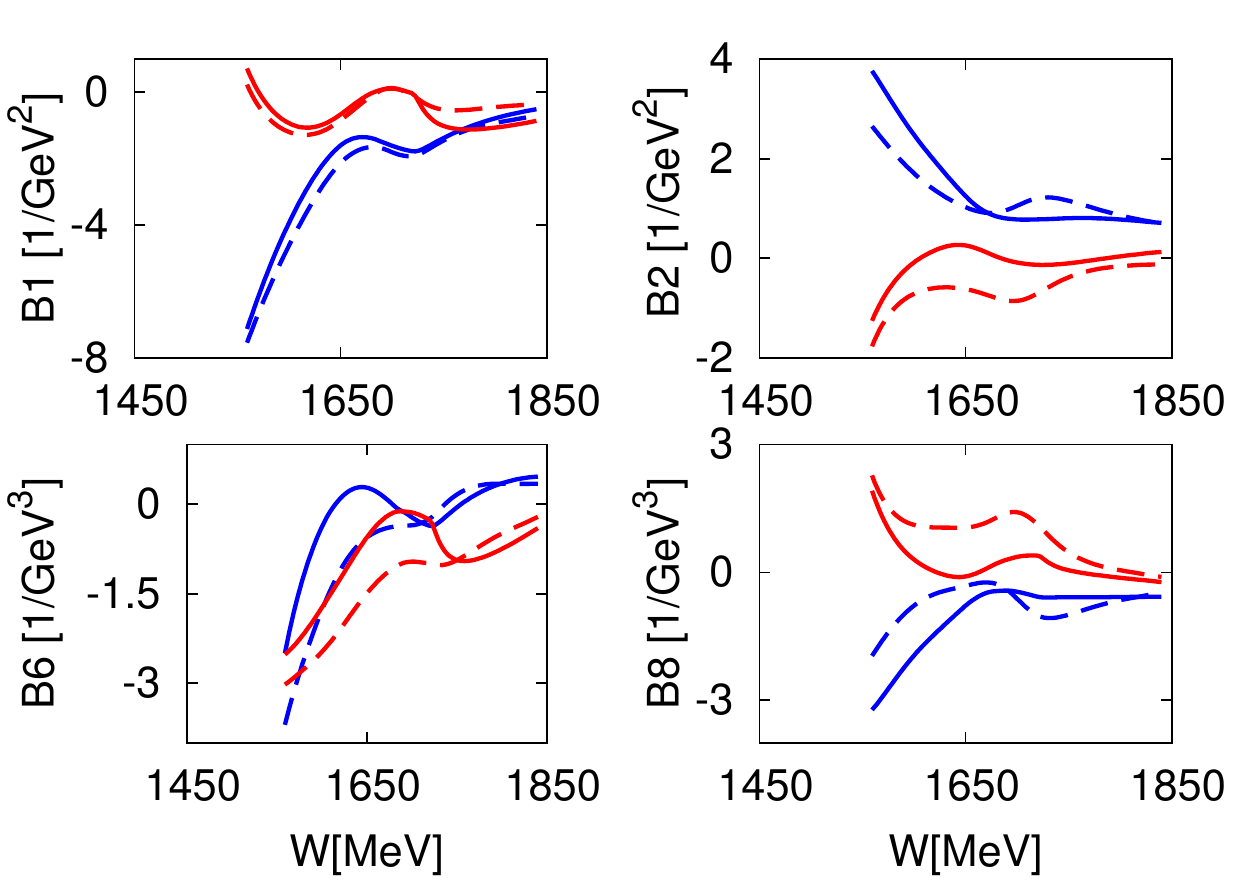}
\vspace{3mm} \caption{\label{FigINV-0200} Invariant amplitudes
$B_1,B_2, B_6$ and $B_8$ as functions of the total c.m. energy $W$
at $t=-0.2$~GeV$^2$ (left) and $-0.5\,$GeV$^{2}$ (right). The blue
and red curves show real and imaginary parts of the amplitudes of
solutions II (dashed) and III (solid) respectively.}
\end{center}
\end{figure}

\begin{figure}[htb]
\begin{center}
\includegraphics[width=8.cm]{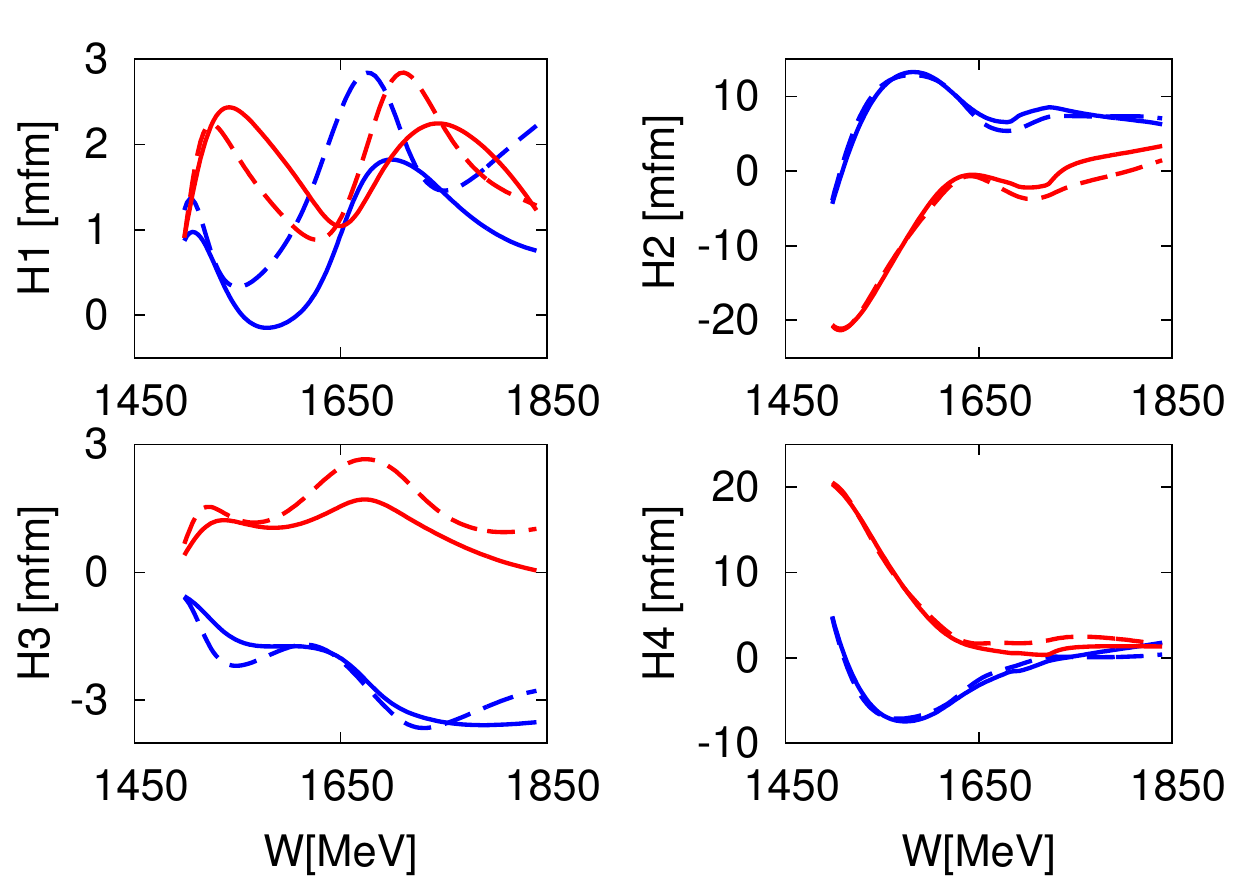}
\includegraphics[width=8.cm]{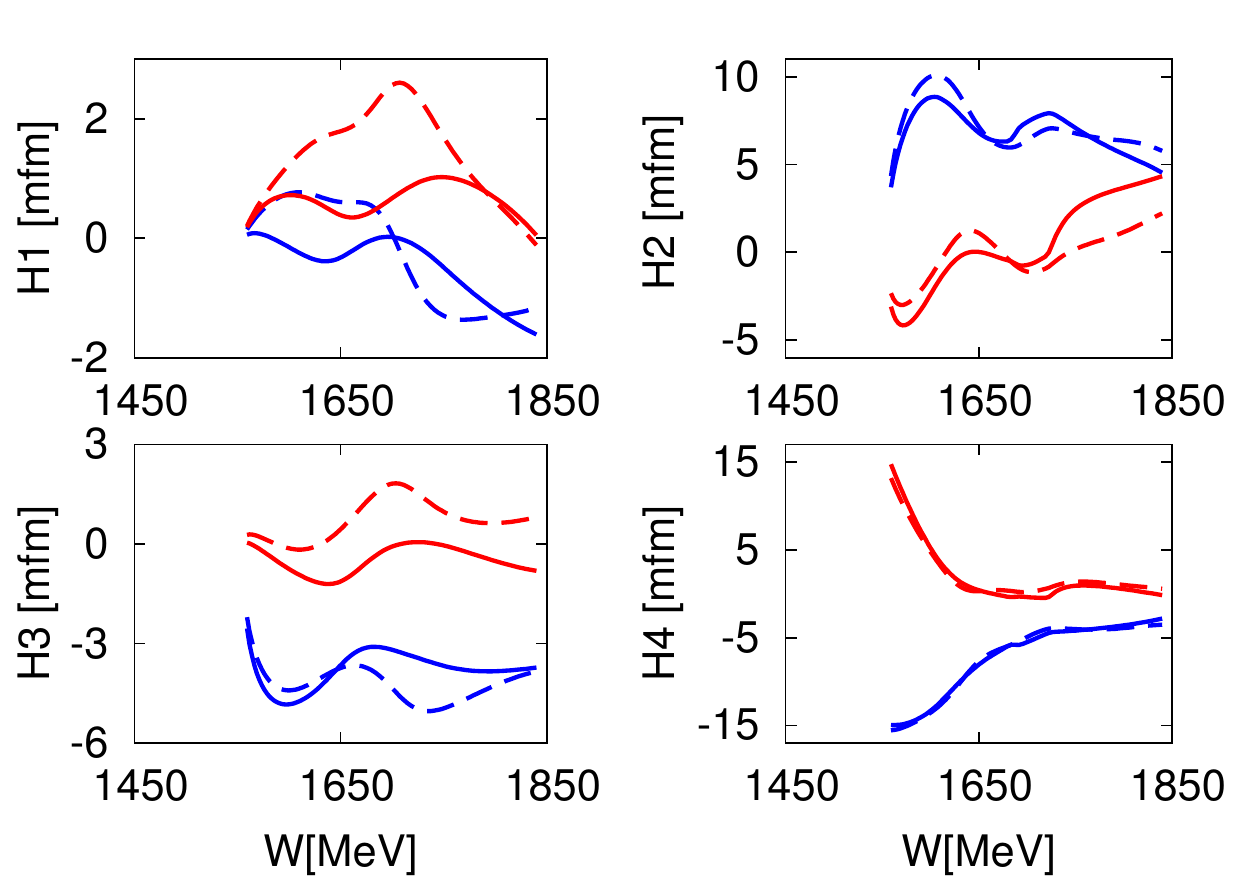}
\vspace{3mm} \caption{\label{FigFTHel} Helicity amplitudes $H_1$,
$H_2$, $H_3$ and $H_4$ as functions of the total c.m. energy W at $t
= -0.2$~GeV$^2$ (left) and $t=-0.5$~GeV$^2$ (right). The blue and
red curves show real and imaginary parts of the amplitudes of
solutions II (dashed) and III (solid) respectively.}
\end{center}
\end{figure}

In order to explore the model dependence of our solution we start
the analysis with two different MAID solutions, which we denote:
Solution II (EtaMAID-2016 \cite{Kashevarov2016}) and Solution III
(EtaMAID-2017 \cite{Kashevarov2017}). To start an analysis, we
randomize the starting values by 50\% around the solutions II and
III. In Fig.~\ref{FigINV-0200} we compare the invariant amplitudes
from initial solutions II and III at $t=-0.2$~GeV$^2$ and
$-0.5\,$GeV$^{2}$ and in Fig.~\ref{FigFTHel} we compare the helicity
amplitudes.
%
The two EtaMAID solutions II and III are significantly different
from each others. This can be seen, especially, where amplitudes
become small. For the largest amplitudes, $H_2$ and $H_4$, where the
$S$-wave multipole dominates, in particular for low energies, the
two models become much more similar.

\begin{figure}[htb]
\begin{center}
\includegraphics[width=8.cm]{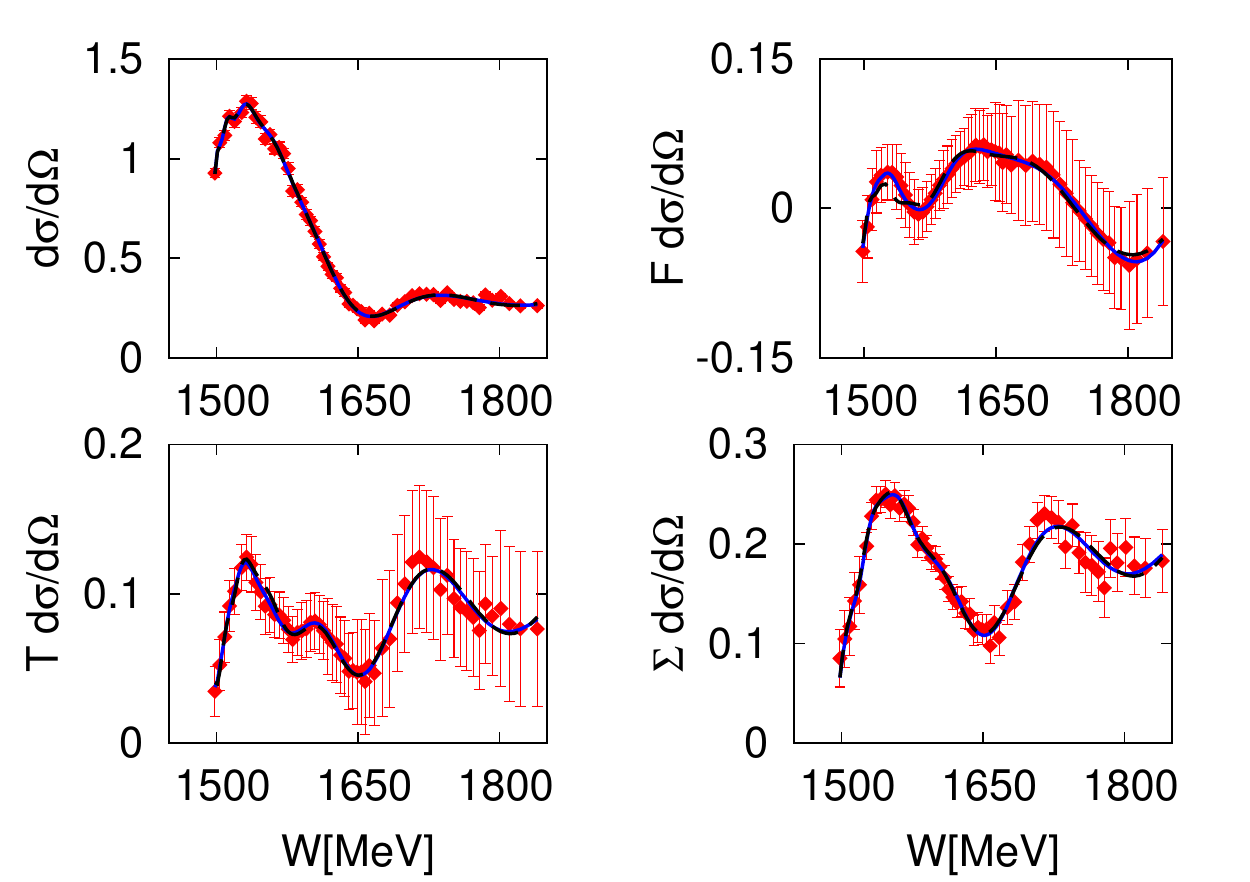}
\includegraphics[width=8.cm]{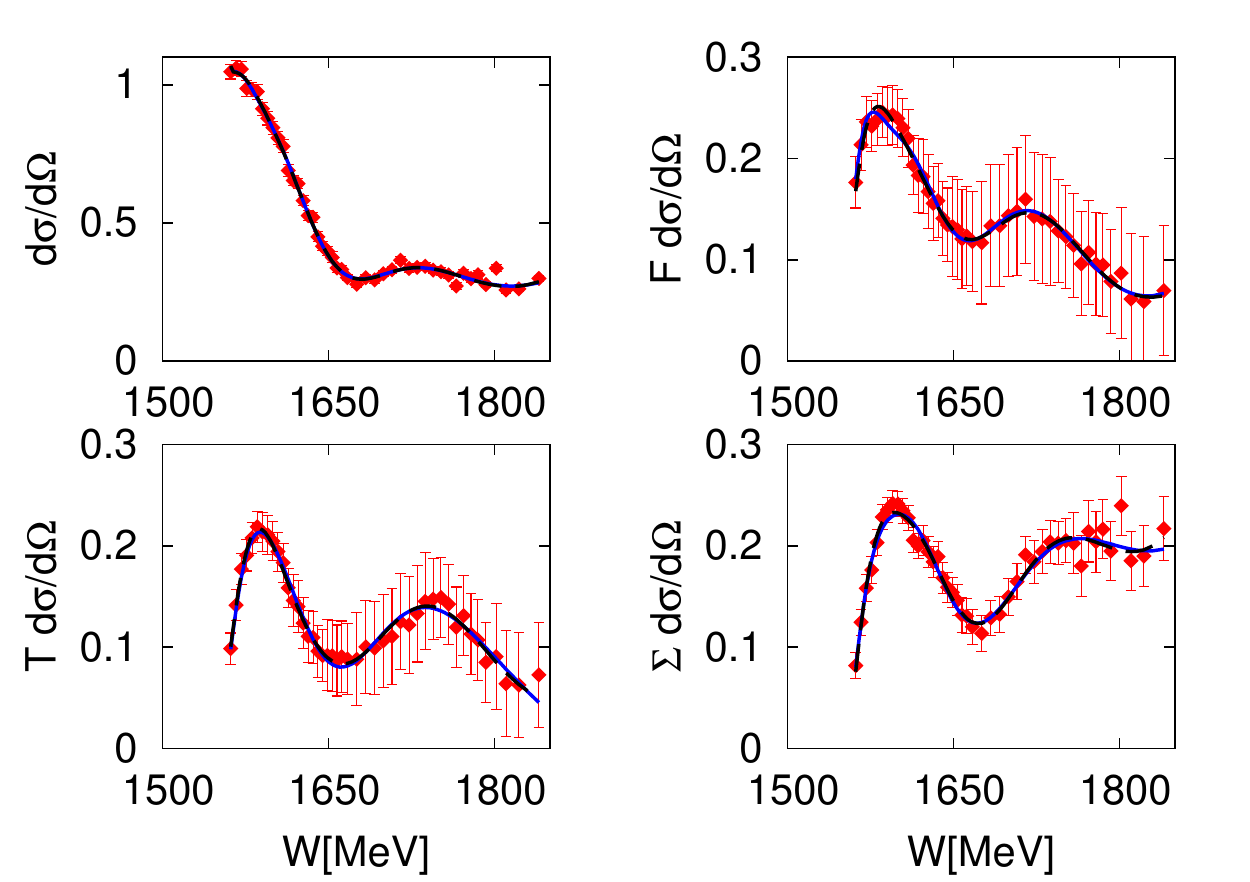}
\vspace{3mm} \caption{\label{FigFTDataFit} Pietarinen fit of the
interpolated data at $t = -0.2\,$GeV$^2$ and $t=-0.5\,$GeV$^2$. The
dashed (black) and solid (blue)  curves are the results starting
with solutions II and III, respectively and are on top of each
other.}
\end{center}
\end{figure}

These two starting solutions are expanded in Pietarinen series as
described before and the Pietarinen coefficients $\{b_{i,j}^k\}$ are
fitted to the fixed-$t$ data. The results of the fits are shown in
Fig.~\ref{FigFTDataFit}. With both initial solutions II and III, the
fits agree very well with the data and can not be distinguished in
the plots. Generally, we want to remind, that in an amplitude
representation of the observables a continuum ambiguity exists.
Therefore even identical fits can still have very different
underlying amplitudes. However, due to the Pietarinen representation
of both amplitudes, fixed-$t$ analyticity is obeyed by both
solutions. The latter one is not generally true for any phase
rotated set of amplitudes.

\begin{figure}[htb]
\begin{center}
\includegraphics[width=8.0cm]{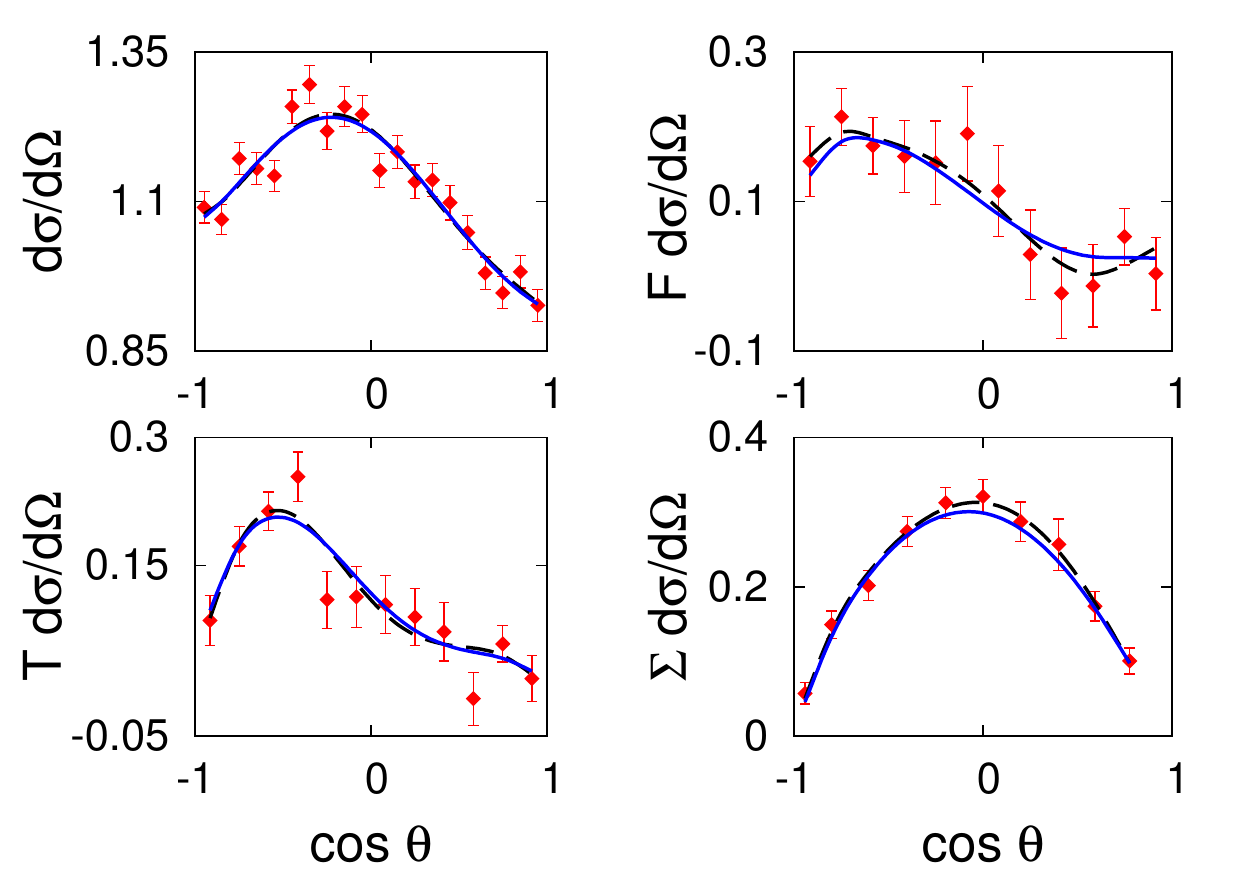}
\includegraphics[width=8.0cm]{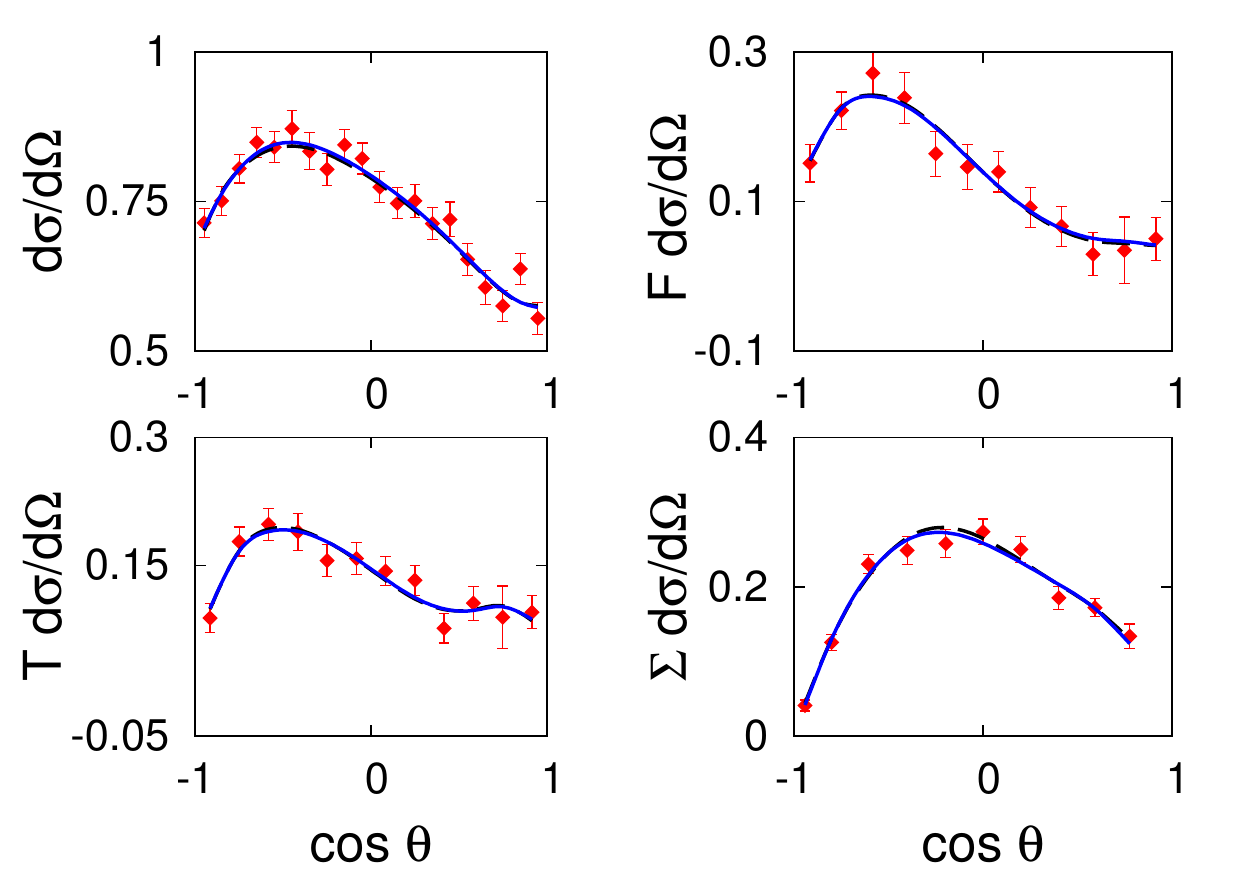}
\includegraphics[width=8.0cm]{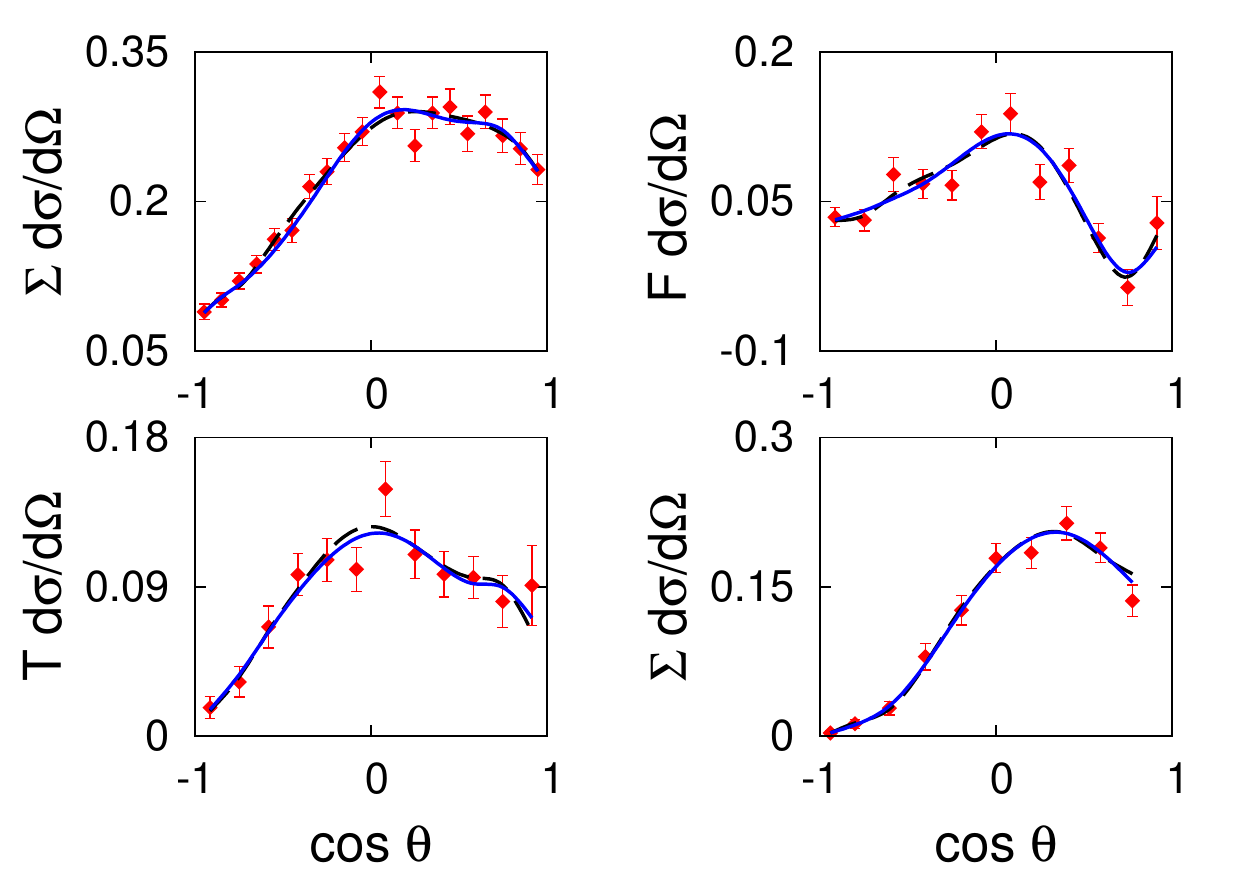}
\includegraphics[width=8.0cm]{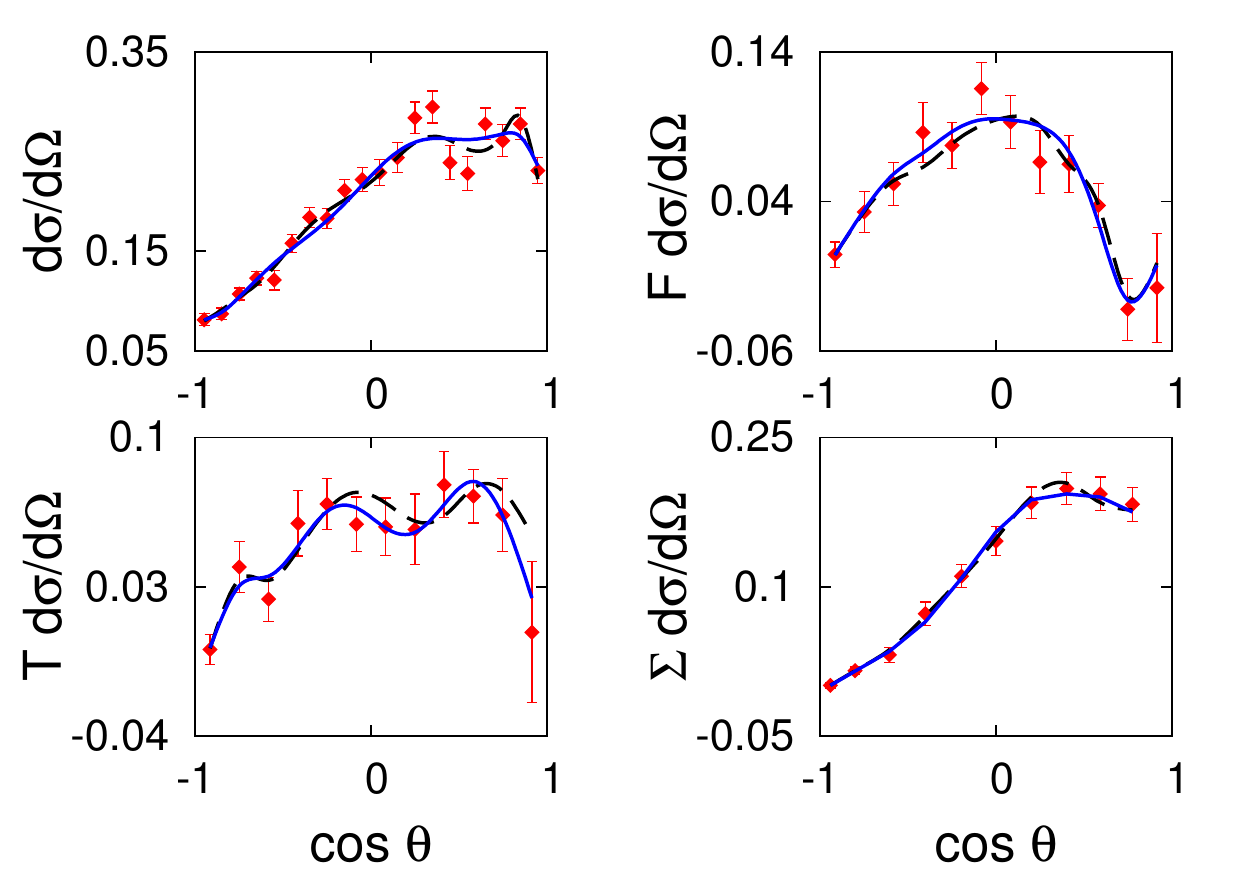}
\vspace{3mm} \caption{\label{FigSEexp1}  Single-energy fit to the
experimental data at $E=818$~MeV ($W=1554$~MeV)  (top left) and
$E=898$~MeV ($W=1602$~MeV) (top right) and at $E=1191$~MeV
($W=1765$~MeV) (bottom left) and $E=1335$~MeV ($W=1840$~MeV) (bottom
right), using the analytical constraint of solutions II and III,
drawn as dashed and solid lines, respectively.}
\end{center}
\end{figure}

Fig.~\ref{FigFTDataFit} is the first step of the iterative
minimization scheme outlined in Fig.~\ref{Fig:Scheme}. The solutions
of the invariant amplitudes allow us to evaluate the helicity
amplitudes at each energy, which is needed for the next step to
perform the SE PWA in first iteration. This is shown in
Fig.~\ref{FigSEexp1} at four different energies in the range of
$W=1554-1840$~MeV. The fit parameters in this minimization step are
the complex electric and magnetic multipoles up to $L_{max}=5$.

\begin{figure}[htb]
\includegraphics[width=8.0cm]{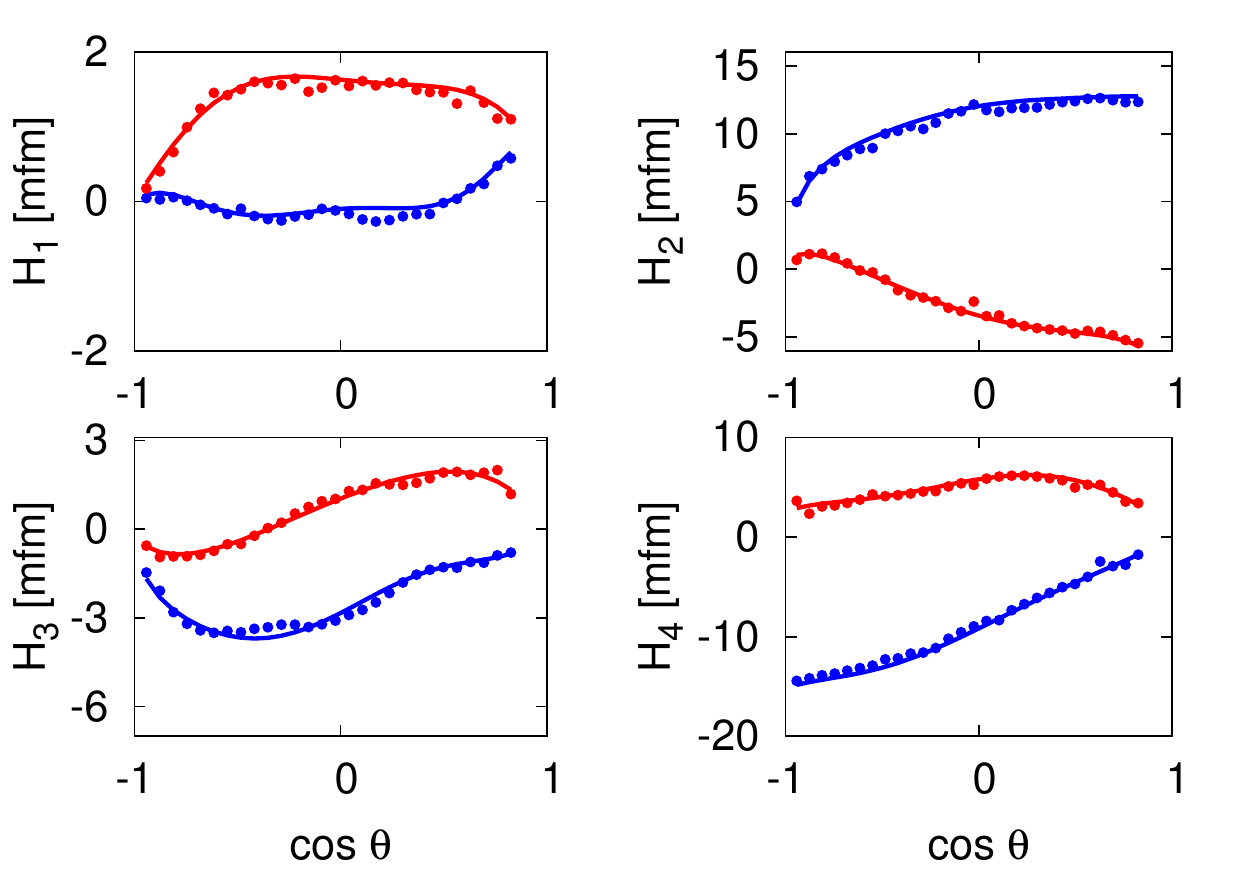}
\includegraphics[width=8.0cm]{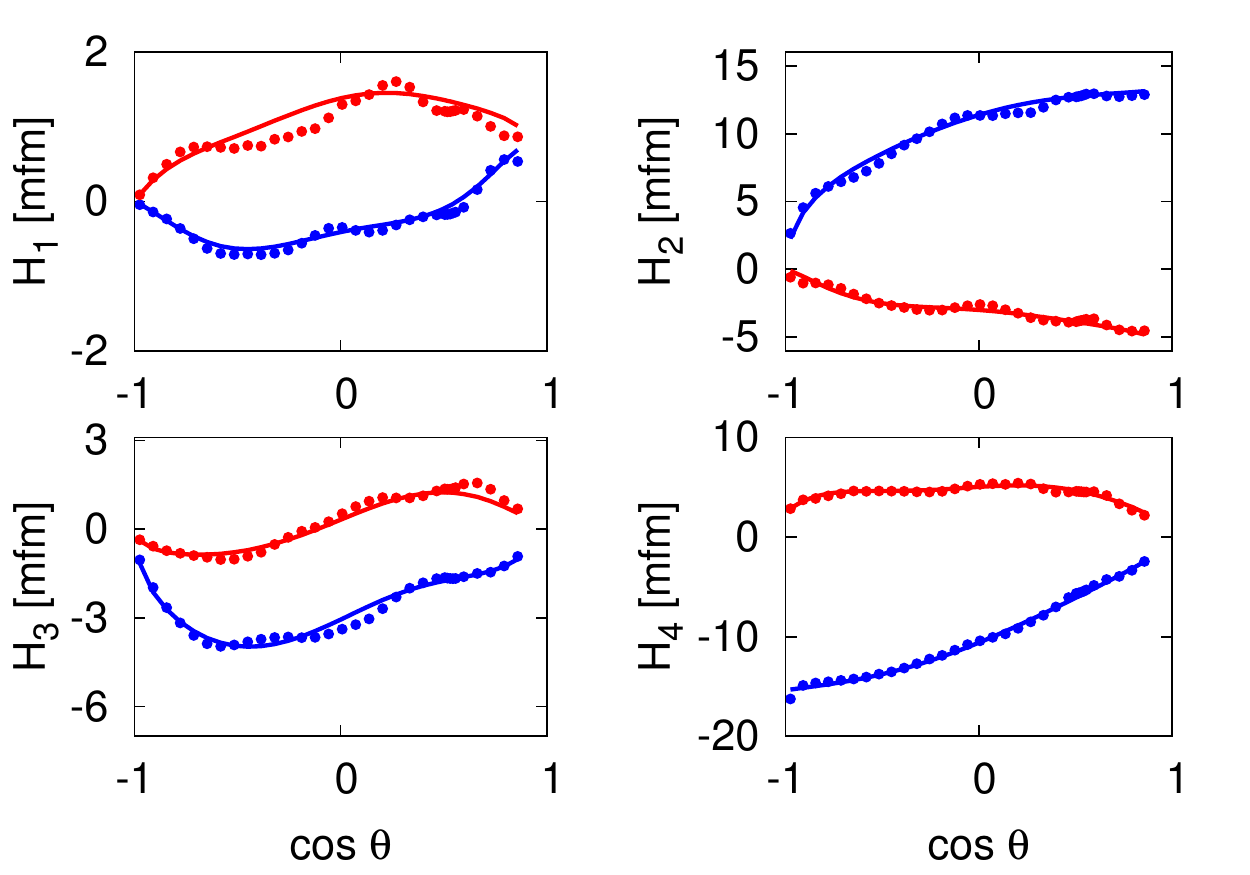}
\includegraphics[width=8.0cm]{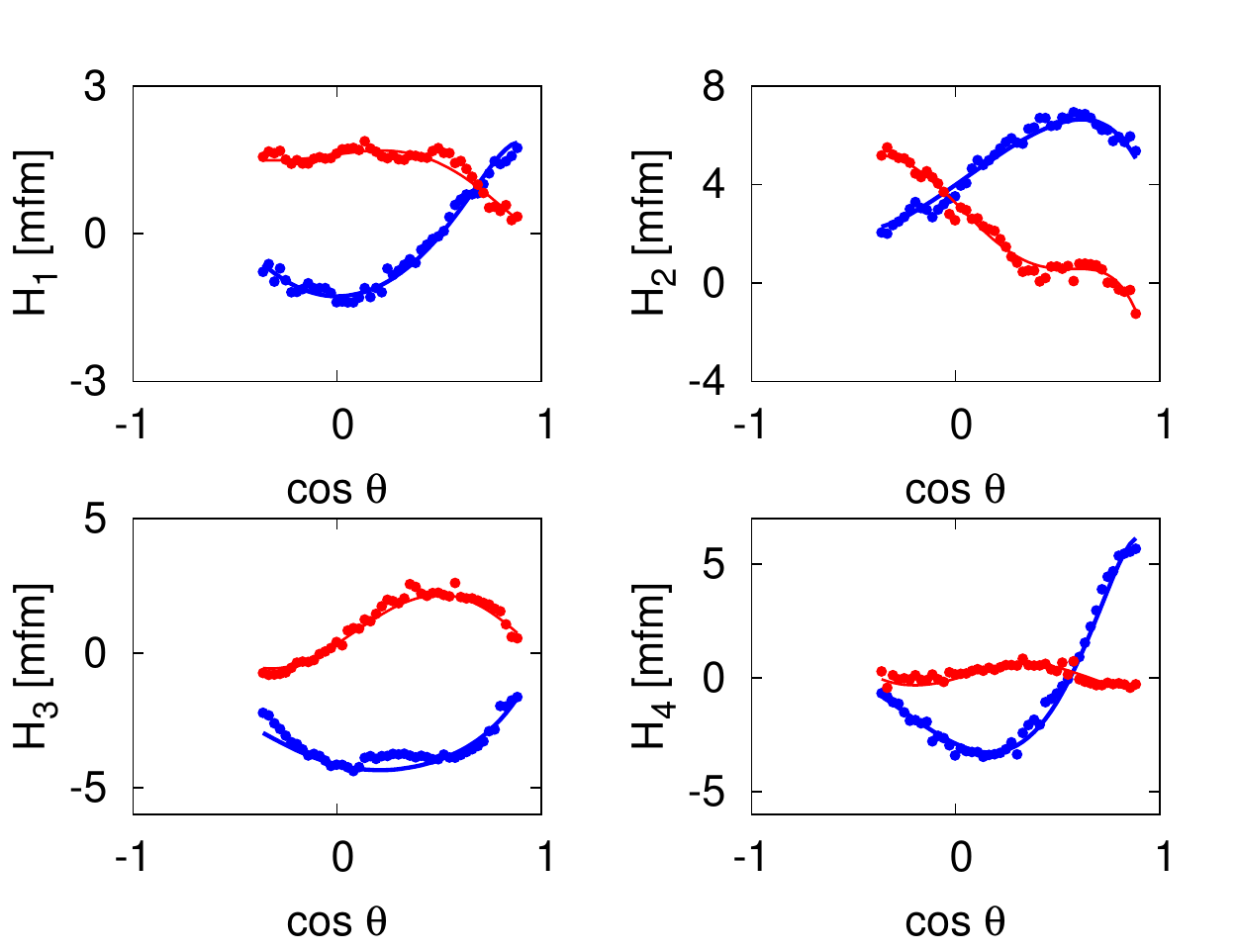}
\includegraphics[width=8.0cm]{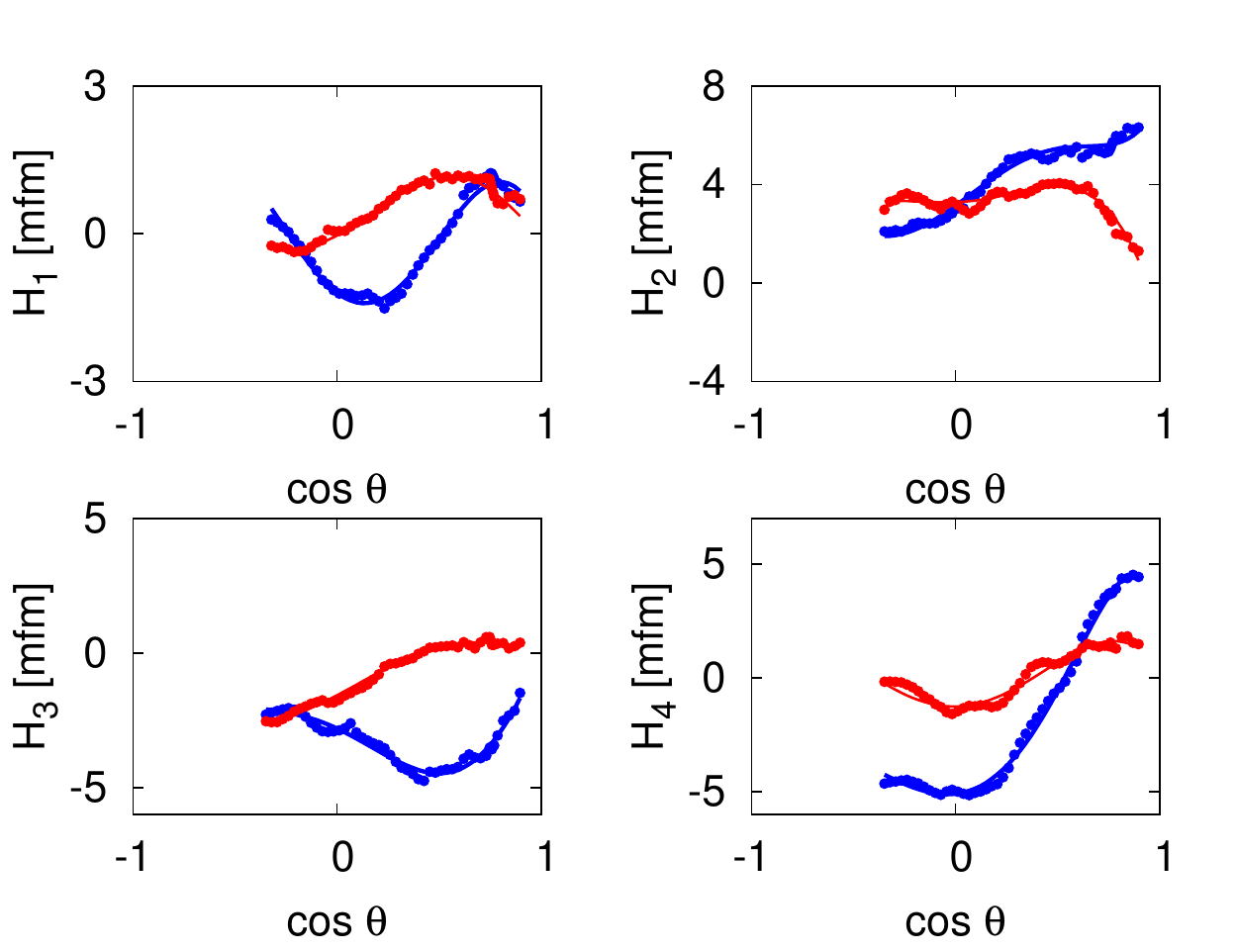}
\vspace{3mm} \caption{\label{FigHelSE16h} Helicity amplitudes
obtained by FT AA (in first iteration) using solutions II (left) and
III (right) as constraints at fixed energies  $W=1602$~MeV (top) and
$W=1840$~MeV (bottom). The blue and red points show the result of
the fixed-$t$ analysis and are independent from point to point. The
blue and red solid lines are obtained from the single-energy partial
wave analysis and are continuous over the angular range. At the
higher energy, the kinematical range is restricted due to the
restrictions in the $t$ value, $-1.0\le t \le 0$.}
\end{figure}

With this first iteration we can now compare helicity amplitudes
from the FT AA analysis with those calculated with the multipoles
obtained in the SE PWA analysis. In Fig.~\ref{FigHelSE16h} we make
this comparison at $W=1602$~MeV as a function of the scattering
angle. The amplitudes obtained at fixed $t$ values are principally
discontinuous as the FT AA is independent at each different $t$
value and therefore also at each different angle $\theta$, while the
multipole expansion of the helicity amplitudes are continuous
functions of $\theta$ and shown as solid curves.

\begin{figure}[htb]
\includegraphics[width=8.0cm]{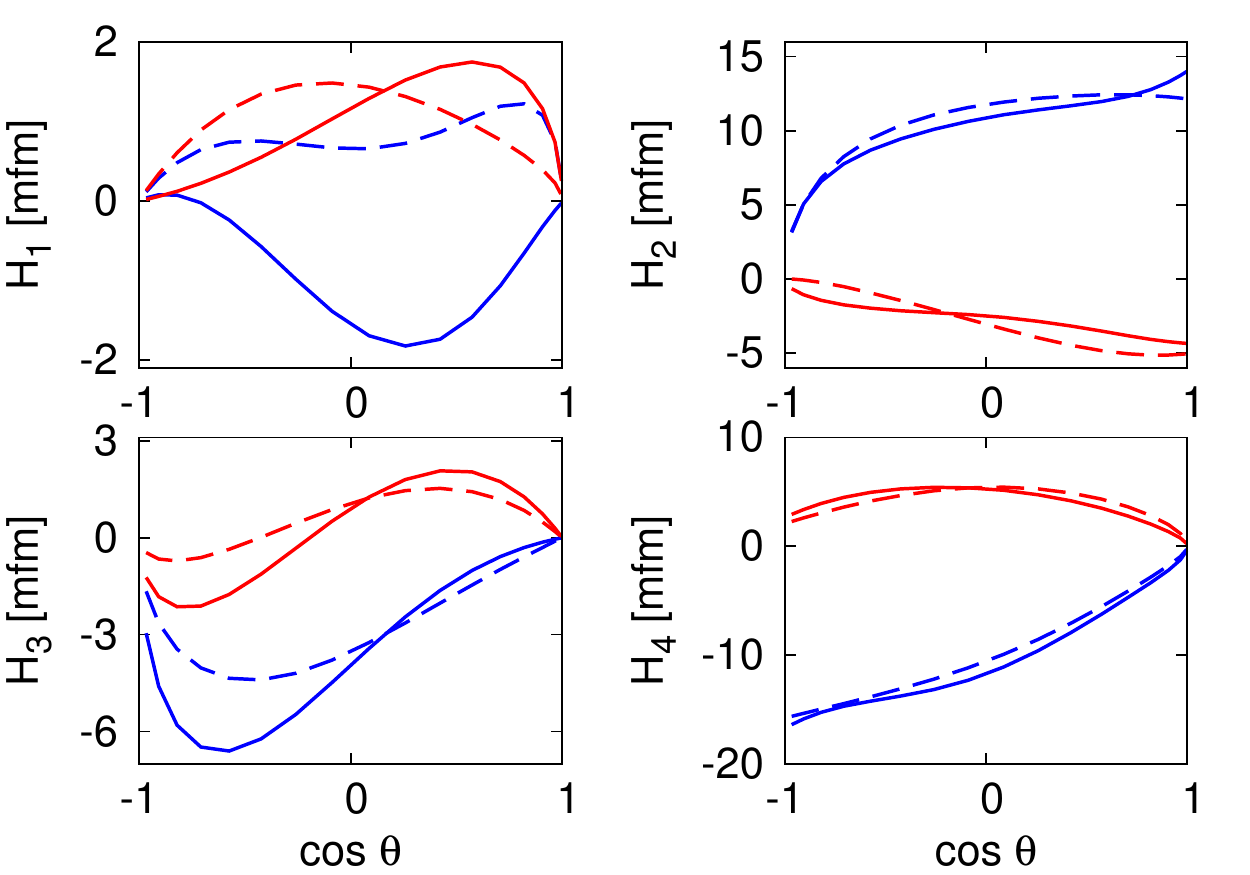}
\includegraphics[width=8.0cm]{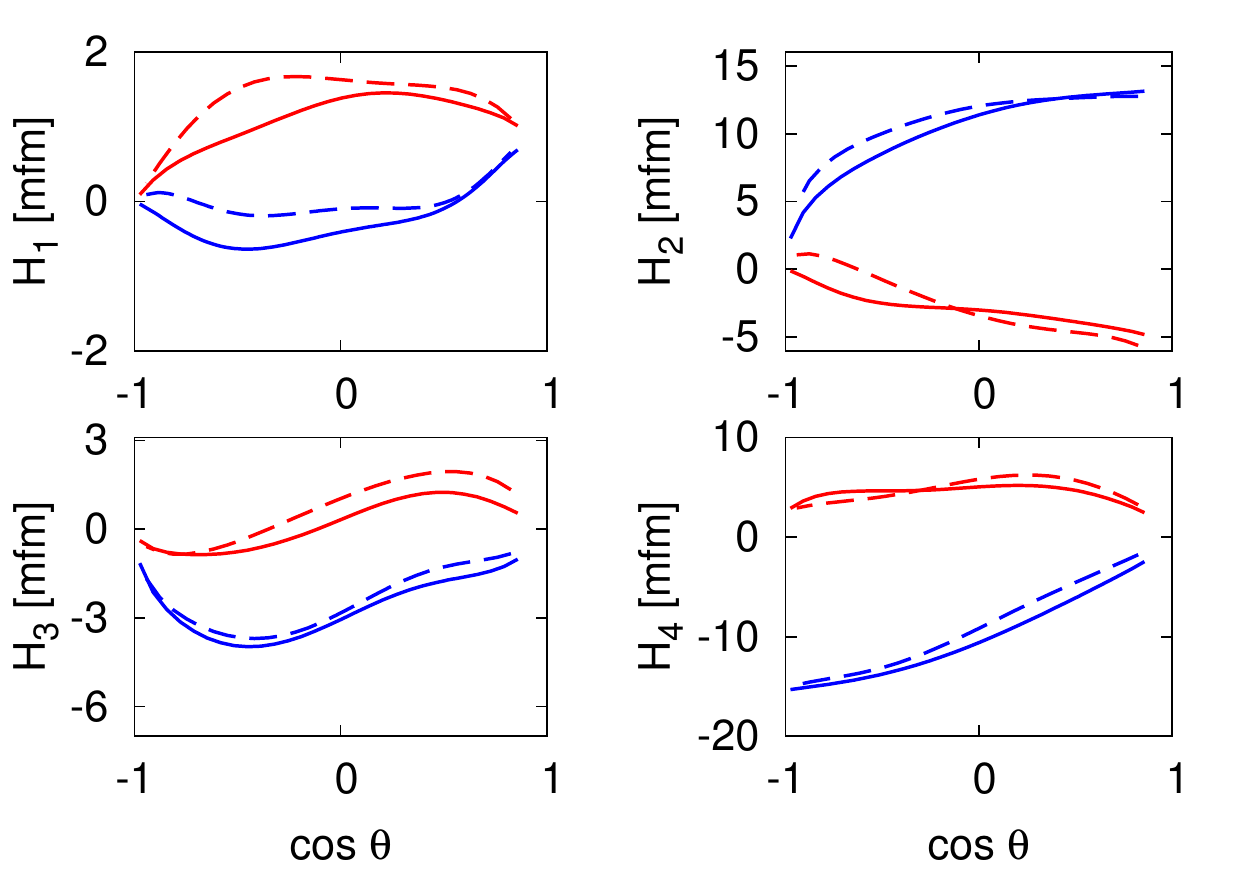}
\includegraphics[width=8.0cm]{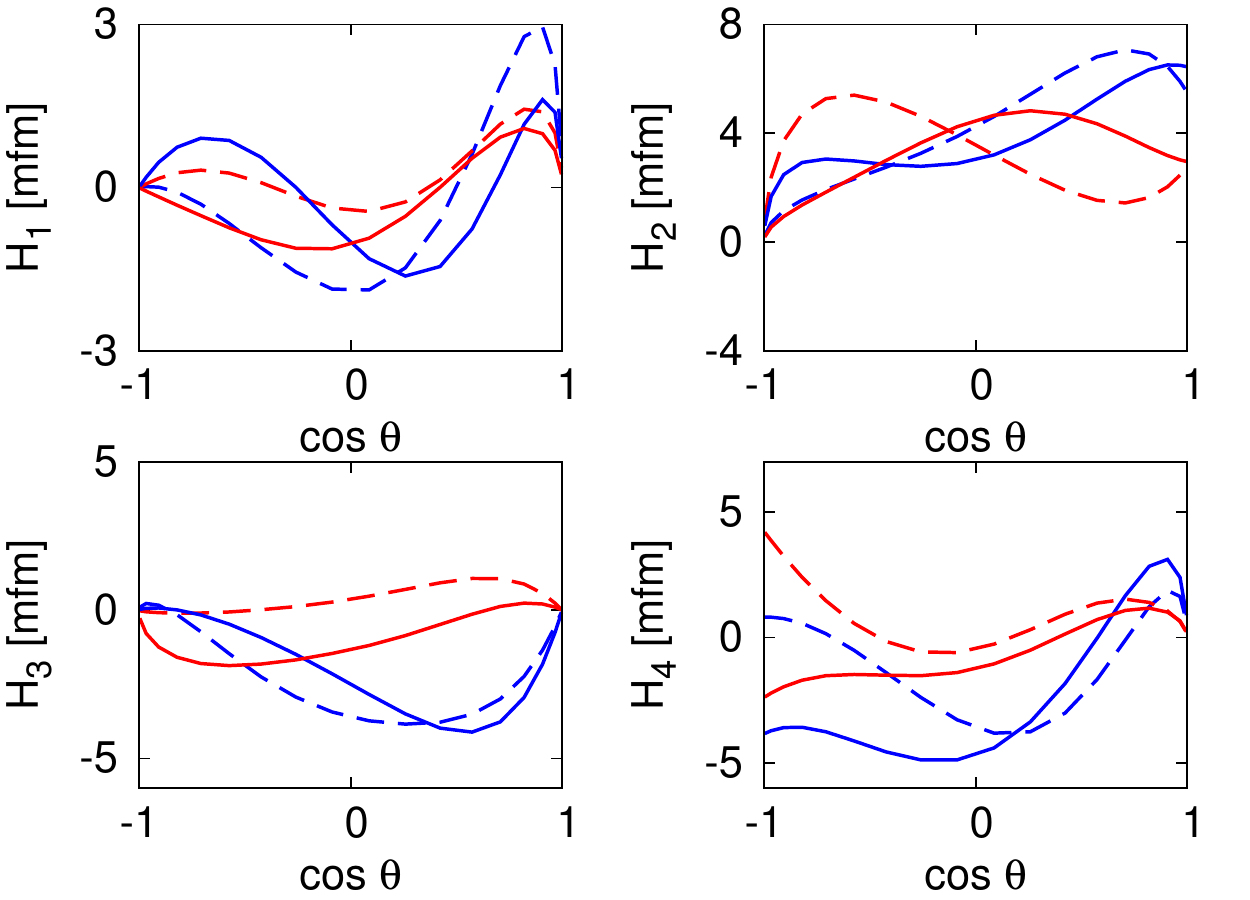}
\includegraphics[width=8.0cm]{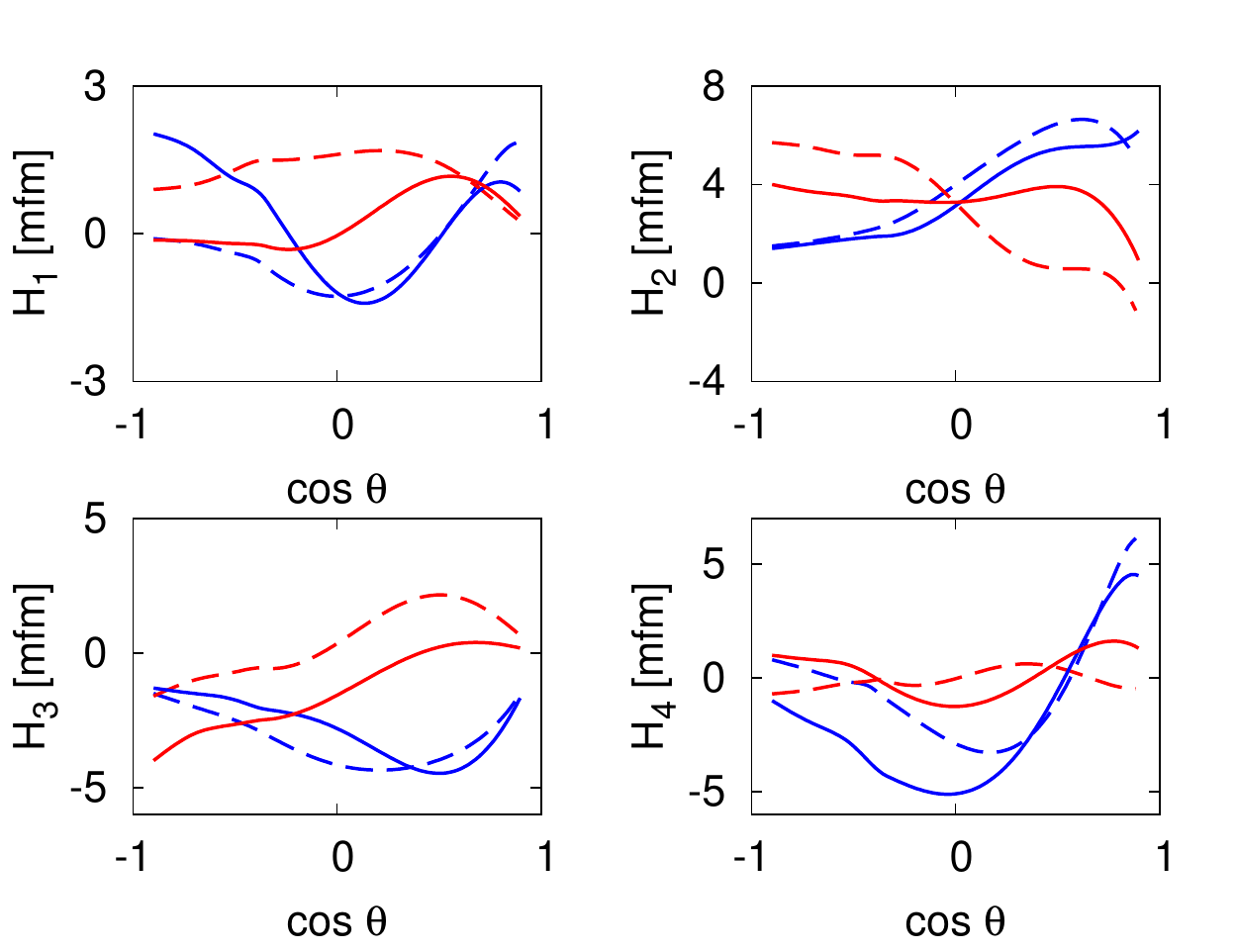}
\vspace{3mm} \caption{\label{FigHelSE} Evolution of the helicity
amplitudes at fixed energies $W=1602$~MeV (top) and $W=1840$~MeV
(bottom) from the initial (left panel) to the final step (right
panel) after three iterations using solutions II (dashed) and III
(solid) as a constraint. The real parts of the amplitudes are shown
in blue and the imaginary parts in red. At the higher energy, for
the iterated solutions, the kinematical range is restricted due to
the restrictions in the $t$ value, $-1.0\le t \le 0$. }
\end{figure}

The iterative procedure outlined in Fig.~\ref{Fig:Scheme} quickly
converges after three iterations and the evolution of the helicity
amplitudes obtained from the SE PWA is shown in Fig.~\ref{FigHelSE}.
The amplitudes $H_1$ and $H_3$ show larger spreading between the two
solutions and also a larger change during the iterative procedure
compared to the amplitudes $H_2$ and $H_4$. The reason for this is
that the changes of the four amplitudes are of the same size, but
the magnitude of $H_2$ and $H_4$ are much larger than for $H_1$ and
$H_3$. Therefore the effects are much more visible in $H_1$ and
$H_3$. For these two amplitudes it can also clearly be seen, that
the iterative procedure brings the two different initial solutions
II and III much closer together after three iterations.

\begin{figure}[htbp]
\begin{center}
\includegraphics[width=7.0cm]{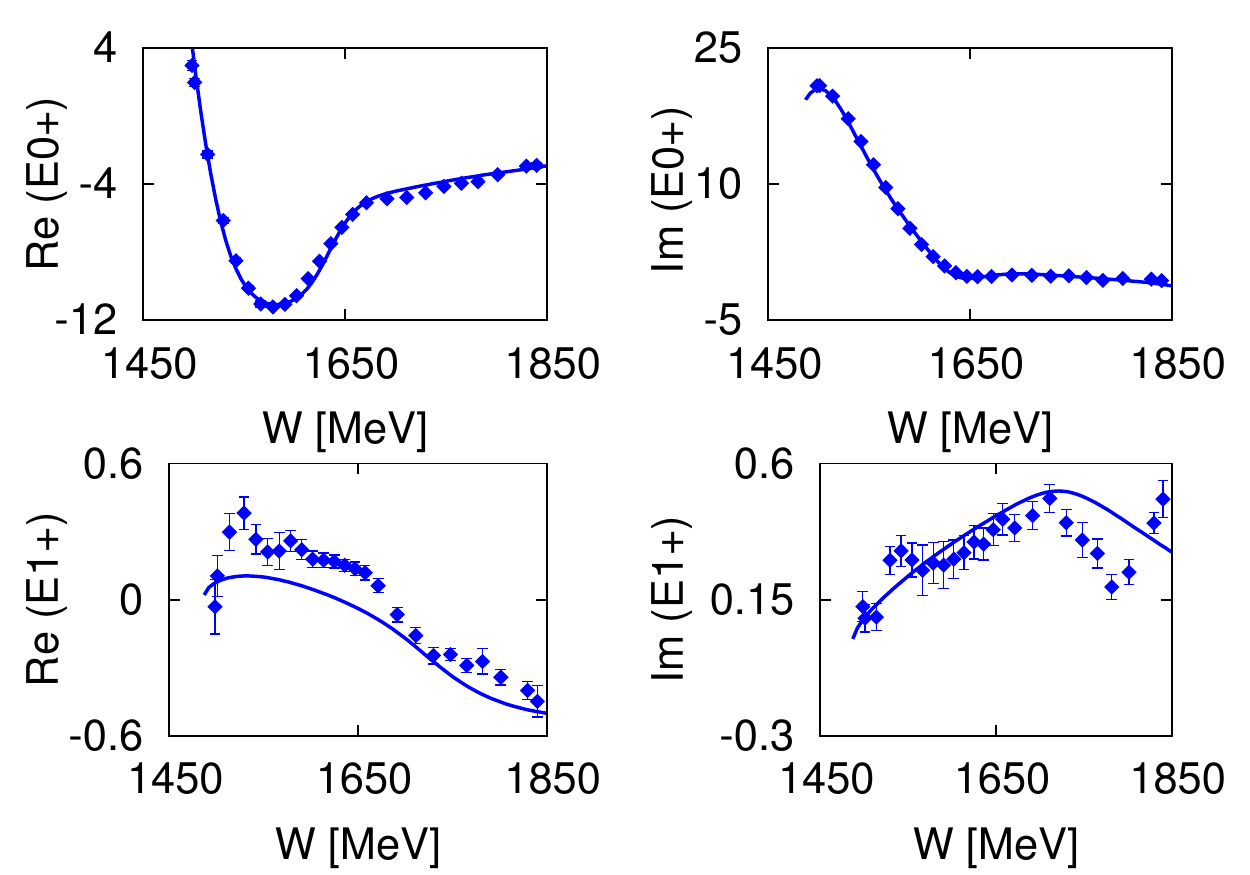}
\includegraphics[width=7.0cm]{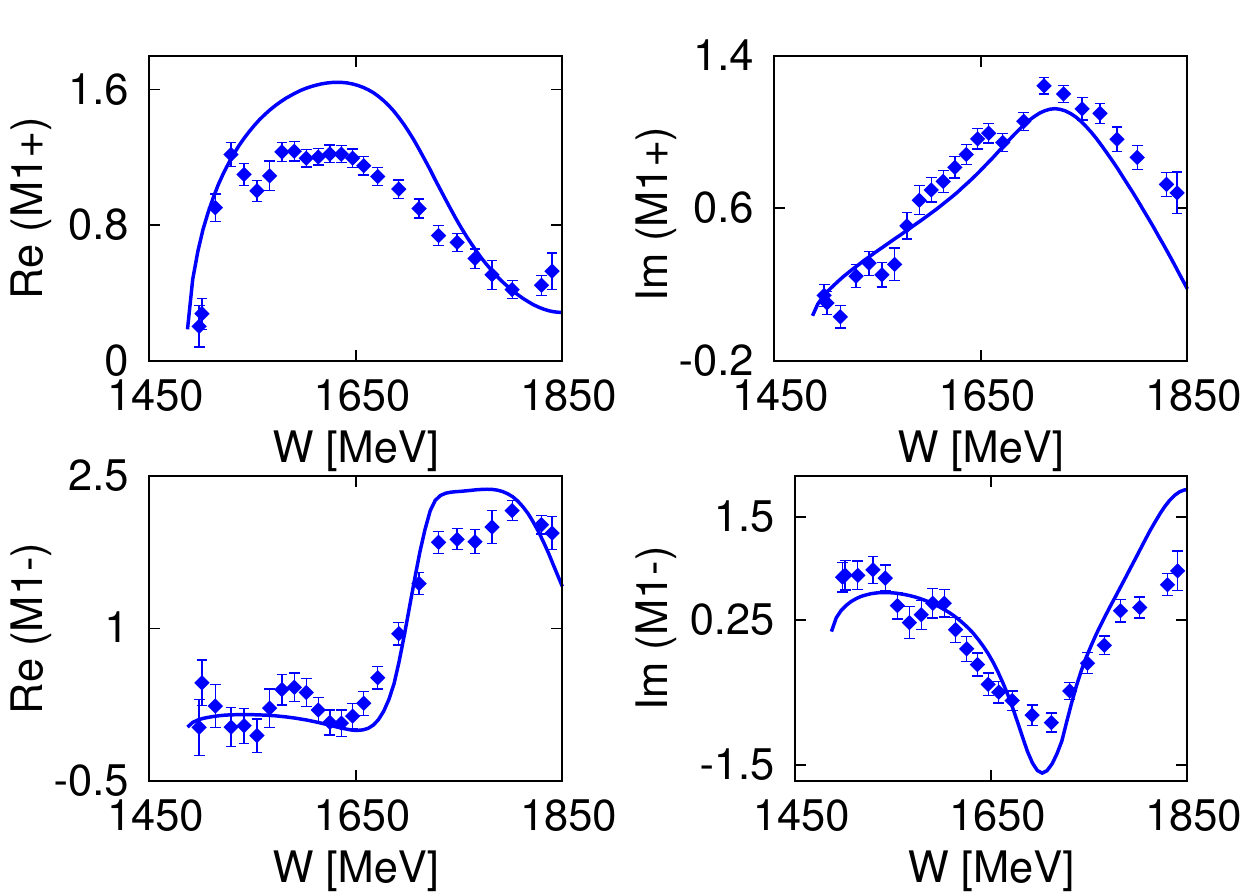}
\includegraphics[width=7.0cm]{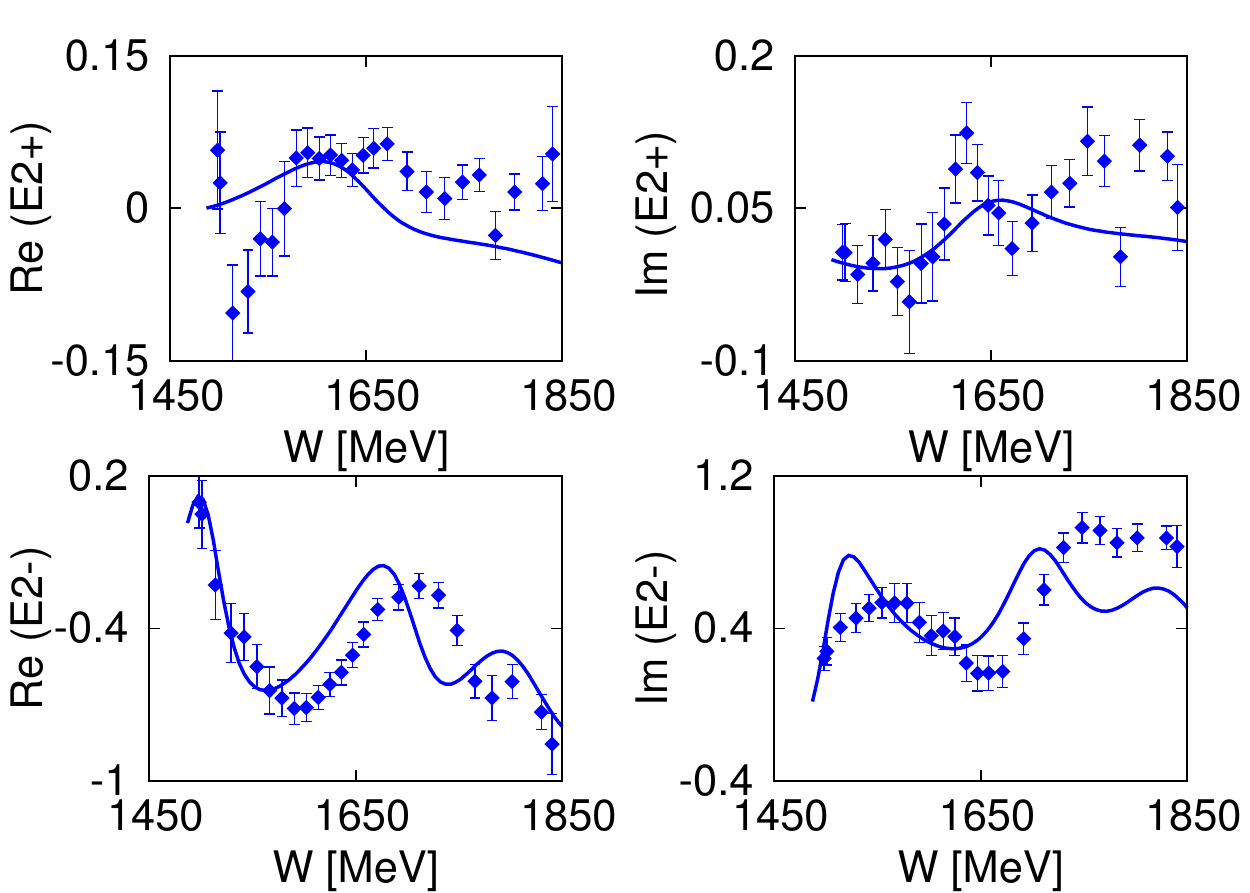}
\includegraphics[width=7.0cm]{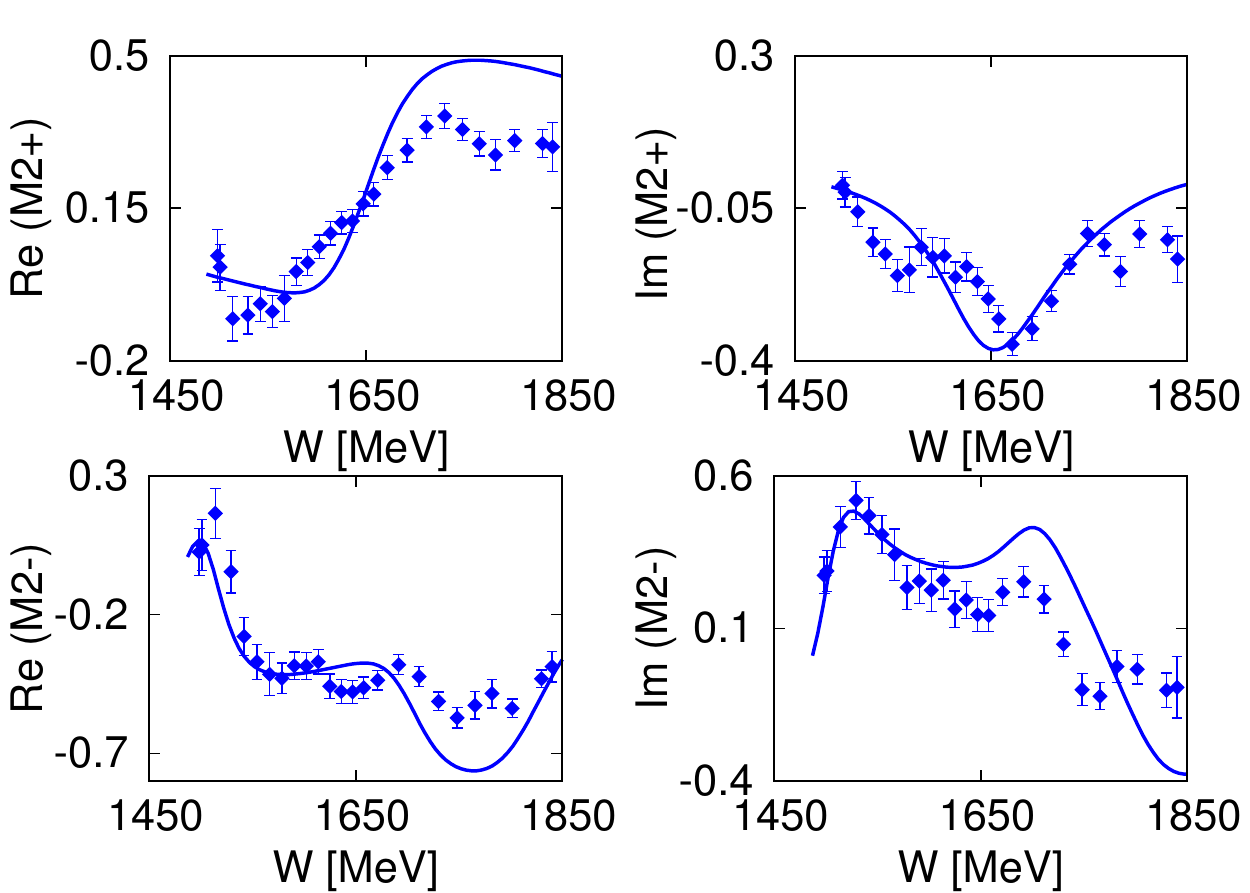}
\vspace{3mm} \caption{\label{FigMultSE16a} Evolution of the $S$-,
$P$- and $D$-wave multipoles from the initial solutions II (full
lines) to the final step (blue dots) after three iterations.}
\end{center}
\end{figure}

\begin{figure}[h]
\begin{center}
\includegraphics[width=7.0cm]{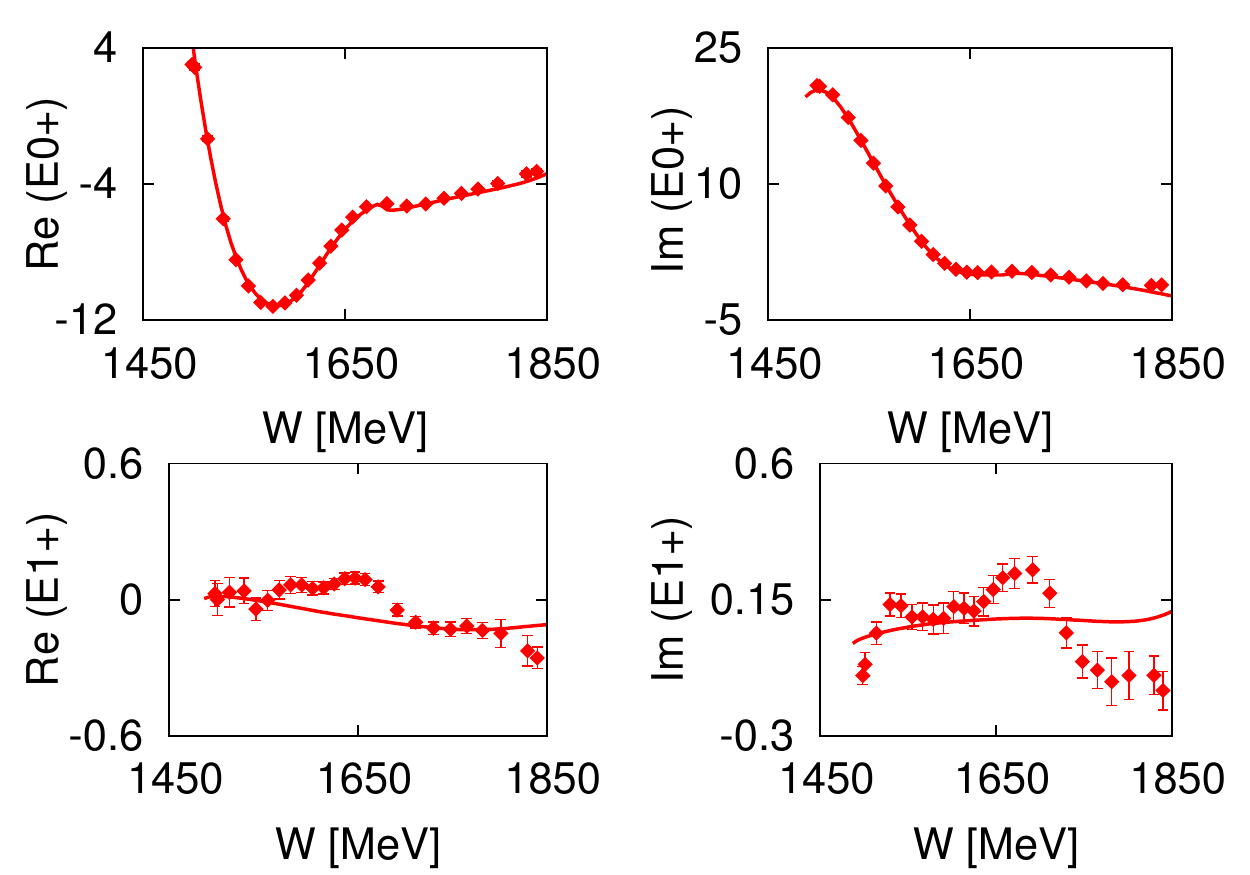}
\includegraphics[width=7.0cm]{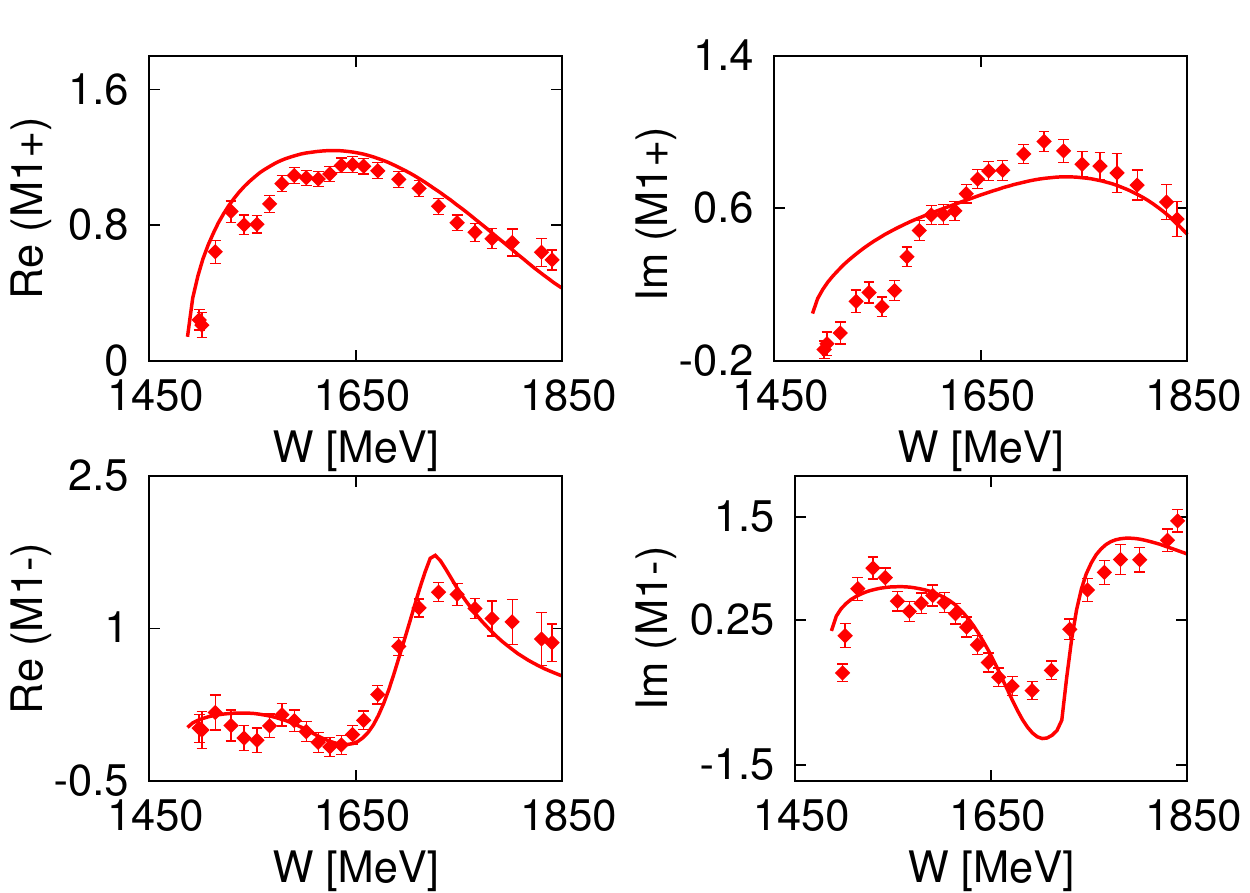}
\includegraphics[width=7.0cm]{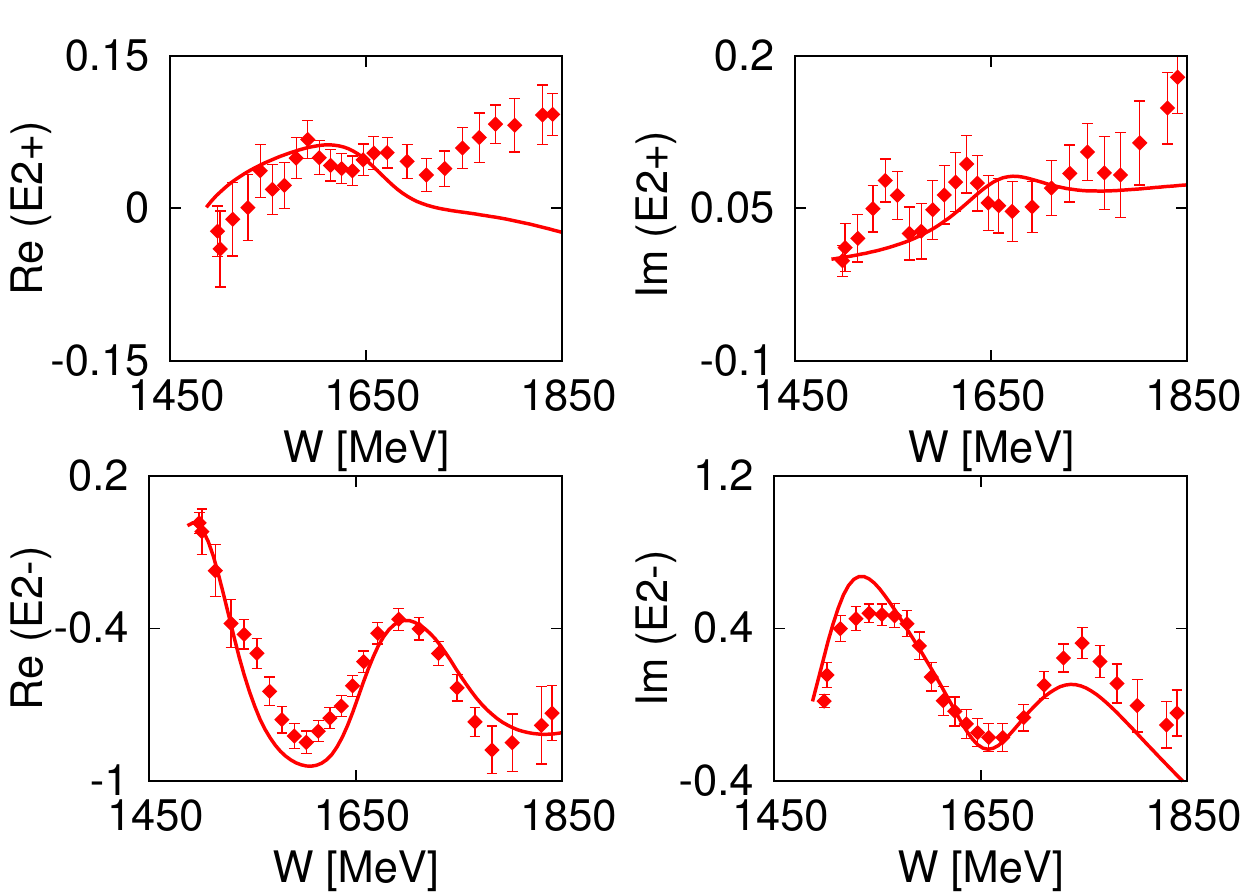}
\includegraphics[width=7.0cm]{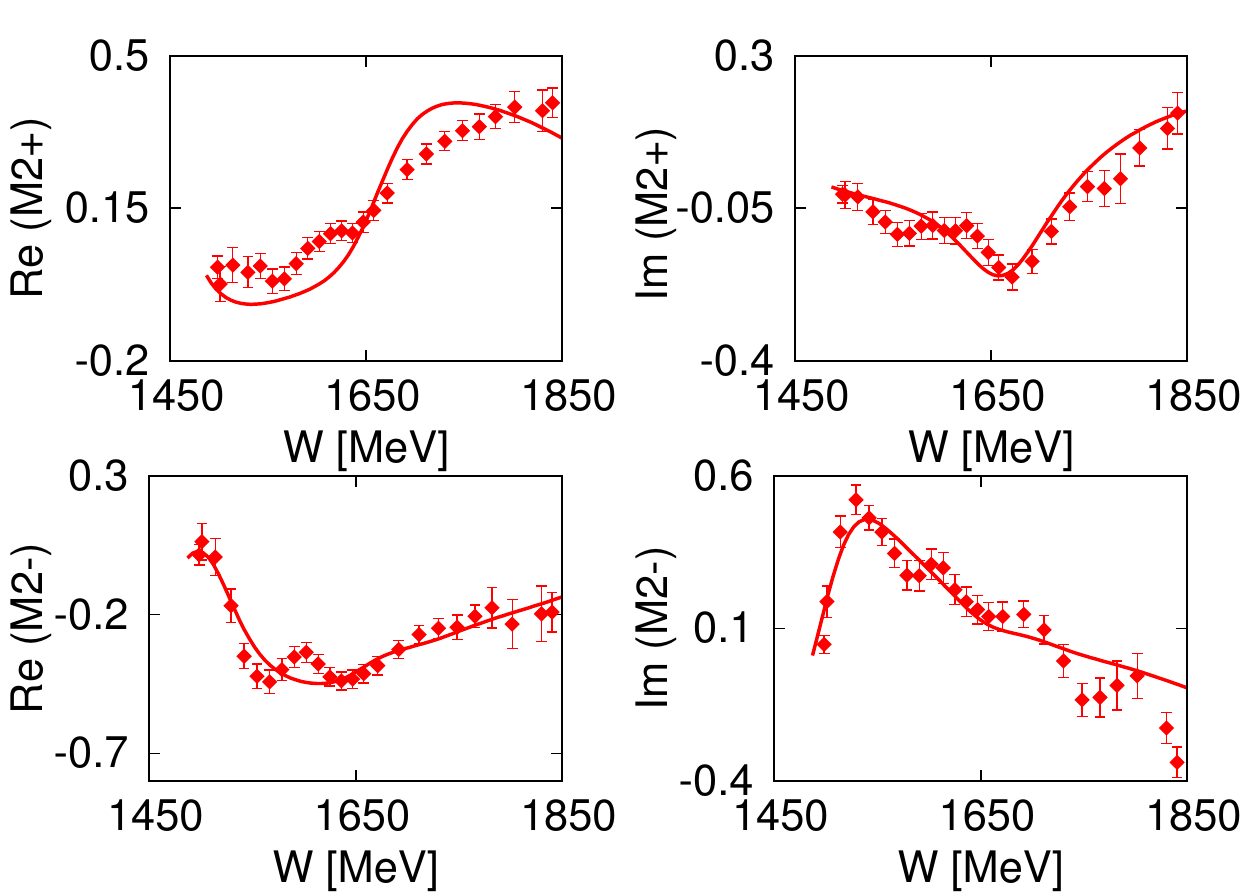}
\vspace{3mm} \caption{\label{FigMultSE16H} Evolution of the $S$-,
$P$- and $D$-wave multipoles from the initial solutions III (full
lines) to the final step (red dots) after three iterations.}
\end{center}
\end{figure}

\begin{figure}[htb]
\begin{center}
\includegraphics[width=8.0cm]{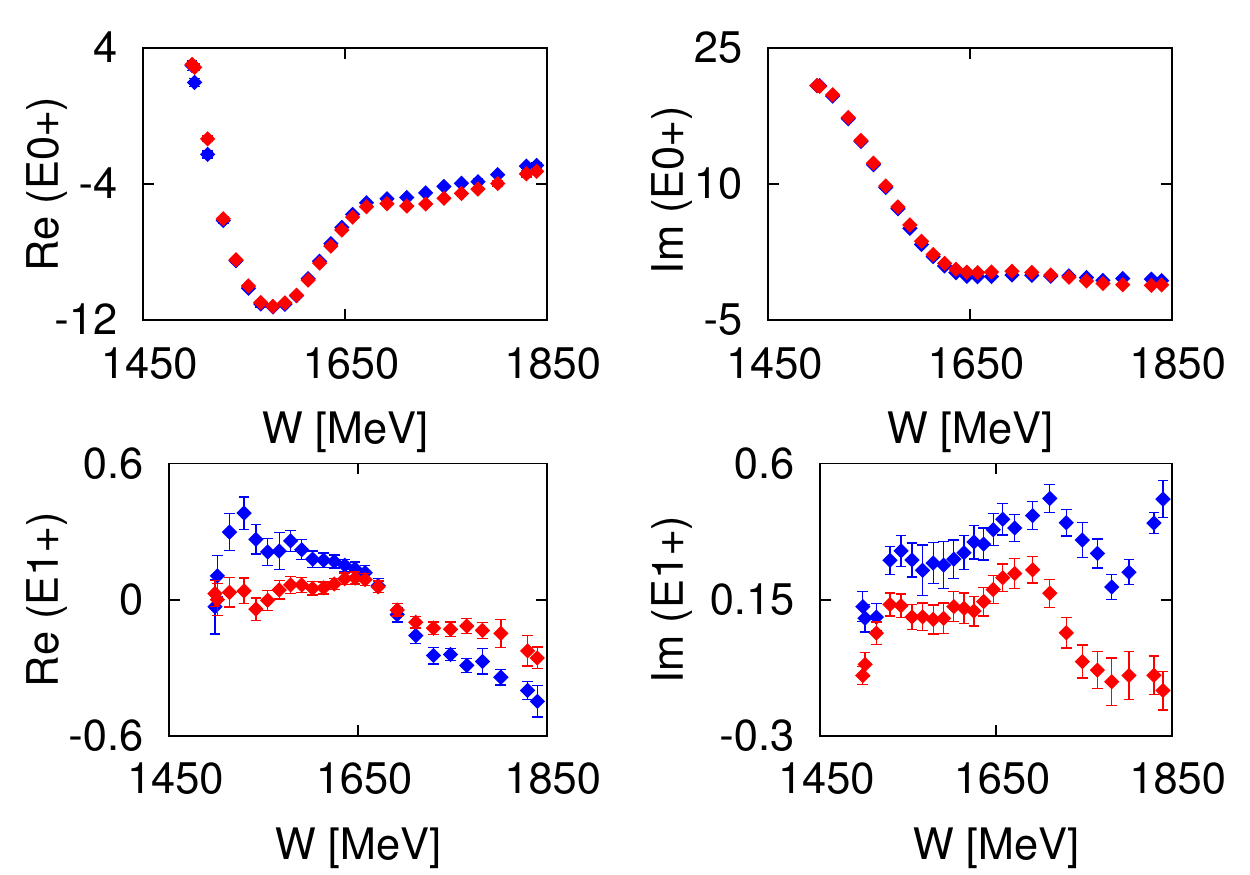}
\includegraphics[width=8.0cm]{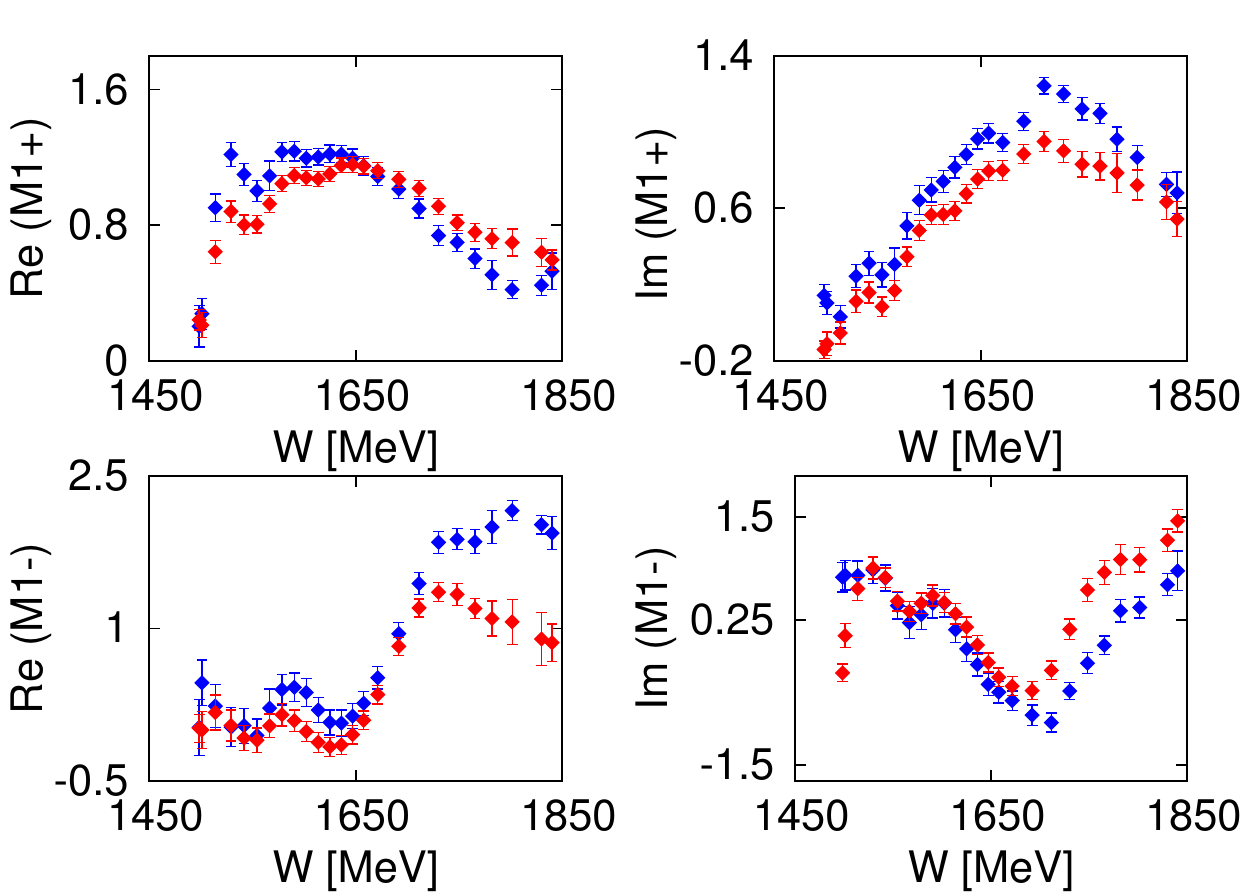}
\includegraphics[width=8.0cm]{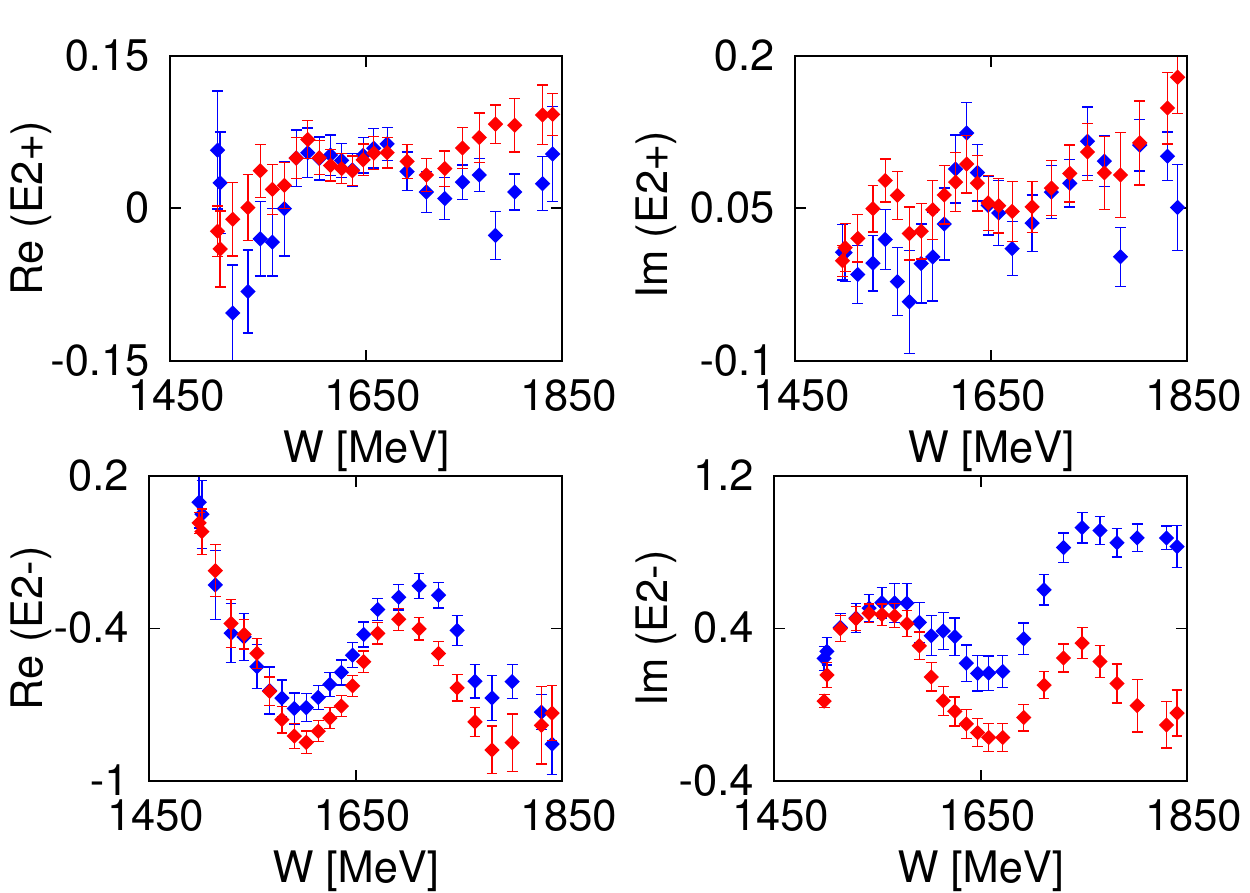}
\includegraphics[width=8.0cm]{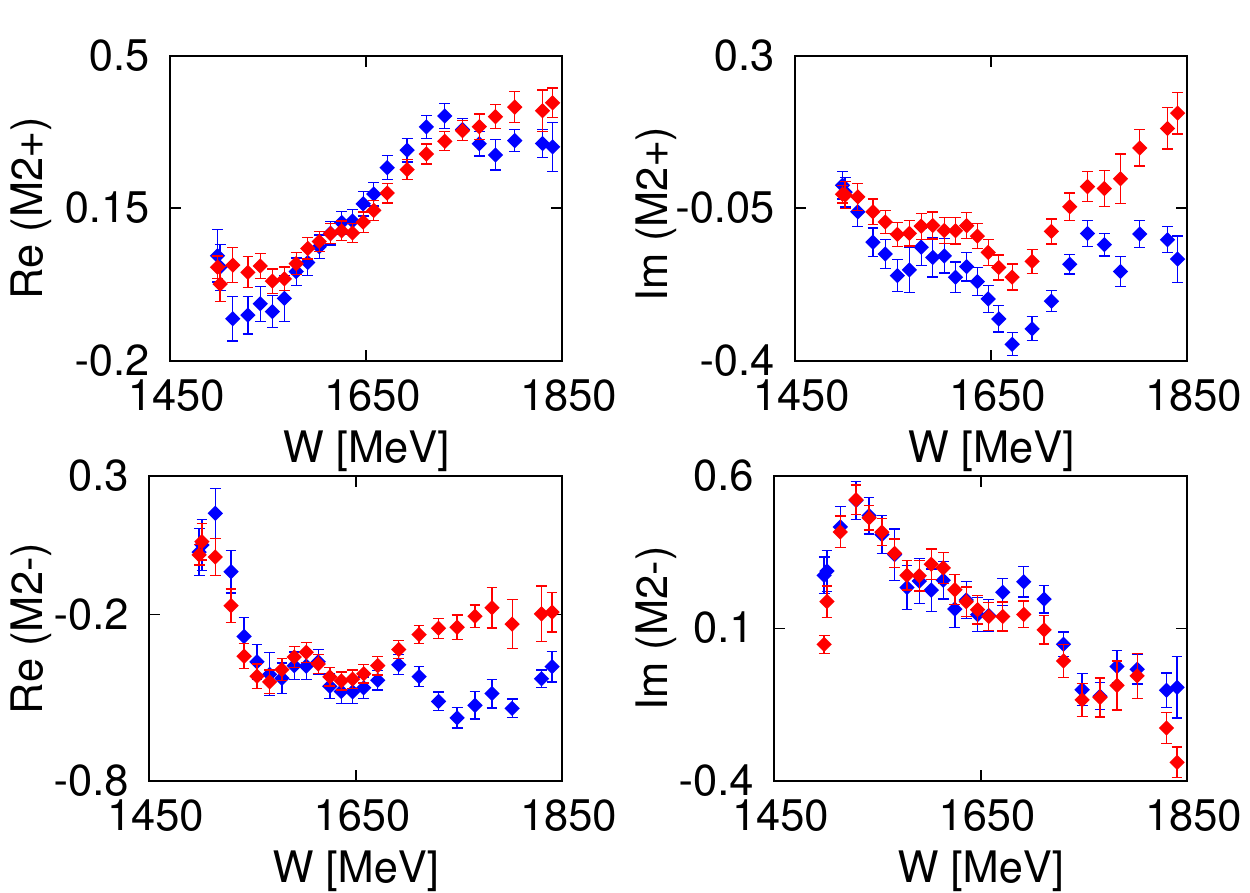}
\vspace{3mm} \caption{\label{FigMultSE} Real and imaginary parts of
the $S$-, $P$- and $D$-wave multipoles obtained in the final step
after three iterations using analytical constraints from helicity
amplitudes obtained from initial solutions II (blue) and III (red).}
\end{center}
\end{figure}

But also after the iterative procedure has converged, there remain
residual differences between the final helicity amplitudes generated
from solution II and from solution III. This results in different
sets of partial waves (multipoles) when solutions II and III are
used. In Figs.~\ref{FigMultSE16a} and \ref{FigMultSE16H} these two
sets of multipoles are compared with the multipoles from their
corresponding initial solutions. In both sets the dominant $S$ wave
almost does not change and agrees very well. For all other partial
waves an evolution from the initial to the final solution can be
observed. In a few cases, especially from solution III, e.g.
Re$\,M_{1+}$, and real and imaginary parts of $E_{2-}$ and $M_{2-}$
change only a little bit. A direct comparison of the two SE PWA is
shown in Fig.~\ref{FigMultSE}. Again, almost no difference can be
observed for the $S$ wave and for most other partial waves the two
solutions are consistent within their statistical uncertainties.
However, for some partial waves considerable differences in certain
kinematical regions show us clearly non-unique solutions, e.g. for
Im$\,E_{1+}$, Im$\,E_{2-}$ and Re$\,M_{2-}$, just to mention the
most obvious deviations.


\begin{figure}[htb]
\begin{center}
\vspace{-1.8cm}
\includegraphics[width=7.5cm]{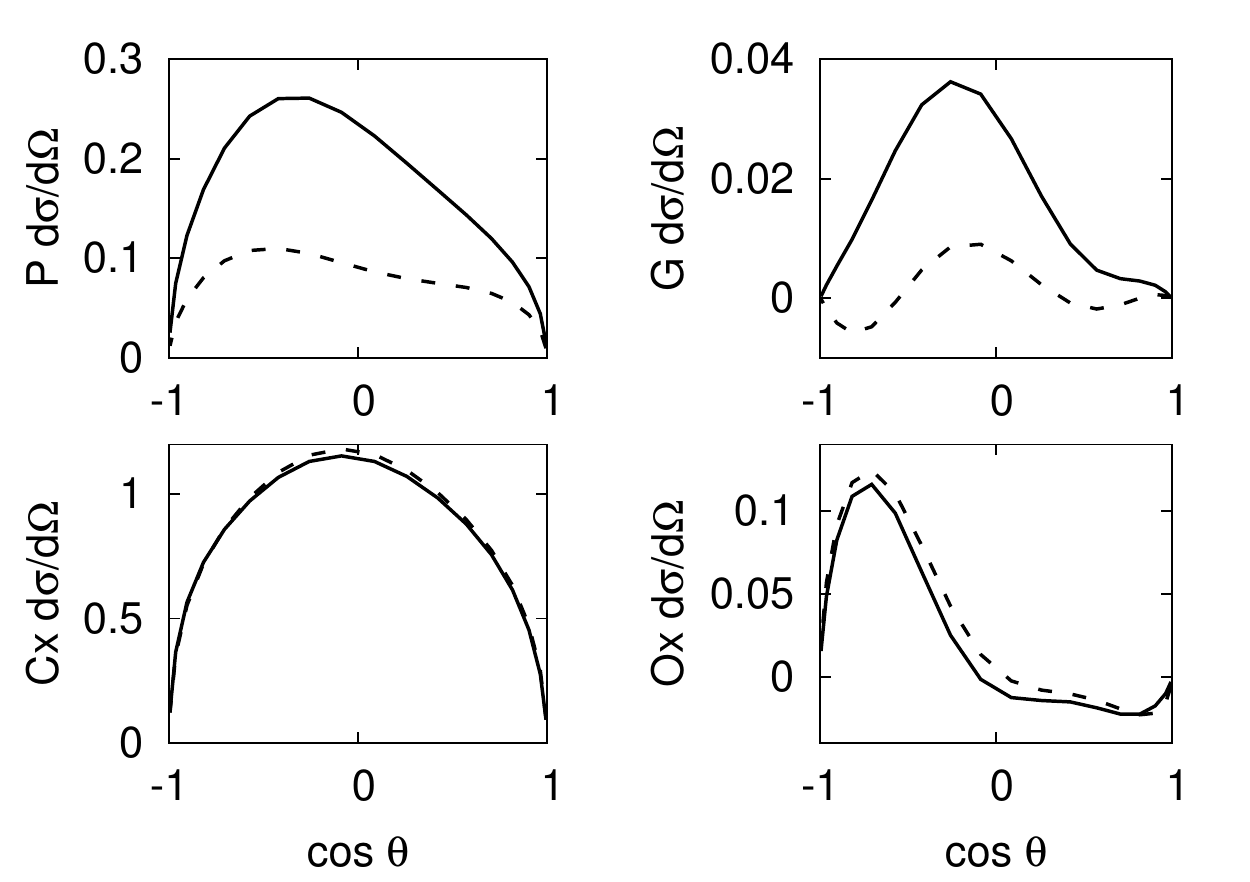}
\includegraphics[width=7.5cm]{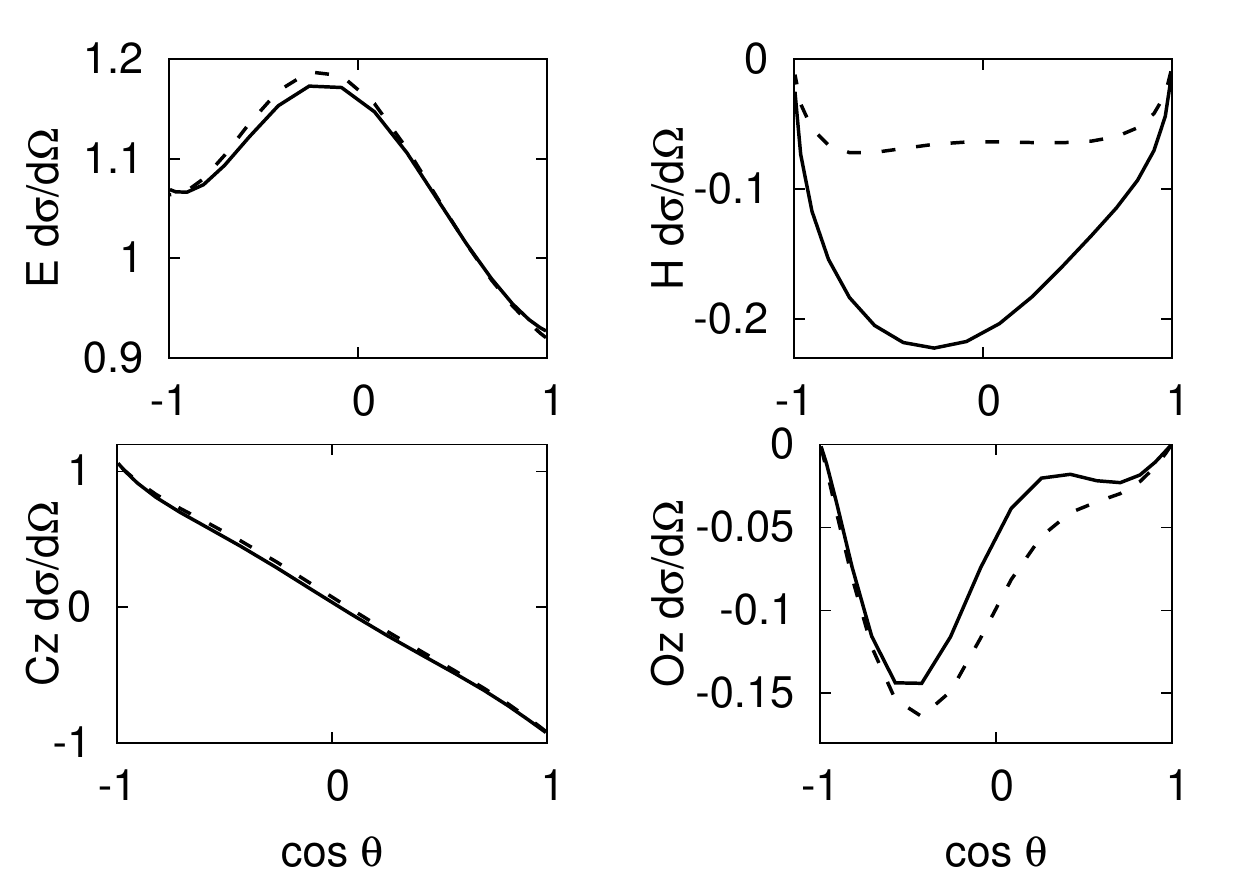}
\includegraphics[width=7.5cm]{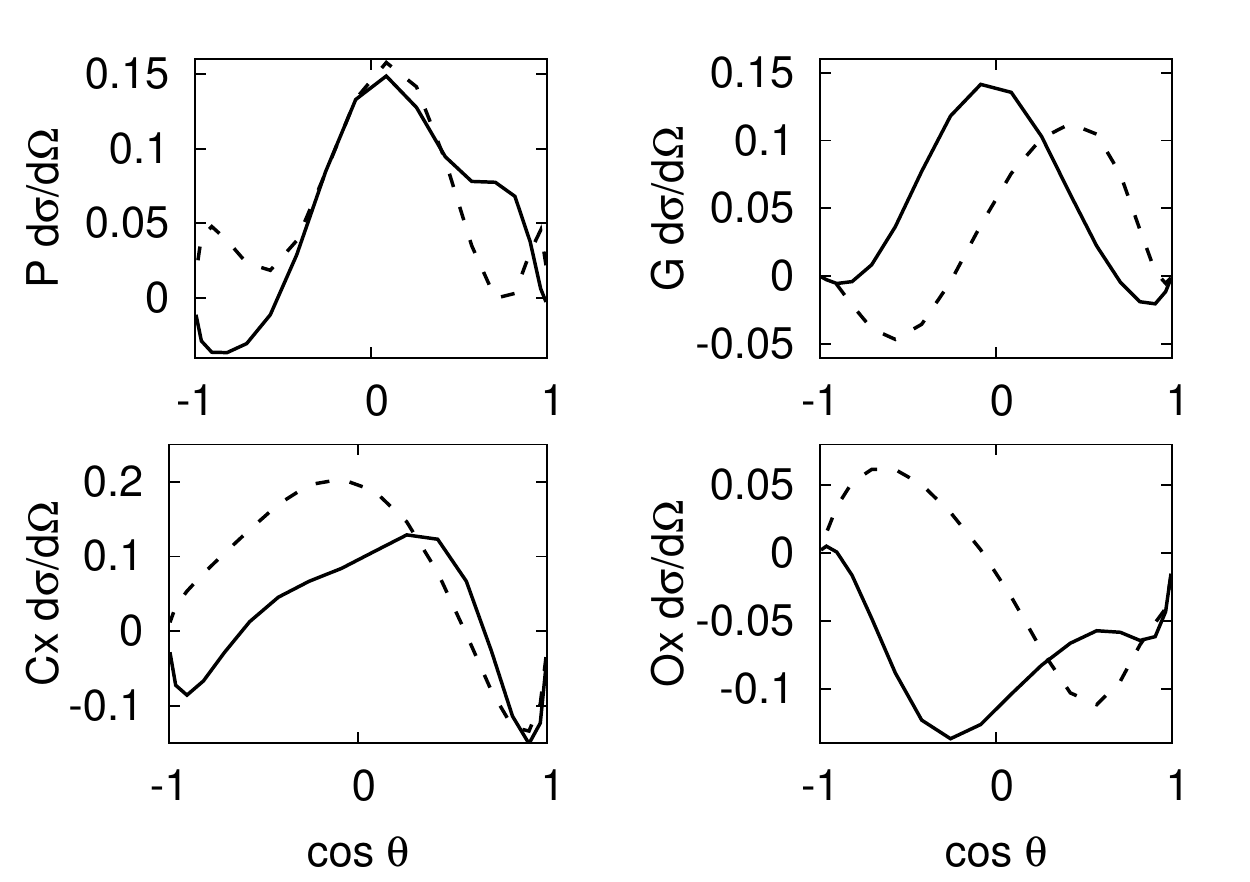}
\includegraphics[width=7.5cm]{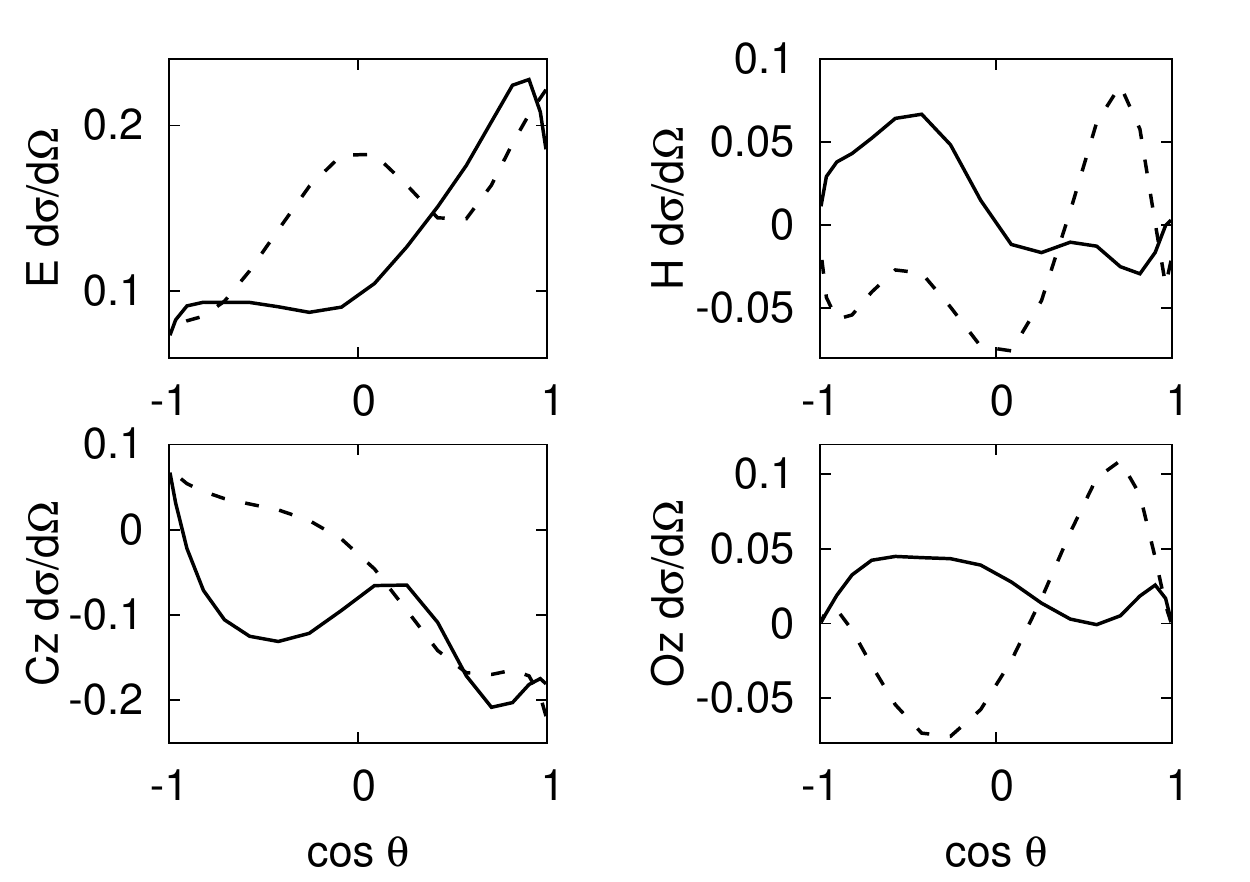}
\caption{\label{FigPredictions} Predictions from solutions II
(dashed) and III (solid) for polarization observables that are not
fitted $\{P,G,C_x,O_x\}$ (left) and $\{E,H,C_z,O_z\}$ (right) at 2
energies, $W=1554$~MeV (top) and $W=1765$~MeV (bottom). Together
with our current experimental data base $\{\sigma_0,\Sigma,T,F\}$,
each additional group of 4 observables would make a complete
experiment. }
\end{center}
\end{figure}

There are different possible reasons for the non-unique partial wave
amplitudes shown in Fig.~\ref{FigMultSE}. First of all, our data
base is currently limited to only four observables, which give a
rather incomplete data set. This was already observed in our study
with pseudo data and is demonstrated in Fig.~\ref{FigPredictions},
where we give predictions for non-fitted polarization observables of
a complete set. Certainly in a complete experiment with a well
chosen set of 8 observables we would have a guarantee that all
possible observables would be well described, if a complete set is
well fitted.
If we now compare the predictions of non-fitted observables in the
experimental data analysis (Fig.~\ref{FigPredictions}) with the
predictions in the pseudo data analysis
(Fig.~\ref{Predictions-pseudo1}), we find large discrepancies in the
observables $P, G$ and $H$. This arises most likely from the much
larger uncertainties in the experimental data and from also possible
systematic errors. Presumably, an inclusion of data from just one of
them would again lead to a very good description of all observables
with a data set of less than 8 observables and without any recoil
polarization.

\clearpage

\section{Summary and Conclusions}
 \label{sec:conclusions}

Partial wave series of meson photoproduction in the resonance region
away from threshold converge rather slowly. Low partial waves,
$L\lesssim 4$ are usually dominated by baryon resonance excitations,
while higher partial waves obtain still significant contributions by
Born terms and $t$-channel vector meson exchanges. Therefore, an
unconstrained truncated partial wave analysis with variation of all
partial waves up to a maximal angular momentum $L_{max}$ will not
converge to unique solutions. With $L_{max}=5$ a consistent data
base can well be fitted, but an increase by one more angular
momentum can produce strong changes in lower partial waves. Further
increase of partial waves will certainly lead to even better
description of the data, but it finally leads to the continuum
ambiguity, where only the four underlying amplitudes are determined,
but the partial waves become non unique. Any phase transformation
with energy and angle dependent phases will lead to a new (infinite)
series of partial waves, where all of such sets can describe the
data base equally well.

Therefore, any PWA must be constrained. The most often used
constraint is taken from a model that is fitted to the data in an
energy dependent way. Such a model will serve both for initial
solutions in the SE fits and for higher partial waves. Such
approaches are certainly strongly model dependent.

In this work we have applied analytical constraints from fixed-$t$
dispersion relations in partial wave analysis of eta photoproduction
data from threshold $W_{\eta,thr}=1487$~MeV up to $W=1850$~MeV.
Following the ideas in the Karlsruhe-Helsinki $\pi N$ PWA, we have
developed a new method for partial wave analysis of meson
photoproduction data. In an iterative procedure we perform fixed-$t$
amplitude analyses using the Pietarinen expansion method and
single-energy partial wave analyses with a mutual support and
constraint of each others, leading to a convergent partial wave
solution after about three iterations.

First, in an application with pseudo data, generated by the EtaMAID
model, we have demonstrated that the method works and converges to
the underlying partial waves of the model, at least in the ideal
case of a complete set of 8 observables. The analysis of an
incomplete set of the 4 observables consisting of the differential
cross section, the single-spin target asymmetry $T$ and the beam
target double polarization asymmetry $F$, leads to ambiguities which
can be resolved adding another one or two selected observables.

Second,  we have analyzed recent experimental data, which are still
limited to the incomplete set of 4 observables described above. We
find solutions for multipoles up to $L = 5$ which are continuous in
energy. However, remaining ambiguities are found in particular at
higher energies. New data from ELSA, JLab and MAMI expected in the
near future will help to resolve these ambiguities.

\begin{acknowledgments}
This work was supported by the Deutsche Forschungsgemeinschaft (SFB
1044).
\end{acknowledgments}

\newpage
\begin{appendix}

\section{Expansion of invariant amplitudes in terms of CGLN amplitudes}
\label{app:AtoF}

The covariant amplitude $A_i$ can be expressed by the CGLN
amplitudes $F_i$ as follows:
\begin{eqnarray}
A_1 &=&  {\mathcal N} \bigg\{ \frac{W+m_N}{W-m_N}\,{F}_1 -
(E_f+m_N)\,\frac{{F}_2}{q} + \frac{m_N(t-\mu^2)}{(W-m_N)^2}\,\frac{{F}_3}{q} \\
&& + \frac{m_N(E_f+m_N)\,(t-\mu^2)}{W^2-m_N^2}\,\frac{{F}_4}{q^2} \bigg\}\, ,\nonumber \\
A_2 &=& \frac{{\mathcal N}}{W-m_N} \left\{ \frac{{F}_3}{q} -
(E_f+m_N)\,\frac{{F}_4}{q^2} \right\}\,,
\vspace{0.3cm} \nonumber \\
A_3 &=&  \frac{{\mathcal N}}{W-m_N}  \bigg\{ F_1 + (E_f+m_N)
\frac{F_2}{q} + \left( W+m_N + \frac{t-\mu^2}{2(W-m_N)} \right) \frac{F_3}{q}\nonumber \\
 && + \left( W-m_N + \frac{t-\mu^2}{2(W+m_N)}\right)\, (E_f+m_N)\,\frac{F_4}{q^2} \bigg\}\,, \nonumber \\
A_4 &=&  \frac{{\mathcal N}}{W-m_N} \bigg\{F_1 +
(E_f+m_N)\,\frac{F_2}{q} + \frac{t-\mu^2}{2(W-m_N)}\,\frac{F_3}{q} +
\frac{t-\mu^2}{2(W+m_N)}\, (E_f+m_N)\,\frac{F_4}{q^2}
\bigg\}\,,\nonumber
\end{eqnarray}
where ${\mathcal N} = 4\pi/\sqrt{(E_i+m_N)\,(E_f+m_N)}$.\\

\section{Expansion of CGLN amplitudes in terms of invariant amplitudes
}\label{app:FtoA} The CGLN amplitudes are obtained from the
invariant amplitudes $A_i$ by the following equations
\cite{Dennery:1961zz,Berends:1967vi}:
\begin{eqnarray}
\label{eq:CGLN1} {F}_1  &=& \frac{W-m_N}{8\pi\,W}\,
\sqrt{(E_i+m_N)(E_f+m_N)}
\big( A_1 +(W-m_N)\,A_4 - \frac{2m_N\nu_B}{W-m_N}\,(A_3-A_4)\big)\,,\nonumber \\
\label{eq:CGLN2} {F}_2  &=& \frac{W+m_N}{8\pi\,W}\,q\,
\sqrt{\frac{E_i-m_N}{E_f+m_N}}
\big( -A_1 + (W+m_N)\,A_4 - \frac{2m_N\nu_B}{W+m_N}\,(A_3-A_4)\big)\,, \nonumber \\
\label{eq:CGLN3} {F}_3  &=&
\frac{W+m_N}{8\pi\,W}\,q\,\sqrt{(E_i-m_N)(E_f+m_N)}
\big((W-m_N)\,A_2 + A_3-A_4\big)\,,  \nonumber \\
\label{eq:CGLN4} {F}_4  &=& \frac{W-m_N}{8\pi\,W}\,q^2
\,\sqrt{\frac{E_i+m_N}{E_f+m_N}} \big(-(W+m_N) \,A_2 +A_3 -
A_4\big)\,,
\end{eqnarray}
with $\nu_B=(t-\mu^2)/(4m_N)$.

\section{Observables expressed in CGLN amplitudes}

Spin observables expressed in CGLN amplitudes
\begin{eqnarray}
\sigma_{0}   & = &  \,\mbox{Re}\,\left\{ \fpf{1}{1} + \fpf{2}{2} +
\sin^{2}\theta\,(\fpf{3}{3}/2
                   + \fpf{4}{4}/2 + \fpf{2}{3} + \fpf{1}{4} \right. \nonumber \\
             &   &   \mbox{} \left. + \cos\theta\,\fpf{3}{4}) - 2\cos\theta\,\fpf{1}{2} \right\} \rho \\
\check{\Sigma} & = & -\sin^{2}\theta\;\mbox{Re}\,\left\{
\left(\fpf{3}{3} +\fpf{4}{4}\right)/2
                   + \fpf{2}{3} + \fpf{1}{4} + \cos\theta\,\fpf{3}{4}\right\}\rho \\
\check{T}      & = &  \sin\theta\;\mbox{Im}\,\left\{\fpf{1}{3} -
\fpf{2}{4} + \cos\theta\,(\fpf{1}{4} - \fpf{2}{3})
                   - \sin^{2}\theta\,\fpf{3}{4}\right\}\rho \\
\check{P}      & = & -\sin\theta\;\mbox{Im}\,\left\{ 2\fpf{1}{2} +
\fpf{1}{3} - \fpf{2}{4} - \cos\theta\,(\fpf{2}{3} -\fpf{1}{4})
                   - \sin^{2}\theta\,\fpf{3}{4}\right\}\rho \\
\check{E}      & = &  \,\mbox{Re}\,\left\{ \fpf{1}{1} + \fpf{2}{2} -
2\cos\theta\,\fpf{1}{2}
                   + \sin^{2}\theta\,(\fpf{2}{3} + \fpf{1}{4}) \right\}\rho \\
\check{F}      & = &  \sin\theta\;\mbox{Re}\,\left\{\fpf{1}{3} - \fpf{2}{4} - \cos\theta\,(\fpf{2}{3} - \fpf{1}{4})\right\}\rho \\
\check{G}      & = &  \sin^{2}\theta\;\mbox{Im}\,\left\{\fpf{2}{3} + \fpf{1}{4}\right\}\rho \\
\check{H}      & = &  \sin\theta\;\mbox{Im}\,\left\{2\fpf{1}{2} +
\fpf{1}{3} - \fpf{2}{4}
                   + \cos\theta\,(\fpf{1}{4} - \fpf{2}{3})\right\}\rho \\
\check{C}_{x'} & = &  \sin\theta\;\mbox{Re}\,\left\{\fpf{1}{1} -
\fpf{2}{2} - \fpf{2}{3} + \fpf{1}{4}
                   - \cos\theta\,(\fpf{2}{4} - \fpf{1}{3})\right\}\rho \\
\check{C}_{z'} & = & \,\mbox{Re}\,\left\{2\fpf{1}{2} -
\cos\theta\,(\fpf{1}{1} + \fpf{2}{2})
                   + \sin^{2}\theta\,(\fpf{1}{3} + \fpf{2}{4})\right\}\rho \\
\check{O}_{x'} & = & \sin\theta\;\mbox{Im}\,\left\{\fpf{2}{3} - \fpf{1}{4} + \cos\theta\,(\fpf{2}{4} - \fpf{1}{3})\right\}\rho \\
\check{O}_{z'} & = & - \sin^{2}\theta\;\mbox{Im}\,\left\{\fpf{1}{3} + \fpf{2}{4}\right\}\rho\\
\check{L}_{x'} & = & - \sin\theta\;\mbox{Re}\,\left\{\fpf{1}{1} -
\fpf{2}{2} - \fpf{2}{3} + \fpf{1}{4}
                   + \sin^{2}\theta\,(\fpf{4}{4} - \fpf{3}{3})/2 \right. \nonumber \\
             &   & \mbox{} \left. + \cos\theta\,(\fpf{1}{3} - \fpf{2}{4})\right\}\rho \\
\check{L}_{z'} & = &  \,\mbox{Re}\,\left\{2\fpf{1}{2} -
\cos\theta\,(\fpf{1}{1} + \fpf{2}{2})
                   + \sin^{2}\theta\,(\fpf{1}{3} + \fpf{2}{4} + \fpf{3}{4}) \right. \nonumber \\
             &   & \mbox{} \left. + \cos\theta\sin^{2}\theta\,(\fpf{3}{3} + \fpf{4}{4})/2 \right\}\rho \\
\check{T}_{x'} & = & -\sin^{2}\theta\;\mbox{Re}\,\left\{\fpf{1}{3} +
\fpf{2}{4} + \fpf{3}{4}
                   + \cos\theta\,(\fpf{3}{3} + \fpf{4}{4})/2 \right\}\rho \\
\check{T}_{z'} & = &  \sin\theta \;\mbox{Re}\, \left\{\fpf{1}{4} -
\fpf{2}{3}
                   + \cos\theta\,(\fpf{1}{3} - \fpf{2}{4}) \right. \nonumber \\
             &   & \mbox{} \left. + \sin^{2}\theta\,(\fpf{4}{4} - \fpf{3}{3})/2 \right\}\rho \\
&& \mbox{with}\;\, \check{\Sigma}={\Sigma}\,\sigma_0\;\, \mbox{etc.
and} \;\, \rho=q/k \,.
\end{eqnarray}

\section{Polarization observables in terms of helicity amplitudes}

The 16 polarization observables of pseudoscalar photoproduction fall
into four groups, single spin with unpolarized cross section
included, beam-target, beam-recoil and target-recoil observables.
The simplest representation of these observables is given in terms
of helicity amplitudes.

\begin{table}[ht]
\caption{Spin observables expressed by helicity amplitudes in the
notation of Barker~\cite{Barker} and Walker~\cite{Walker:1968xu}. A
phase space factor $q/k$ has been omitted in all expressions. The
differential cross section is given by $\sigma_0$ and the spin
observables $\check{O}_i$ are obtained from the spin asymmetries
$A_i$ by $\check{O}_i=A_i\,\sigma_0$.}
\begin{center}
\begin{tabular}{|c|c|c|c|}
\hline
 Observable & Helicity Representation  & Type  \\
\hline
$\sigma_0$     & $\frac{1}{2}(|H_1|^2 + |H_2|^2 + |H_3|^2 + |H_4|^2)$  &  \\
$\check{\Sigma}$ & Re$(H_1 H_4^* - H_2 H_3^*)$                           &  ${\cal S}$ \\
$\check{T}$      & Im$(H_1 H_2^* + H_3 H_4^*)$                           &   (single spin) \\
$\check{P}$      & $-$Im$(H_1 H_3^* + H_2 H_4^*)$                        &   \\
\hline
$\check{G}$      & $-$Im$(H_1 H_4^* + H_2 H_3^*)$                        &   \\
$\check{H}$      & $-$Im$(H_1 H_3^* - H_2 H_4^*)$                        &  ${\cal BT} $  \\
$\check{E}$      & $\frac{1}{2}(-|H_1|^2 + |H_2|^2 - |H_3|^2 + |H_4|^2)$ &   (beam--target)\\
$\check{F}$      & Re$(H_1 H_2^* + H_3 H_4^*)$                           &   \\
\hline
$\check{O_{x'}}$    & $-$Im$(H_1 H_2^* - H_3 H_4^*)$                        &   \\
$\check{O_{z'}}$    & Im$(H_1 H_4^* - H_2 H_3^*)$                           &  ${\cal BR}$ \\
$\check{C_{x'}}$    & $-$Re$(H_1 H_3^* + H_2 H_4^*)$                        &   (beam--recoil) \\
$\check{C_{z'}}$    & $\frac{1}{2}(-|H_1|^2 - |H_2|^2 + |H_3|^2 + |H_4|^2)$ &   \\
\hline
$\check{T_{x'}}$    & Re$(H_1 H_4^* + H_2 H_3^*)$                           &   \\
$\check{T_{z'}}$    & Re$(H_1 H_2^* - H_3 H_4^*)$                           &  ${\cal TR}$  \\
$\check{L_{x'}}$    & $-$Re$(H_1 H_3^* - H_2 H_4^*)$                        &   (target--recoil)\\
$\check{L_{z'}}$    & $\frac{1}{2}(|H_1|^2 - |H_2|^2 - |H_3|^2 + |H_4|^2)$  &   \\
\hline
\end{tabular}
\end{center}
\end{table}

\end{appendix}
\clearpage



\end{document}